\documentclass[final,leqno]{siamltex}
\usepackage{amsfonts,amsmath,empheq}
\usepackage{graphicx}
\usepackage{url}
\renewcommand{\exp}{{\rm e}^} 
\newtheorem{remark}{Remark}

\newcounter{step}
\newcommand{\brk}[1]{\left(#1\right)}          
\newcommand{\Brk}[1]{\left[#1\right]}          
\newcommand{\Abs}[1]{\left\vert #1 \right\vert}        
\newcommand{\Norm}[1]{\left\Vert #1 \right\Vert} 
\newcommand{\pd}[2]{\dfrac{\partial#1}{\partial#2}}
\newcommand{\deriv}[2]{\dfrac{d#1}{d#2}}

\title{Macroscopic limit of a one-dimensional model for
  aging fluids}

\author{David Benoit\footnotemark[2]  
\and Claude Le Bris \footnotemark[2] \footnotemark[3]
\and Tony Leli\`evre\footnotemark[2]}

\begin{document}

\maketitle

\renewcommand{\thefootnote}{\fnsymbol{footnote}}

\footnotetext[2]{  Ecole des
  Ponts and INRIA, 6 \& 8 Av. B. Pascal,
  77455 Marne-la-Vall\'ee, France.
  (\{benoitd,lebris,lelievre\}@cermics.enpc.fr)}
\footnotetext[3]{ Corresponding author}

\renewcommand{\thefootnote}{\arabic{footnote}}

\begin{abstract}
We study a one-dimensional equation arising in the
multiscale modeling of
some non-Newtonian fluids. At a given shear rate, the equation provides
the instantaneous mesoscopic response of the fluid, allowing to compute the
corresponding stress. In a simple setting, we study the well-posedness of the equation and
next the long-time behavior of its solution. In the limit of a response
of the fluid much faster than the time variations of the ambient shear rate, we
 derive some equivalent macroscopic differential equations that relate the
 shear rate and the stress. Our analytical conclusions are confronted to
 some numerical experiments. The latter quantitatively confirm our derivations.
\end{abstract}



\pagestyle{myheadings}
\thispagestyle{plain}
\markboth{MODELING OF AGING FLUIDS}{D. BENOIT, C. LE BRIS \& T.
  LELI\`EVRE}

\section{Introduction}
\label{sec:introduction}

\medskip

\paragraph{Generalities} The present contribution is devoted to the mathematical analysis of the equation
\begin{align} 
\label{edp:p}
  \pd{p}{t}(t,\sigma)  + {\dot\gamma}(t) \pd{p}{\sigma}(t,\sigma) =
    -\chi(\sigma) p(t,\sigma) + \brk{\int \chi(\sigma) p(t,\sigma) d\sigma} \delta_0 (\sigma).  
\end{align}
Equation~\eqref{edp:p} arises in the modeling of some non-Newtonian
fluid flows. Some details on the modeling will be given below. The
variable~$\sigma$ is  one-dimensional, varies on the real line~$\mathbb{R}$, and models a quantity
homogeneous to a stress (actually to 
a certain entry of the stress tensor). The variable~$t$
of course denotes the time, and equation~\eqref{edp:p} is supplied with some
initial condition $p(t=0,.)=p_0(.)$. The unknown real-valued function~$p(t,\sigma)$, solution
to~\eqref{edp:p}, satisfies the two properties: it is nonnegative
\begin{equation}
  \label{eq:p-positive}
  p(t,\sigma)\geq 0,\quad \hbox{\rm for all}\quad t\geq 0\quad\hbox{\rm and}\quad\sigma\in\mathbb{R},
\end{equation}
and normalized to one
\begin{equation}
  \label{eq:norm}
  \int_{-\infty}^{+\infty} p(t,\sigma)\,d\sigma=1,
\end{equation}
for all times~$t\geq 0$. The function~$p$ 
models the density of probability to have a certain
(elementary microscopic) stress~$\sigma$, at time~$t$, at a macroscopic space
position~$x$. The actual, deterministic stress within the fluid is thus given by $\displaystyle \tau=\int\sigma\,p(t,\sigma)\,d\sigma$, of Equation~\eqref{def:tau} below. Equation~\eqref{edp:p} is thus implicitly parameterized by
this position~$x$ (thus the multiscale nature of the problem, as will be
seen below). The function~${\dot\gamma}$, also a function of  time, is assumed
given. It models the shear rate at the position~$x$, under which we
wish to compute, using~\eqref{edp:p}, the mesoscopic response of the
fluid. The notation~${\dot\gamma}$ is traditional in Fluid Mechanics, hence its use here.
 In~\eqref{edp:p}, we denote by~$\chi$
the characteristic function
$$\displaystyle\chi={\rm 1\mskip-4mu l}_{\mathbb{R}\backslash[-\sigma_c,\sigma_c]}$$
where $\sigma_c>0$ a scalar positive parameter, fixed once and
for all. It models some local threshold value of the stress, which plays a
crucial role in the modeling. As is usual, we denote by $\delta_0$ the
Dirac mass at zero.  From the definitions of~$\chi$ and~$\delta_0$, it is
immediately seen that, at least formally (and this will indeed be made rigorous, see Lemmata~\ref{lem:max} and~\ref{lem:mass} in Section~\ref{sec:properties} below), equation~\eqref{edp:p} preserves in time the two
properties~\eqref{eq:p-positive} and~\eqref{eq:norm}. 

 Two quantities are typically computed
using the solution~$p$ to~\eqref{edp:p}: first the so-called
\emph{fluidity} 
\begin{equation} 
\label{def:f}
  f(t)=\int \chi(\sigma)  p(t,\sigma)\,d\sigma
\end{equation}
and next the (real-valued)  stress 
\begin{equation}
\label{def:tau}
  \tau(t)=\int \sigma p(t,\sigma) d\sigma.
\end{equation}
Our purpose in this article is to mathematically study
equation~\eqref{edp:p} (in terms of existence and uniqueness of the
solution~$p$, properties and long-time behavior of that solution) and
to derive a macroscopic equation equivalent to this equation. By
\emph{macroscopic} equation,  we
mean an equation (actually a differential equation, or a system of
differential equations) that directly relates the shear rate~${\dot\gamma}$,
the fluidity~$f$ and the stress~$\tau$ without the explicit need to
compute~$p$.  In these macroscopic equations, the scalar $f(t)$ 
will be the inverse of the mechanical relaxation time, thus its name
``fluidity''. We will be able, in particular, to obtain a macroscopic
equation which is close to models that have been proposed for aging
fluids~\cite{derec-01,Picard2002}, see the discussion at the end of Section~\ref{sec:passage}.

\medskip

\paragraph{Some elements on the modeling} 
Equation~\eqref{edp:p} is the simplest possible form of an equation
describing the mesoscopic behavior of a complex fluid, such as a concentrated
suspension, or more generically  a soft amorphous material, with properties
intermediate between those of a fluid and those of a solid. These materials exhibit a highly non-Newtonian behavior and may give
rise to a macroscopic yield stress.
 
At low stress, such a material behaves in an elastic way. But above a
certain stress threshold, here denoted by the critical value~$\sigma_c$,
one observes a relaxation toward a completely relaxed state. This
behavior is modeled by equation~\eqref{edp:p}. The probability of
finding the fluid in the state of stress~$\sigma$ at time~$t$ evolves in
time for two different reasons: the term $\displaystyle {\dot\gamma}(t)
\pd{p}{\sigma}$ models  the modification of the stress induced by the
existence of the shear rate, while the term  $\displaystyle-\chi
p+\brk{\int \chi p} \delta_0 $ encodes the relaxation  toward zero of the part of the
stress above the
threshold~$\sigma_c$. From a probabilistic viewpoint, the stochastic
process $(\Sigma_t)_{t \ge0}$ associated to the Fokker-Planck equation~\eqref{edp:p} evolves
deterministically when $|\Sigma_t| \le \sigma_c$ and jumps
to zero with an exponential rate $1$ when $|\Sigma_t| \ge
\sigma_c$. The process $(\Sigma_t)_{t \ge0}$ belongs to the class of
{\em piecewise-deterministic Markov processes}, which have 
been introduced in the probabilistic literature in the 1980's for
biological modeling for example. In particular, coupling arguments have
been proposed to study the longtime behavior of such processes (see \cite{Bardet2013}). 
We argue on the Fokker-Planck equation and proceed differently.
The argument we are using here to study the longtime behavior is purely
deterministic in nature, and is based on a delay equation related to
the Fokker-Planck equation~\eqref{edp:p}.

We would like to mention that for more realistic models, a third
phenomenon is
typically at play, in addition to the stress induced by the ambient fluid,
and to the relaxation to zero. All states of stress are not independent of one
another, and they may also depend on the state of stress at neighboring
points within the fluid. A certain redistribution of the stress  therefore always occurs. This redistribution can be encoded
in various ways, depending on some more detailed elements of
modeling. In the so-called H\'ebraud-Lequeux model introduced in the
seminal article~\cite{Hebraud1998} (and then extensively studied
mathematically in the works~\cite{Cances2005, cances2006, LeBris}), the
redistribution is performed by a diffusion term in the stress space, at
the given location~$x$ in the ambient physical space, and the
complete equation thus writes
\begin{align} 
\label{eq:hl}
   \pd{p}t  + {\dot\gamma}(t) \pd{p}{\sigma} =  -\chi p + \brk{\int \chi p} \delta_0 (\sigma) + \alpha \brk{\int \chi p} \frac{\partial^2 p}{\partial \sigma^2}
\end{align}
where $\alpha>0$ is some parameter. In an alternative model introduced
by Bocquet and coll. in~\cite{Bocquet2009}, the redistribution is
achieved by some type of local ``convolution'' in the physical
space. The equation (we recall, set at the physical location~$x$) writes
\begin{eqnarray} 
\label{eq:limpartb}
  \pd {p}t (t,x,\sigma) + {\dot\gamma}(t,x) \pd{p}{\sigma}  (t,x,\sigma)
=  -\chi(\sigma) p  (t,x,\sigma)+ 
\brk{\int d \sigma' \chi(\sigma') p(t,x,\sigma')} \delta_0(\sigma).
\nonumber\\
+\int \int d x' d \sigma^\prime 
\chi(\sigma') \brk{ p(t,x',\sigma') p(t,x,\sigma - G(x,x') \sigma^\prime)-p(t,x',\sigma') p(t,x,\sigma) }.
\end{eqnarray}
with a function $G(x,x')$ related to the Green function of some
local Stokes-type problem.

The equation~\eqref{edp:p} which we study in the present article ignores
the redistribution phenomenon, which amounts to taking $\alpha=0$
in~\eqref{eq:hl} or $G\equiv 0$ in \eqref{eq:limpartb}. 
In the absence of such a simplification, we are unable to proceed with the main result of this article, which is the derivation of the macroscopic equation from our multiscale model. 
The well posedness result contained in our Theorem~\ref{th:exu}, on the other hand, also holds for~\eqref{eq:hl} whith $\alpha >0$ and has indeed been established some years ago in~\cite{Cances2005}.
Some more detailed comments on the modeling, as well as some formal
foundations of the model based on a system of interacting particles are
presented in~\cite{these}.

\medskip

\paragraph{Plan of our contribution} Our article is organized as follows. To start with, we study 
in Section~\ref{sec:ss} the stationary solutions to~\eqref{edp:p}. We
next show in Section~\ref{sec:exist-uniq} existence and uniqueness of
the solutions to the time-dependent
equation~\eqref{edp:p}. Our result is stated in Theorem~\ref{th:exu}. Section~\ref{sec:properties} follows,
establishing some useful properties of the solution. In order to understand the
macroscopic equivalent of equation~\eqref{edp:p} for a given shear
rate~${\dot\gamma}(t)$, which we assume varies slowly as compared to  the characteristic time of
equation~\eqref{edp:p}, we need to understand the long-time behavior of
the solution to~\eqref{edp:p}. We therefore study this behavior in
Sections~\ref{sec:tlc} and~\ref{sec:long-time}, respectively in the case
of a constant shear rate~${\dot\gamma}(t)\equiv {\dot\gamma}_\infty$, and in the case of a slowly varying
shear rate~${\dot\gamma}(\epsilon t)$. The results are stated in
Theorems~\ref{th:tlc} and~\ref{th:tle}. We are then in position to derive, in
Section~\ref{sec:passage}, the macroscopic differential equations
equivalent to~\eqref{edp:p} in this limit, namely system~\eqref{eq:mac2}. Our final section,
Section~\ref{sec:numerics}, presents some numerical experiments which
confirm and illustrate our theoretical results.


\section{Stationary states}
\label{sec:ss}
We study in this section the stationary states of~\eqref{edp:p}. We
therefore assume that ${\dot\gamma}(t)\equiv{\dot\gamma}_\infty$ is a fixed scalar
and consider the solutions $p_\infty:\mathbb{R} \to \mathbb{R}$ to the following equation
\begin{align} \label{edp:py}
   {\dot\gamma}_\infty \deriv{p_\infty}{\sigma} = -\chi p_\infty + \brk{\int \chi p_\infty} \delta_0 \mbox{ in } \mathcal{D}'(\mathbb{R}).
\end{align}
Here and in the following, for a  subset $I \subset \mathbb{R}^d$,
$\mathcal{D}'(I)$ denotes the set of distributions on $I$.
By
convention, since the time-dependent version of the equation is linear and formally preserves
positiveness and the integral over the real line, we are only interested
in the stationary solutions $p_\infty$ that additionally satisfy~\eqref{eq:p-positive} and~\eqref{eq:norm}, that is,
\begin{equation}
  \label{eq:normalizations}
  p_\infty\in L^1(\mathbb{R}), \quad p_\infty\geq 0,\quad\hbox{\rm a.e. and}\quad \int p_\infty=1.
\end{equation}

\medskip

We have the following result:

\medskip

\begin{lemma} \label{lem:ss}
  When ${\dot\gamma}_\infty=0$, the solutions to~\eqref{edp:py}-\eqref{eq:normalizations} are exactly
  all nonnegative normalized densities with compact support in
  $[-\sigma_c,\sigma_c]$. When ${\dot\gamma}_\infty \neq 0$, there exists a
  unique solution $p_\infty$ to~\eqref{edp:py}-\eqref{eq:normalizations}. 
\end{lemma}

\medskip

\begin{proof}
In the case ${\dot\gamma}_\infty=0$, the equation~\eqref{edp:py} implies
$p_\infty=0$ in
$\mathcal{D}'(\mathbb{R}\backslash[-\sigma_c,\sigma_c])$, hence the
result. Up to a change of $p_\infty$ into $\sigma\rightarrow
p_\infty(-\sigma)$, we may, without loss of generality, consider only
the case ${\dot\gamma}_\infty>0$ for our proof. 
We first note that  $p_\infty$ defined by
\begin{align} \label{eq:py}
    p_\infty(\sigma) = 
  \begin{cases}
      0 & \mbox{if }  \sigma<0 \\
      \frac1{\sigma_c+{\dot\gamma}_\infty}& \mbox{if } 0<\sigma\le\sigma_c \\
      \frac1{\sigma_c+{\dot\gamma}_\infty}\exp{-(\sigma-\sigma_c)/{\dot\gamma}_\infty} 
      & \mbox{if }  \sigma_c<\sigma.
    \end{cases}
  \end{align}
is a solution to~\eqref{edp:py}, hence the existence result. 

\medskip

We now show uniqueness. By linearity, we assume that $p_\infty\in L^1(\mathbb{R})$ is a solution of~\eqref{edp:py} with $\int_{\mathbb{R}} p_\infty = 0$ and show that $p_\infty=0$.
Equation~\eqref{edp:py} implies ${\dot\gamma}_\infty\deriv{p_\infty}\sigma
+p_\infty=0$ in
$\mathcal{D}'(\mathbb{R}\backslash[-\sigma_c,\sigma_c])$. Because
$p_\infty\in L^1(\mathbb{R})$, this leads to
$p_\infty=\alpha \exp{-\frac\sigma{{\dot\gamma}_\infty}}$ a.e. on $(\sigma_c,\infty)$ with $\alpha$ a scalar
and  $p_\infty=0$ a.e. on $(-\infty,-\sigma_c)$.

In the case $\alpha=0$, this implies $\chi p_\infty=0$. Consequently,
\eqref{edp:py} rewrites  $\deriv{p_\infty}\sigma=0$ on the whole real
line. This readily implies $p_\infty=0$ since $\int_{\mathbb{R}}
p_\infty = 0$. 

In the case $\alpha\neq0$, we obtain that~\eqref{edp:py} writes
\begin{align*}
  {\dot\gamma}_\infty\deriv{p_\infty}\sigma = \alpha \brk{\int_{\sigma_c}^\infty  \exp{-\frac{\sigma'}{{\dot\gamma}_\infty}}d\sigma'} \delta_0\mbox{ in }\mathcal{D}'(-\infty,\sigma_c)
\end{align*}
so that $p_\infty =\frac\alpha{{\dot\gamma}_\infty}
\brk{\int_{\sigma_c}^\infty
  \exp{-\frac{\sigma'}{{\dot\gamma}_\infty}}d\sigma'}\, {\rm 1\mskip-4mu l}_{\mathbb{R}_+}(\sigma)$ a.e. on $(-\infty,\sigma_c)$. Using that $p_\infty=\alpha \exp{-\frac\sigma{{\dot\gamma}_\infty}}$ a.e. on $(\sigma_c,\infty)$ and $\int p_\infty=0$, we find $\alpha=0$ and thus $p_\infty=0$. This concludes the proof.
\hfill\end{proof}

\section{Existence and uniqueness}
\label{sec:exist-uniq}

This section is devoted to the proof of the following result:

\medskip

\begin{theorem}[Existence and uniqueness] 
\label{th:exu}
  Consider ${\dot\gamma}$ a function of  time that satisfies
\begin{align} \label{prop:gamma}
  {\dot\gamma} \in L_{\mathrm{loc}}^1(0,\infty) \text{ and } {\dot\gamma}\ge
  m_{\dot\gamma} \mbox{ a.e. where }
  m_{\dot\gamma}>0 \mbox{ is a fixed scalar}.
\end{align} 
Denote by $\gamma(t)=\int_0^t{\dot\gamma}(s)\,ds$. Consider $p_0\in
L^1(\mathbb{R})$.
Then, for any $T>0$, there exists a unique $p$ in ${C}^0([0,T);L^1)$
such that $p(0,\sigma)=p_0(\sigma)$ for almost all $\sigma\in\mathbb{R}$
and such that \eqref{edp:p} holds for $p$ in the sense of distributions on $\brk{0,T}\times\mathbb{R}$.
In addition, introduce
\begin{align} \label{def:A}
  A(t)=\int \chi(\sigma) p_0(\sigma-\gamma(t))\exp{-\int_0^t \chi(\sigma-\gamma(t)+\gamma(u))du}   d\sigma
\end{align}
and $\phi$ defined by induction on $k\in \mathbb{N}^*$ as follows 
\begin{align} \label{def:phi}
  &\phi(t)= A(t) \nonumber\\
  &+\begin{cases}
   0 & \mbox{when } t\in \brk{0,\gamma^{-1}(\sigma_c)}\\
    \int_0^{\gamma^{-1}(\gamma(t)-\sigma_c)} \phi(s) \exp{-t+\gamma^{-1}(\gamma(s)+\sigma_c)}ds &\mbox{when }t\in \brk{\gamma^{-1}(k\sigma_c),\gamma^{-1}((k+1)\sigma_c)}.
 \end{cases}
\end{align}
Both $A$ and $\phi$ belong to $L^\infty(0,T)$.  Then, the solution $p$ to~\eqref{edp:p} is
explicitly given by
\begin{eqnarray} \label{for:p}
  p(t,\sigma)& =&
     p_0(\sigma-\gamma(t)) \exp{-\int_0^t \chi(\sigma-\gamma(t)+\gamma(u))du} \nonumber\\
&&+ \dfrac{\phi\circ\gamma^{-1}(\gamma(t)-\sigma)}{{\dot\gamma}\circ\gamma^{-1}(\gamma(t)-\sigma)}
\exp{-\int_{\gamma^{-1}(\gamma(t)-\sigma)}^t \chi(\sigma -\gamma(t) + \gamma(u))du}
{\rm 1\mskip-4mu l}_{\brk{0,\gamma(t)}}(\sigma)
\end{eqnarray}
and $f$ defined by~\eqref{def:f} is equal to~$\phi$:
\begin{align} \label{eq:fphi}
  f=\phi \mbox{ a.e. on } (0,T).
\end{align}
\end{theorem}

\begin{remark}
  The above results also hold for ${\dot\gamma}<-m_{\dot\gamma}$
  negative. However, it is
  unclear how to extend these results if ${\dot\gamma}$ is allowed to vanish. 
\end{remark}

\medskip

\begin{remark}
  We will see in the next section that if the initial condition $p_0$ is
  nonnegative and normalized, then this property is preserved in time
  for the solution~$p$.
\end{remark}

\medskip

\begin{proof}
We first note that
\begin{align*} 
  \gamma(t)=\int_0^t {\dot\gamma}(s)ds
\end{align*}
is a strictly increasing continuous function of the time because of~\eqref{prop:gamma}. 
Throughout this proof, we assume
$T>\gamma^{-1}(\sigma_c)$. When $T\leq\gamma^{-1}(\sigma_c)$, the
arguments are similar and actually simpler. We first show uniqueness,
then $p$ given by~\eqref{for:p} belongs to ${C}^0([0,T);L^1)$ and is a solution of \eqref{edp:p} in $\mathcal{D}'((0,T)\times\mathbb{R})$.

\refstepcounter{step}
\subparagraph*{Step \arabic{step}: Uniqueness}
Equation~\eqref{edp:p} is linear, we therefore consider a  solution $p
\in {C}^0([0,T);L^1)$
associated to the zero initial condition~$p_0=0$ and we intend to show that $p=0$.
Denote by
\begin{align*} 
  \tilde p(t,\xi) = p(t,\xi+\gamma(t)) \exp{\int_0^t \chi(\xi+\gamma(u))du} .
\end{align*}
We now show
\begin{align} \label{eq:tildep}
  \pd{\tilde p}t  (t,\xi)= f(t) \delta_{-\gamma(t)}(\xi)
\exp{\int_0^t \chi(-\gamma(t)+\gamma(u))du} 
\quad \mbox{ in } \mathcal{D}'(\brk{0,T}\times\mathbb{R})
\end{align}
with $f$ defined from~$p$ by~\eqref{def:f}.
We have, for all $\psi\in \mathcal{D}(\brk{0,T}\times\mathbb{R})$,
(where $\mathcal{D}(\brk{0,T}\times\mathbb{R})$ denotes the set of ${C}^\infty$
functions with compact support in $\brk{0,T}\times\mathbb{R}$)
\begin{align}
  -\int_0^T\int_{\mathbb{R}} \tilde p \pd\psi t 
&= -\int_0^T\int_{\mathbb{R}} p(t,\xi+\gamma(t)) \exp{\int_0^t \chi(\xi+\gamma(u))du} \pd\psi t(t,\xi) d\xi dt \nonumber\\
&=-\int_0^T\int_{\mathbb{R}} p(t,\sigma)  \pd\psi t(t,\sigma-\gamma(t))
\exp{\int_0^t \chi(\sigma-\gamma(t)+\gamma(u))du}  d\sigma dt.
\label{fv:tildep}
\end{align}
For $n\in \mathbb{N}$, denote now by~$\rho^n$ a mollifier, $\chi^n = \rho^n * \chi$ and
\begin{align*}
  \eta^n(t,\sigma) = \psi(t,\sigma-\gamma(t)) \exp{\int_0^t \chi^n(\sigma-\gamma(t)+\gamma(u))du}.
\end{align*}
The fact that $p$ is solution to \eqref{edp:p} in $\mathcal{D}'((0,T)\times\mathbb{R})$ yields
 \begin{align*}
  -\int_0^T\int_{\mathbb{R}} p \brk{\pd{\eta^n}t + {\dot\gamma} \pd{\eta^n}\sigma - \chi \eta^n} 
&= \int_0^T f(t) \eta^n(t,0) dt. 
 \end{align*}
This rewrites
\begin{align}\label{fvn:tildep}
  -\int_0^T\int_{\mathbb{R}} p(t,\sigma)  \pd\psi t(t,\sigma-\gamma(t))
\exp{\int_0^t \chi^n(\sigma-\gamma(t)+\gamma(u))du}  d\sigma dt \nonumber\\
+ \int_0^T\int_{\mathbb{R}} p(t,\sigma)  \eta^n(t,\sigma)\brk{\chi-\chi^n}(\sigma)  d\sigma dt\nonumber\\
=\int_0^T f(t) \psi(t,-\gamma(t)) \exp{\int_0^t \chi^n(-\gamma(t)+\gamma(u))du}dt.
\end{align}
As $n$ goes to infinity, $\chi^n$ converges to $\chi$ in $L_{\mathrm{loc}}^1(\mathbb{R})$ and for almost all $t\in (0,T), \sigma\in\mathbb{R}$, 
\begin{align*}
  \int_0^t \chi^n(\sigma-\gamma(t)+\gamma(u))du \rightarrow \int_0^t \chi(\sigma-\gamma(t)+\gamma(u))du.
\end{align*}
Because $p$ and $\psi$ respectively belong to $L^\infty( (0,T),L^1)$
and $\mathcal{D}((0,T)\times\mathbb{R})$, all terms of
\eqref{fvn:tildep} are bounded from below and from above by an
integrable function independent of $n$. We apply the dominated convergence theorem to pass to the limit in \eqref{fvn:tildep}
\begin{align*}
-\int_0^T\int_{\mathbb{R}} p(t,\sigma)  \pd\psi t(t,\sigma-\gamma(t))
\exp{\int_0^t \chi(\sigma-\gamma(t)+\gamma(u))du}  d\sigma dt \nonumber\\
=\int_0^T f(t) \psi(t,-\gamma(t)) \exp{\int_0^t \chi(-\gamma(t)+\gamma(u))du}dt,
\end{align*}
hence~\eqref{eq:tildep}, using~\eqref{fv:tildep}.

Define $t^*=\gamma^{-1}(\sigma_c)$. We now show that  $\tilde p=0$
in $L^\infty(0,t^*;L^1)$, this will prove that $p=0$ in $L^\infty(0,t^*;L^1)$.

From~\eqref{eq:tildep}, we deduce that $\pd{\tilde p}t = 0$ in $\mathcal{D}'(\tilde\Omega)$ with
\begin{align*}
 \tilde\Omega = \brk{0,t^*}\times\brk{(-\infty,-\gamma(t^*))\cup(0,\infty)}.
\end{align*}
Using that $p_0=0$, we find $\tilde p = 0$ in $\mathcal{D}'(\tilde \Omega)$ and therefore in $L^\infty(0,t^*;L^1((-\infty,-\gamma(t^*))\cup(0,\infty)))$. This implies
$p=0$ in $L^\infty(0,t^*;L^1\brk{(-\infty,-\gamma(t^*))\cup(\gamma(t^*),\infty)})$.

In particular, since $\gamma(t^*)=\sigma_c$, we have, for all $t\in\brk{0,t^*}, \ f(t)=\int \chi p(t,\cdot)=0$ and thus, the equation~\eqref{eq:tildep} reads $\pd{\tilde p}t=0$ in $\mathcal{D}'(\brk{0,t^*}\times \mathbb{R})$. We deduce that $\tilde p=0$ and therefore that $p=0$ in $L^\infty(0,t^*;L^1)$.

Taking $t^*$ as initial time, we find $p=0$ in $L^\infty(t^*,2t^*;L^1)$ with the previous arguments. Iterating, we obtain $p=0$ in $L^\infty(0,T;L^1)$.
This concludes the proof of uniqueness.

Our next two steps are respectively devoted to proving that $p$ defined
by~\eqref{for:p} belongs to ${C}([0,T);L^1)$ and that it satisfies equation~\eqref{edp:p}.

\refstepcounter{step}
\subparagraph*{Step \arabic{step}: Regularity of expression~\eqref{for:p}}
\label{s:espace}
First, the function $A$ defined by \eqref{def:A} belongs to $L^\infty(0,T)$ (with $\Norm{A}_{L^\infty(0,T)}\le \Norm{p_0}_{L^1}$) and therefore $\phi$ defined by the recurrence relation  \eqref{def:phi} also belongs to $L^\infty(0,T)$.
This implies, for almost all $t\in[0,T)$,
\begin{align*}
  \int \Abs{p(t,\sigma)}d\sigma \le \Norm{p_0}_{L^1} + \frac{\gamma(T)}{m_{\dot\gamma}}\Norm{\phi}_{L^\infty(0,T)},
\end{align*}
that is $p$ defined by \eqref{for:p} belongs to $L^\infty(0,T;L^1)$.
Denote 
\begin{align*}
  p_{11}(t,\sigma)&=p_0(\sigma-\gamma(t))\\
  p_{12}(t,\sigma)&=\int_0^t \chi(\sigma-\gamma(t)+\gamma(u))du\\
  p_{21}(t,\sigma)&=\dfrac{\phi\circ\gamma^{-1}(\gamma(t)-\sigma)}{{\dot\gamma}\circ\gamma^{-1}(\gamma(t)-\sigma)} {\rm 1\mskip-4mu l}_{\brk{0,\gamma(t)}}(\sigma) {\rm 1\mskip-4mu l}_{\mathbb{R}_+^*}(t)\\
   p_{22}(t,\sigma)&=\int_{\gamma^{-1}(\gamma(t)-\sigma)}^t \chi(\sigma -\gamma(t) + \gamma(u))du 
\end{align*}
so that $p=p_{11}  \exp{-p_{12}} + p_{21} \exp{-p_{22}}$.
We now check that $p_{11}$ and $p_{21}$ belong to ${C}^0([0,T);L^1)$ and
$p_{12}$ and $p_{22}$ belong to ${C}^0([0,T);L^\infty)$. Using that
$x\mapsto \exp{-x}$ is 1-Lipschitz on $[0,\infty)$, this will prove that $p\in{C}^0([0,T);L^1)$ .  

Consider $\epsilon>0,\ t\in[0,T)$ and $h$ such that $t+h\in[0,T)$. 
By density of $\mathcal{D}(\mathbb{R})$ in $L^1(\mathbb{R})$, there exists  $p_\epsilon\in \mathcal{D}(\mathbb{R})$ such that 
\begin{align*}
 \Norm{p_\epsilon-p_0}_{L^1}&<\epsilon.
\end{align*}
We obtain 
\begin{align*}
 \Norm{p_{11}(t+h,\cdot)-p_{11}(t,\cdot)}_{L^1} &= \int_{\mathbb{R}} \Abs{p_0(\sigma - \gamma(t+h) ) - p_0(\sigma - \gamma(t))} d\sigma\\
&\le 2  \Norm{p_\epsilon-p_0}_{L^1} +  \int_{\mathbb{R}} 
 \Abs{p_\epsilon(\sigma - \gamma(t+h) ) - p_\epsilon(\sigma - \gamma(t))} d\sigma\\
&\le 2 \epsilon  + \int_{\mathbb{R}} g_\epsilon^h(\sigma)d\sigma
\end{align*}
with $g_\epsilon^h(\sigma)=\Abs{p_\epsilon(\sigma - \gamma(t+h) ) - p_\epsilon(\sigma - \gamma(t))} $.
 Moreover, by continuity of $p_\epsilon$ and $\gamma$, we have, for all $\sigma\in\mathbb{R}$,
\begin{align*}
   g_\epsilon^h(\sigma)\rightarrow 0 \mbox{ as } h\rightarrow 0 
\end{align*}
and $g_\epsilon^h\le 2\Norm{p_\epsilon}_{L^\infty}$ on a bounded
interval (depending on the support of $p_\epsilon$ and
on~$\gamma(T)$). Using Lebesgue dominated convergence Theorem, we deduce that there exists $\eta_\epsilon >0$ such that for all $h\in\brk{-\eta_\epsilon, \eta_\epsilon}$, 
\begin{align*}
 \int_{\mathbb{R}} g_\epsilon^h(\sigma)d\sigma\le\epsilon
\end{align*}
so that
\begin{align*}
 \Norm{p_{11}(t+h,\cdot)-p_{11}(t,\cdot)}_{L^1}&\le3\epsilon.
\end{align*}
This yields $p_{11}\in{C}^0([0,T);L^1)$.

 We now turn to $p_{12}$. For almost all $\sigma\in\mathbb{R}$, we have
\begin{align*}
&  \Abs{p_{12} (t+h,\sigma) - p_{12}(t,\sigma)}\\
&\quad\le \int_0^t \Abs{\chi(\sigma-\gamma(t+h)+\gamma(u)) - \chi(\sigma-\gamma(t)+\gamma(u))} du \\
&\qquad+  \int_t^{t+h} \chi(\sigma-\gamma(t+h)+\gamma(u))du\\
&\quad \le \frac1{m_{\dot\gamma}} \int_\sigma^{\sigma+\gamma(t)} \Abs{\chi(v-\gamma(t+h)) -\chi(v-\gamma(t))  }dv +h\\
&\quad \le \frac1{m_{\dot\gamma}} \int_{\mathbb{R}} \Abs{\chi(v-\gamma(t+h)) -\chi(v-\gamma(t))  }dv +h.
\end{align*}
This leads to
\begin{align*}
  \Norm{p_{12}(t+h,\cdot)-p_{12}(t,\cdot)}_{L^\infty} \le  \frac2{m_{\dot\gamma}} \brk{\gamma(t+h)-\gamma(t)} + h
\end{align*}
which yields $p_{12}\in{C}^0([0,T);L^\infty)$.

For any $t\in(0,T)$ and $h$ such that $t+h\in[0,T)$, we have
\begin{align*}
&   \int_{\mathbb{R}} \Abs{p_{21}(t+h,\sigma) - p_{21}(t,\sigma)} d\sigma\\
&\quad\le \int_0^{\gamma(t)}  \Abs{
\dfrac{\phi\circ\gamma^{-1}(\gamma(t+h)-\sigma)}{{\dot\gamma}\circ\gamma^{-1}(\gamma(t+h)-\sigma)} 
-\dfrac{\phi\circ\gamma^{-1}(\gamma(t)-\sigma)}{{\dot\gamma}\circ\gamma^{-1}(\gamma(t)-\sigma)}} d\sigma \\
&\qquad+\Abs{\int_{\gamma(t)}^{\gamma(t+h)} \dfrac{\phi\circ\gamma^{-1}(\gamma(t+h)-\sigma)}{{\dot\gamma}\circ\gamma^{-1}(\gamma(t+h)-\sigma)} d\sigma}.
\end{align*}
Since $\phi$ belongs to $L^\infty(0,T)$, $\phi\circ\gamma^{-1}$ belongs to $L^\infty(0,\gamma(T))\subset L^1(0,\gamma(T))$ and we introduce a sequence $\theta_n$ in $\mathcal{D}(0,\gamma(T))$ such
that 
\begin{align*}
  \theta_n \rightarrow \phi\circ\gamma^{-1} \mbox{ in } L^1(0,\gamma(T))
\end{align*}
and obtain
\begin{align*}
&   \int_{\mathbb{R}} \Abs{p_{21}(t+h,\sigma) - p_{21}(t,\sigma)} d\sigma\\
&\quad\le \frac2{m_{\dot\gamma}} \Norm{\theta_n- \phi\circ\gamma^{-1}}_{L^1(0,\gamma(T))}\\
&\qquad+\int_0^{\gamma(t)}  \Abs{
\dfrac{\theta_n(\gamma(t+h)-\sigma)}{{\dot\gamma}\circ\gamma^{-1}(\gamma(t+h)-\sigma)} 
-\dfrac{\theta_n(\gamma(t)-\sigma)}{{\dot\gamma}\circ\gamma^{-1}(\gamma(t)-\sigma)}} d\sigma\\ 
&\qquad+\frac{\gamma(t+h)-\gamma(t)}{m_{\dot\gamma}} \Norm{\phi}_{L^\infty(0,T)}.
\end{align*}
Using the dominated convergence theorem for the second term, this implies that $\int_{\mathbb{R}} \Abs{p_{21}(t+h,\sigma) - p_{21}(t,\sigma)} d\sigma $ vanishes with~$h$.
We have obtained that $p_{21}$ belongs to ${C}^0((0,T);L^1)$. The continuity holds also at $t=0$ because for all $h\in(0,T)$
\begin{align*}
  \int_{\mathbb{R}} \Abs{p_{21}(h,\sigma) - 0} d\sigma
&\le \int_0^{\gamma(h)}  \Abs{
\dfrac{\phi\circ\gamma^{-1}(\gamma(h)-\sigma)}{{\dot\gamma}\circ\gamma^{-1}(\gamma(h)-\sigma)} } d\sigma\\
&\le \frac{\gamma(h)}{m_{\dot\gamma}} \Norm{\phi}_{L^\infty(0,T)}.
\end{align*}

Finally, the function $p_3=p_{12}-p_{22}$ is  in ${C}^0((0,T);L^\infty)$ (and so is $p_{22}$). Indeed, for any $t\in[0,T)$ and $h$ such that $t+h\in[0,T)$, for almost all $\sigma\in\mathbb{R}$,
\begin{align*}
&  \Abs{p_3(t+h,\sigma) - p_3(t,\sigma)}\\
&\quad\le \int_0^{\gamma^{-1}(\gamma(t)-\sigma)} \Abs{\chi(\sigma-\gamma(t+h)+\gamma(u)) - \chi(\sigma-\gamma(t)+\gamma(u))} du \\
&\qquad+  \int_{\gamma^{-1}(\gamma(t)-\sigma)}^{\gamma^{-1}(\gamma(t+h)-\sigma)} \chi(\sigma-\gamma(t+h)+\gamma(u))du\\
\end{align*}
so that, with the change of variable $v=\sigma+\gamma(u)$,
\begin{align*}
&  \Norm{p_3(t+h,\cdot)-p_3(t,\cdot)}_{L^\infty}\\
&\quad \le \frac1{m_{\dot\gamma}}\brk{ \int_{\mathbb{R}} \Abs{\chi(v-\gamma(t+h)) -\chi(v-\gamma(t))  }dv + \int_{\gamma(t)}^{\gamma(t+h)}dv}\\
&\quad \le \frac3{m_{\dot\gamma}} \brk{\gamma(t+h)-\gamma(t)}.
\end{align*}
This concludes the proof of the continuity in time of~$p$, with values in~$L^1$.

\refstepcounter{step}
\subparagraph*{Step \arabic{step}: Expression~\eqref{for:p} satisfies \eqref{edp:p}}
We first show that $p$ defined by~\eqref{for:p} satisfies
\begin{align} \label{eq:pphi}
     \pd{p}t  + {\dot\gamma}(t) \pd{p}{\sigma} =  -\chi p + \phi \delta_0 (\sigma) 
\quad \mbox{ in } \mathcal{D}'(\brk{0,T}\times\mathbb{R}).
\end{align}
For all $\psi\in \mathcal{D}(\brk{0,T}\times\mathbb{R})$,
\begin{align*}
&-\int_0^T \int_{\mathbb{R}} p \brk{\pd\psi t + {\dot\gamma} \pd\psi\sigma-\chi \psi} \\
&= -\int_0^T \int_{\mathbb{R}} 
p_0(\sigma-\gamma(t)) \exp{-\int_0^t \chi(\sigma-\gamma(t)+\gamma(u))du}
\brk{\pd\psi t + {\dot\gamma} \pd\psi\sigma - \chi \psi}(t,\sigma)d\sigma dt \\
&\quad- \int_0^T \int_0^{\gamma(t)}
 \dfrac{\phi\circ\gamma^{-1}(\gamma(t)-\sigma)}{{\dot\gamma}\circ\gamma^{-1}(\gamma(t)-\sigma)}
\exp{-\int_{\gamma^{-1}(\gamma(t)-\sigma)}^t \chi(\sigma -\gamma(t) + \gamma(u))du}
\\
&\quad \phantom{- \int_0^T \int_0^{\gamma(t)}} \times
\brk{\pd\psi t + {\dot\gamma} \pd\psi\sigma - \chi \psi}(t,\sigma) d\sigma dt \\
&= -\int_0^T \int_{\mathbb{R}} 
p_0(\xi) \exp{-\int_0^t \chi(\xi+\gamma(u))du}
\brk{\pd\psi t + {\dot\gamma} \pd\psi\sigma - \chi  \psi}(t,\xi+\gamma(t))d\xi dt \\ 
&\quad- \int_0^T \int_0^t \phi(v)
\exp{-\int_v^t \chi( -\gamma(v) + \gamma(u))du}
\\
&\quad \phantom{- \int_0^T \int_0^t} \times
\brk{\pd\psi t + {\dot\gamma} \pd\psi\sigma - \chi \psi}(t,-\gamma(v) + \gamma(t)) dv dt 
\end{align*}
where we have made the changes of variables $\xi=\sigma-\gamma(t)$ and $v=\gamma^{-1}(\gamma(t)-\sigma)$. Introducing 
\begin{align*}
  \nu(t,\xi) &= \psi(t,\xi+\gamma(t)) \exp{-\int_0^t \chi(\xi+\gamma(u))du}\\
\text{ and }  \mu(t,v) &= \psi(t,-\gamma(v) + \gamma(t)) \exp{-\int_v^t \chi( -\gamma(v) + \gamma(u))du}
\end{align*}
this rewrites
\begin{align*}
  &-\int_0^T \int_{\mathbb{R}} p \brk{\pd\psi t + {\dot\gamma} \pd\psi\sigma-\chi \psi} \\
&\quad = -\int_{\mathbb{R}} p_0(\xi) \Brk{\int_0^T \pd\nu t(t,\xi)dt}d\xi
-\int_0^T \phi(v) \Brk{\int_v^T \pd\mu t(t,v)dt} dv\\
&\quad=\int_{\mathbb{R}} p_0(\xi) \Brk{\nu(0,\xi) - \nu(T,\xi)}d\xi\\
&\quad + \int_0^T \phi(v) \Brk{\mu(v,v)-\mu(T,v)} dv = \int_0^T \phi(v) \psi(v,0) dv
\end{align*}
thus~\eqref{eq:pphi}. 

\medskip

We finally show that $\phi=\int \chi p$ a.e. on $(0,T)$, where $p$ is
defined by~\eqref{for:p}. This will prove that  \eqref{edp:p} holds in $D'(\brk{0,T}\times\mathbb{R})$.
First, for almost all $t\in\brk{0,\gamma^{-1}(\sigma_c)}$,
definition~\eqref{for:p} of $p$ implies
$\int \chi(\sigma) p(t,\sigma) \, d\sigma=A(t)$. The definition \eqref{def:phi} of $\phi$ implies $\phi(t) =
A(t)$, thus $\phi(t) = \int \chi(\sigma) p(t,\sigma) \, d\sigma$ for such a time~$t$. 
We next take~$t\in\brk{\gamma^{-1}(\sigma_c),T}$. We have 
\begin{align*}
    f(t)&=A(t) + \int_{\sigma_c}^{\gamma(t)} 
\dfrac{\phi\circ\gamma^{-1}(\gamma(t)-\sigma)}{{\dot\gamma}\circ\gamma^{-1}(\gamma(t)-\sigma)}
\exp{-\int_{\gamma^{-1}(\gamma(t)-\sigma)}^t \chi(\sigma -\gamma(t) + \gamma(u))du}
 d\sigma\\
&=A(t) + \int_{\sigma_c}^{\gamma(t)}  
\dfrac{\phi\circ\gamma^{-1}(\gamma(t)-\sigma)}{{\dot\gamma}\circ\gamma^{-1}(\gamma(t)-\sigma)}
\exp{-t + \gamma^{-1}(\gamma(t)-\sigma+\sigma_c)}d\sigma\\ 
&=A(t) + \int_0^{\gamma^{-1}(\gamma(t)-\sigma_c)} \phi(s) \exp{-t+\gamma^{-1}(\gamma(s)+\sigma_c)}ds\\
&=\phi(t), 
\end{align*}
where we have respectively simplified the exponential term, made the
change of variables $s=\gamma^{-1}(\gamma(t)-\sigma)$ and used the
definition~\eqref{def:phi} of $\phi$. This concludes the proof of this
step, and thus that of Theorem~\ref{th:exu}.
\hfill\end{proof}

\section{Properties of the solution}
\label{sec:properties}

In this section we prove various properties of the solution of
\eqref{edp:p} the existence and uniqueness of which has been established
in Theorem \ref{th:exu}. We therefore assume throughout this section
that, as for Theorem~\ref{th:exu}, $p_0$ in $L^1(\mathbb{R})$ and ${\dot\gamma}$ satisfies \eqref{prop:gamma}.

\medskip

\begin{lemma}[Maximum principle] \label{lem:max}
  If $p_0(\sigma)\ge0$ for almost all $\sigma\in\mathbb{R}$ then
  $p(t,\sigma)\ge0$ for almost all $t\in[0,T), \sigma\in\mathbb{R}$.
\end{lemma}

\medskip

\begin{proof}
  Using the definition~\eqref{def:A} of $A$, we first have $A(t)\ge0$ for almost all $t\in[0,T)$. Because of recurrence relation~\eqref{def:phi} on $\phi$, we then find that $\phi(t)\ge0$ for almost all $t\in[0,T)$. Consequently, the expression~\eqref{for:p} on $p$  gives the result.
\hfill\end{proof}

\medskip

\begin{lemma}[Mass conservation] \label{lem:mass}
  If $\int_{\mathbb{R}} p_0=1$ then $\int_{\mathbb{R}} p(t,\cdot)=1$ for all $t\in [0,T)$.
\end{lemma}

\medskip

\begin{proof}
  Denote 
  \begin{align}
    h(t) = \int_{\mathbb{R}} p(t,\cdot)
  \end{align}
which is continuous since $p$ belongs to ${C}^0([0,T);L^1)$.
Using~\eqref{for:p}, we obtain that $h$ reads
\begin{align*}
  h(t) &= 
\int_{\mathbb{R}}     p_0(\sigma-\gamma(t)) \exp{-\int_0^t \chi(\sigma-\gamma(t)+\gamma(u))du} d\sigma
\\&\quad
+ \int_0^{\gamma(t)} \dfrac{\phi\circ\gamma^{-1}(\gamma(t)-\sigma)}{{\dot\gamma}\circ\gamma^{-1}(\gamma(t)-\sigma)}
\exp{-\int_{\gamma^{-1}(\gamma(t)-\sigma)}^t \chi(\sigma -\gamma(t) + \gamma(u))du}
d\sigma.
\end{align*}
For all $t\in\Brk{0,\gamma^{-1}(\sigma_c)}$, $h$ rewrites
\begin{align*}
  h(t) &= \int_{\mathbb{R}} p_0(\xi) \exp{-\int_0^t \chi(\xi+\gamma(u))du} d\xi
+  \int_0^{\gamma(t)} \dfrac{\phi\circ\gamma^{-1}(\gamma(t)-\sigma)}{{\dot\gamma}\circ\gamma^{-1}(\gamma(t)-\sigma)} d\sigma\\
&= \int_{\mathbb{R}}     p_0(\xi) \exp{-\int_0^t \chi(\xi+\gamma(u))du} d\xi
+  \int_0^t \phi(s)ds.
\end{align*}
Differentiating $h$ in the sense of distributions in time, we obtain 
\begin{align*}
  \dot h(t)& = - \int_{\mathbb{R}} \chi(\xi+\gamma(t)) p_0(\xi) \exp{-\int_0^t \chi(\xi+\gamma(u))du} d\xi + \phi(t)\\
&= - A(t)+\phi(t)\\
&=0,
\end{align*}
using the definitions~\eqref{def:A} of $A$ and~\eqref{def:phi} of $\phi$. 
Using that $h$ is continuous on $[0,T)$ with $h(0)=1$, we find $h=1$  on $[0,\gamma^{-1}(\sigma_c)]$.
For all $t\in(\gamma^{-1}(\sigma_c),T)$, $h$ rewrites
\begin{align*}
  h(t) 
&=  \int_{\mathbb{R}} p_0(\xi) \exp{-\int_0^t \chi(\xi+\gamma(u))du} d\xi
+\int_0^{\sigma_c} \dfrac{\phi\circ\gamma^{-1}(\gamma(t)-\sigma)}{{\dot\gamma}\circ\gamma^{-1}(\gamma(t)-\sigma)} d\sigma
\\&\quad
+\int_{\sigma_c}^{\gamma(t)} \dfrac{\phi\circ\gamma^{-1}(\gamma(t)-\sigma)}{{\dot\gamma}\circ\gamma^{-1}(\gamma(t)-\sigma)} \exp{-t + \gamma^{-1}(\gamma(t)-\sigma+\sigma_c)} d\sigma \\
&= \int_{\mathbb{R}} p_0(\xi) \exp{-\int_0^t \chi(\xi+\gamma(u))du} d\xi
+ \int_{\gamma^{-1}(\gamma(t)-\sigma_c)}^t \phi(s)ds
\\&\quad
+ \int_0^{\gamma^{-1}(\gamma(t)-\sigma_c)} \phi(s) \exp{-t+\gamma^{-1}(\gamma(s)+\sigma_c)}ds.
\end{align*}
Differentiating $h$ in $\mathcal{D}'(\gamma^{-1}(\sigma_c),T)$, we obtain
\begin{align*}
  \dot h(t)
&= - A(t)+\phi(t) - \int_0^{\gamma^{-1}(\gamma(t)-\sigma_c)} \phi(s) \exp{-t+\gamma^{-1}(\gamma(s)+\sigma_c)}ds \\
&=0,
\end{align*}
because of definitions~\eqref{def:A} of $A$ and~\eqref{def:phi} of $\phi$. This implies $h=1$  on $[0,T)$, hence the result.
\hfill\end{proof}

\medskip

\begin{lemma}[$L^\infty$-bounds] \label{lem:li}
   Assume $p_0$ satisfies
  \begin{align*}
    p_0\in L^1(\mathbb{R}), \quad p_0\ge0.
  \end{align*}
Then, $\phi$ defined by~\eqref{def:phi} (or, equivalently, $f$,
given~\eqref{eq:fphi}) satisfies
\begin{align} \label{maj:phi}
  \Norm{\phi}_{L^\infty(0,T)} \le \int_{\mathbb{R}} p_0.
\end{align}
  If in addition $p_0$ satisfies
  \begin{align*}
    p_0\in  L^\infty(\mathbb{R}),
  \end{align*}
then $p$ belongs to $L^\infty(0,T;L^\infty)$ and there exists a constant
$C_\infty$ which depends only on $\|p_0\|_{L^1\cap L^\infty}$ and the bound $m_{\dot\gamma}$ in \eqref{prop:gamma} such that
\begin{align} \label{maj:pli}
 \Norm{p}_{L^\infty_T(L^\infty)} \le C_\infty.
\end{align}
\end{lemma}

\medskip

{\em Proof.}
  First, we notice that 
  \begin{equation}
    f(t)=\int_{\mathbb{R}} \chi p(t,\cdot)\le \int_{\mathbb{R}}
    p(t,\cdot) =\int_{\mathbb{R}} p_0,
  \end{equation}
successively using  the definition~\eqref{def:f} of $f$,
Lemma~\ref{lem:max} and Lemma~\ref{lem:mass}. This immediately implies
\begin{align*}
  \Norm{\phi}_{L^\infty(0,T)} \le \int_{\mathbb{R}} p_0.
\end{align*}
Using the lower bound \eqref{prop:gamma} on ${\dot\gamma}$, the
$L^\infty$-bound on $p_0$ and the expression~\eqref{for:p} of $p$ in
terms of~$\phi$, we find that $p$ belongs to $L^\infty(0,T;L^\infty)$ and
\begin{align*}
   \Norm{p}_{L^\infty_T(L^\infty)}\le \Norm{p_0}_{L^\infty}+\frac{\int_{\mathbb{R}} p_0}{m_{\dot\gamma}}. \qquad\endproof
\end{align*}

\medskip

\begin{lemma}[Delay differential equation] \label{lem:dde}
  Assume $p_0\in L^\infty(\mathbb{R})\cap L^1(\mathbb{R})$.
Then, $A$
  and $\phi=f$ respectively defined by~\eqref{def:A}, \eqref{def:phi}
  and \eqref{def:f}, belong to $W^{1,1}(0,T)$ thus are continuous in
  time. Moreover, $A$ and $\phi$ satisfy, for almost all $t\in (0,T)$,
\begin{align}
\dot{A}(t)+A(t)&={\dot\gamma}(t) \Big[p_0(\sigma_c - \gamma(t))
  \exp{-\int_0^t \chi(\sigma_c- \gamma(t) + \gamma(u)) \, du}
 \label{eq:ApointplusA} \\
& \quad \phantom{{\dot\gamma}(t) \Big[} - p_0(-\sigma_c - \gamma(t))
  \exp{-\int_0^t \chi(-\sigma_c- \gamma(t) + \gamma(u)) \, du} \Big]   \nonumber
\end{align}
and for almost all $t\in \brk{\gamma^{-1}(\sigma_c),T}$,
\begin{align} \label{eq:phipointplusphi}
    \dot \phi(t) + \phi(t) -  \frac{{\dot\gamma}(t)}{{\dot\gamma} \circ \gamma^{-1}(\gamma(t)-\sigma_c)} \phi\circ\gamma^{-1}(\gamma(t)-\sigma_c) =
    \dot A(t) + A(t) .
\end{align}
\end{lemma}

\medskip

\begin{proof}
As stated in Theorem~\ref{th:exu}, the functions $A$ and $\phi$ belong to $L^\infty(0,T)\subset L^1(0,T)$. We show that both $\dot A$ and $\dot \phi$ belong to $L^1(0,T)$. 

  The expression~\eqref{def:A} on $A$ rewrites
  \begin{align*}
    A(t)= \int_{\mathbb{R}\backslash[-\sigma_c-\gamma(t),\sigma_c-\gamma(t)]} p_0(\xi)\exp{-\int_0^t \chi(\xi+\gamma(u))du}   d\xi.
  \end{align*}
Differentiating the above expression in the sense of distributions in time, we obtain equation~\eqref{eq:ApointplusA}. Since $p_0$ belongs to $L^\infty(0,T)$ and ${\dot\gamma}$ and $A$ belong to $L^1(0,T),\ \dot A$ belongs to $L^1(0,T)$ and the equation~\eqref{eq:ApointplusA} holds for almost all $t\in(0,T)$.
 
Since $\phi=A$ on $(0,\gamma^{-1}(\sigma_c))$, $\dot \phi$ belongs to $L^1(0,\gamma^{-1}(\sigma_c))$. Differentiating the recurrence relation \eqref{def:phi} in the sense of distributions on  $\brk{\gamma^{-1}(\sigma_c),T}$, we have
\begin{align*}
  \dot \phi(t) &=\dot A(t) + \frac{{\dot\gamma}(t)}{{\dot\gamma} \circ
    \gamma^{-1}(\gamma(t)-\sigma_c)}\phi\circ\gamma^{-1}(\gamma(t)-\sigma_c)\\
& \quad
-\int_0^{\gamma^{-1}(\gamma(t)-\sigma_c)} \phi(s) \exp{-t+\gamma^{-1}(\gamma(s)+\sigma_c)}ds,\\
&=\dot A(t) + \frac{{\dot\gamma}(t)}{{\dot\gamma} \circ
    \gamma^{-1}(\gamma(t)-\sigma_c)}\phi\circ\gamma^{-1}(\gamma(t)-\sigma_c)
+ A(t) - \phi(t).
\end{align*}
Using that $A,\ {\dot\gamma}$ and $\phi$ respectively belong to $W^{1,1}(0,T), L^1(0,T)$ and $L^\infty(0,T)$, the right-hand side of the above equation and thus $\dot \phi$ belong to $L^1(\gamma^{-1}(\sigma_c),T)$. Moreover, the equation~\eqref{eq:phipointplusphi} holds for almost all $t\in \brk{\gamma^{-1}(\sigma_c),T}$. This ends the proof. 
\hfill\end{proof}

\medskip

\begin{lemma}[Existence of $\tau$]
\label{lemma:tau}
  Assume that $p_0$ satisfies
  \begin{align*}
    p_0\in L^1(\mathbb{R}), \quad p_0\ge0 \quad \int \Abs{\sigma} p_0 < \infty.
  \end{align*}
    Then $\sigma p$ belongs to $L^\infty(0,T;L^1)$ so that the average stress $\tau$ defined by~\eqref{def:tau} belongs to $L^\infty(0,T)$.
Moreover, if there exists a scalar $M_{\dot\gamma}$ independent from $T$ such that ${\dot\gamma} \le M_{\dot\gamma}$, then there exists a constant $C_\tau$ independent from $T$ such that
\begin{align} \label{maj:tau}
  \Norm{\int \Abs \sigma p(t,\sigma)d\sigma}_{L^\infty(0,T)} \le C_\tau \brk{1+M_{\dot\gamma} T}.
\end{align}
\end{lemma}

\medskip

\begin{proof}
We multiply the expression~\eqref{for:p} on $p$ by $\Abs\sigma$ and integrate in $\sigma$. This implies
\begin{align} 
&\int \Abs\sigma p(t,\sigma)d\sigma
\nonumber\\& 
\le    \int \Abs\sigma  p_0(\sigma-\gamma(t)) 
d\sigma
+ \int_0^{\gamma(t)} \sigma \frac{\Norm{\phi}_{L^\infty}}{m_{\dot\gamma}} \exp{-\int_{\gamma^{-1}(\gamma(t)-\sigma)}^t \chi(\sigma -\gamma(t) + \gamma(u))du}d\sigma
\nonumber\\&
\le \int \Abs\sigma p_0
+ \brk{\int p_0} \brk{\gamma(T)+\int_0^{\gamma(t)} \frac{\sigma}{m_{\dot\gamma}} \exp{-\int_{\gamma^{-1}(\gamma(t)-\sigma)}^t \chi(\sigma -\gamma(t) + \gamma(u))du}d\sigma},
\label{maj:rafdsigma}\end{align}
using that  ${\dot\gamma}\ge m_{\dot\gamma}$ and then the upper bound \eqref{maj:phi} on $\|\phi\|_{L^\infty(0,T)}$.
We deduce 
\begin{align*}
\int \Abs\sigma p(t,\sigma)d\sigma
 &\le \int \Abs{\sigma} p_0
+\brk{\int p_0} \brk{\gamma(T) +\int_0^{\gamma(T)}\frac{\sigma}{m_{\dot\gamma}} d\sigma},
\end{align*}
so that $\sigma p$ belongs to $L^\infty(0,T;L^1)$.
Moreover, from~\eqref{maj:rafdsigma}, we obtain, for all $t>\gamma^{-1}(\sigma_c)$,
\begin{align} \label{maj:tempsigmap}
 \int \Abs\sigma p(t,\sigma)d\sigma 
&\le \max\brk{\int p_0,\int \Abs{\sigma} p_0}\nonumber\\
&\quad\brk{1+M_{\dot\gamma} T + \int_0^{\sigma_c}\frac{\sigma}{m_{\dot\gamma}} d\sigma
+ \frac{1}{m_{\dot\gamma}}  \int_{\sigma_c}^{\gamma(t)} \sigma \exp{-t + \gamma^{-1}(\gamma(t)-\sigma+\sigma_c)}d\sigma}.
\end{align}
Additionally, with the change of variable $v=t - \gamma^{-1}(\gamma(t)-\sigma+\sigma_c)$, the last integral satisfies
\begin{align*}
  \int_{\sigma_c}^{\gamma(t)} \sigma \exp{-t + \gamma^{-1}(\gamma(t)-\sigma+\sigma_c)}d\sigma 
&= \int_0^{t- \gamma^{-1}(\sigma_c)}
\brk{\gamma(t)-\gamma(t-v)+\sigma_c} \exp{-v} \frac{1}{{\dot\gamma} \circ \gamma^{-1}(\gamma(t)-\sigma+\sigma_c)} dv\\
& \le \int_0^{t- \gamma^{-1}(\sigma_c)} \brk{\int_{t-v}^t M_{\dot\gamma} du+ \sigma_c} \frac{\exp{-v}}{m_{\dot\gamma}}dv\\
& \le \int_0^\infty \brk{M_{\dot\gamma} v + \sigma_c}
\frac{\exp{-v}}{m_{\dot\gamma}} dv<\infty.
\end{align*}
Inserting the above inequality in~\eqref{maj:tempsigmap}
yields~\eqref{maj:tau}.
This concludes the proof.
\hfill\end{proof}
\setcounter{step}{0}

\medskip

Throughout our article, we now assume that 
\begin{align} \label{hyp:p0m}
  p_0 \in L^1(\mathbb{R}) \cap L^\infty(\mathbb{R}) , \quad p_0\ge 0, \quad \int_{\mathbb R} p_0 = 1 \quad\mbox{and } \int_{\mathbb R} \Abs{\sigma} p_0 < \infty,
\end{align}
so that the five Lemmata~\ref{lem:max}, \ref{lem:mass}, \ref{lem:li}, \ref{lem:dde} and~\ref{lemma:tau} hold.

\section{Longtime behavior in the case  ${\dot\gamma}(t) = {\dot\gamma}_\infty$}
\label{sec:tlc}
In this section, we assume that ${\dot\gamma}(t) = {\dot\gamma}_\infty$ where
${\dot\gamma}_\infty>0$ is a given \emph{fixed} constant and we study the longtime
convergence for equation~\eqref{edp:p}. This allows us to lay some ground work which will prove useful for the  general case of a slowly varying shear rate addressed in the next sections.  We prove the following.

\medskip

\begin{theorem} 
\label{th:tlc}
  Assume that ${\dot\gamma}(t) = {\dot\gamma}_\infty$ with ${\dot\gamma}_\infty> 0$ a given constant.
Supply equation~\eqref{edp:p} with an initial condition $p_0$ that satisfies~\eqref{hyp:p0m}. 
Consider $p$ the solution to~\eqref{edp:p} and $p_\infty$ the associated
stationary solution, the existence and uniqueness of those has been respectively established in Theorem~\ref{th:exu} and Lemma~\ref{lem:ss}. 
Then $p$ converges exponentially fast in time to $p_\infty$ for almost all $\sigma\in \mathbb{R}$. 
In addition, $f$ defined by~\eqref{def:f} converges exponentially fast in time to $\int \chi p_\infty$.

More precisely, there exist $b, C_1>0$  
such that, for all $t$ and for almost all $\sigma\in \mathbb{R}$,
\begin{align}
\label{eq:exp-rate}
  \Abs{p(t,\sigma) - p_\infty(\sigma)} \le C_1  \brk{\exp{-t}+ 
\brk{\exp{-bt}\exp{b\frac{\sigma}{{\dot\gamma}_\infty}} + \exp{-t}\exp{\frac{\sigma}{{\dot\gamma}_\infty}} }
{\rm 1\mskip-4mu l}_{(0,{\dot\gamma}_\infty t)}(\sigma)}.
\end{align}
In addition, there exists a positive continuous function $C_2$ of
$\omega\in {\mathbb R}_+^*$ such
that for all~$t \ge 0$,
\begin{align} \label{convc2:f}
  \Abs{f(t) - \int \chi p_\infty} \le C_2(\omega)  \brk{\exp{-bt}+\exp{-t}}.
\end{align}
In the two estimates~\eqref{eq:exp-rate} and~\eqref{convc2:f}, the
rate $b>0$ can be chosen as
\begin{align} \label{max:alpha}
b=-  \max \{x<0 \text{ s.t. } x +1 -\exp{-\omega x}
\cos(\omega\sqrt{\exp{-2\omega x} - (x+1)^2}) = 0  \} - \eta
\end{align}
for any $\eta >0$, where $\omega$ is defined by:
 \begin{align} \label{def:omega}
   \omega=\frac{\sigma_c}{{\dot\gamma}_\infty}.
 \end{align}
\end{theorem}

Notice that  the estimates~\eqref{eq:exp-rate}
and~\eqref{convc2:f} rely on the two natural timescales of the
original problem: the exponential rate $1$ of the jump process to
zero (the hidden coefficient $1$ multiplying the right-hand side of the Fokker-Planck equation~\eqref{edp:p}), and the typical time $\omega$ required for the process to leave the
domain $(-\sigma_c,\sigma_c)$, when the shear rate is ${\dot\gamma}_\infty$.

Let us notice from~\eqref{max:alpha} that we may assume in the
following, without loss of generality, that
\begin{equation}\label{eq:bnequn}
b \neq 1.
\end{equation}
This will be an assumption on $b$ in the forthcoming sections. 
This assumption is exclusively technical.
We require it to simplify some proofs where convolutions
with exponential kernels are involved, in particular to use estimates
such as $\int_0^t \exp{-(t-s)} \exp{-bs} \, ds \le C (\exp{-bt} +
\exp{-t})$ (which is only true if $b \neq 1$).

Before we get to the proof of Theorem~\ref{th:tlc}, to which the rest of
this section is devoted, we need to introduce some notation and make
some preliminaries.

\medskip

Denote by
  $q(t,\sigma)=p(t,\sigma)-p_\infty(\sigma)$ and $q_0(\sigma)=p_0(\sigma)-p_\infty(\sigma)$.
By linearity, and in place of $A$ and $\phi$ introduced in
Theorem~\ref{th:exu}, we similarly introduce
\begin{align} \label{def:B}
  B(t) = \int \chi(\xi + {\dot\gamma}_\infty t) q_0(\xi) \exp{-\frac1{{\dot\gamma}_\infty} \int_\xi^{\xi+{\dot\gamma}_\infty t} \chi(v)dv} d\xi
\end{align}
and
\begin{align} \label{def:g} 
  g(t)= 
  \begin{cases}
   B(t) & \mbox{ for almost all } t\in \brk{0,\omega}\\
   B(t) + \int_0^{t-\omega} g(s) \exp{-t+s+\omega}ds &\mbox{ for almost all }t\in \brk{k\omega,(k+1)\omega}.
 \end{cases}
\end{align}
We then have from Theorem~\ref{th:exu}
\begin{align}
  q(t,\sigma)&=q_0(\sigma - {\dot\gamma}_\infty t) \exp{-\frac1{{\dot\gamma}_\infty} \int_{\sigma-{\dot\gamma}_\infty t}^\sigma \chi(v)dv}
+  \frac1{{\dot\gamma}_\infty} g\left(t-\frac{\sigma}{{\dot\gamma}_\infty}\right) \exp{-\frac1{{\dot\gamma}_\infty} \int_0^\sigma \chi(v)dv} {\rm 1\mskip-4mu l}_{(0,{\dot\gamma}_\infty t)}(\sigma).  \label{ex:q}
\end{align}
Using Lemma~\ref{lem:dde}, $g$ satisfies, for almost all $t>\omega$,
\begin{align} \label{eq:dde}
  \dot g(t)+g(t) - g(t-\omega) = \dot B(t) + B(t).
\end{align}

The equation~\eqref{eq:dde} on $g$ is a delay differential equation with constant coefficients.
The proof of Theorem~\ref{th:tlc} is based upon three lemmata for such a
 delay  differential equation, denoted in generality by 
\begin{equation} \label{dde1}
\left\{
\begin{aligned}
  \dot u(t) +  u(t) -  u(t-\omega)&=\mu(t) \text{ for } t \ge \omega\\
u(t) &= \nu(t) \text{ for } t \in (0,\omega)
\end{aligned}
\right.
\end{equation}
 where $\omega>0$ is a constant and $\mu$ is a locally integrable
 function. Equation~\eqref{dde1} is understood in the sense of  distribution in time.
To such a delay differential equation is classically associated the
unique function~$k(t)$ satisfying the following properties:
  \begin{enumerate}
  \item $k(t)=0, \quad \forall t<0$;
  \item $k(0)=1$;
  \item $k(t)$ is continuous on $[0,\infty)$;
  \item $k(t)$ satisfies for all $t>0$,
    \begin{align} \label{eq:ddek}
      \dot k(t) +  k(t) - k(t-\omega) = 0.
    \end{align}
  \end{enumerate}

\medskip

The three lemmata useful for the proof of Theorem~\ref{th:tlc} are Lemma~\ref{lem:cons}, itself proved using
Lemma~\ref{lem:residue}, and Lemma~\ref{lem:varconst}.  The latter two lemmata, Lemma~\ref{lem:residue} and
Lemma~\ref{lem:varconst}, are borrowed
respectively from \cite{Hale1993} and \cite{Bellman1963}. 
They are valid for more general  
delay differential equations, but for simplicity, we state them here
for our specific  delay differential equation \eqref{dde1}.

\medskip

\begin{lemma} \cite[Equation~(5.10), p.~22]{Hale1993} 
\label{lem:residue}
Consider $k$ defined by~\eqref{eq:ddek} and the associated
properties above. Denote by
\begin{align} \label{def:h}
  h(\lambda)=\lambda+1-\exp{-\omega \lambda}.
\end{align}
Then, for all $\alpha_m\in\mathbb{R}$  such that no root of $h$ has real
part equal to $\alpha_m$, the function~$k$ writes, for all $t>0$,
  \begin{align} \label{form:residue}
    k(t) =  \sum_{j=1}^{k_m} {\rm Res}_{\lambda=\lambda_j} \brk{\frac{\exp{\lambda t}}{h(\lambda)}}
+ \frac1{2\pi i} \lim_{T\rightarrow\infty} \int_{-T}^T \frac{\exp{(\alpha_m+iu) t}}{h(\alpha_m+iu)} du.
  \end{align}
where  $\lambda_1,\ldots,\lambda_{k_m}$ are the roots of $h$ such that
$\Re(\lambda_j)>\alpha_m$ and ${\rm Res}_{\lambda=\lambda_j}$ denotes the residue at $\lambda=\lambda_j$.
\end{lemma}

Intuitively, \eqref{form:residue} is obtained as follows. The
function~$k$ solves $\dot k(t) +  k(t) - k(t-\omega)
=\delta_0$ on ${\mathcal D}'({\mathbb R})$, which by Laplace
transform and using the notation~\eqref{def:h}, yields $h(s)\,{\mathcal
  L} (k)(s)=1$ (where ${\mathcal
  L} (k)$ denotes the Laplace transform of $k$). It remains then to
divide by $h$ and apply a reverse Laplace transform to finally
obtain $k$. The difficulty is of course related to the zeros of the
function~$h$.

\medskip

\begin{lemma} \label{lem:varconst} \cite[Theorem~3.7, p.~75]{Bellman1963}
Consider two functions $\nu\in {C}^0\Brk{0,\omega}$ and $\mu\in
L_{\mathrm{loc}}^1(\omega,\infty)$. Then, there exists a unique
solution $u(t) \in {C}^0({\mathbb R}_+)$ verifying~\eqref{dde1} in
the sense of distribution, and $u(t)$ satisfies: for $t\ge0$, 
\begin{align} \label{eq:varconst}
  u(t) = \nu(\omega)k(t-\omega) -  \int_0^\omega \nu(t_1) k(t-t_1-\omega) dt_1
+\int_\omega^t \mu(t_1) k(t-t_1) dt_1.
\end{align}
\end{lemma}

\begin{remark}
 The result of Lemma~\ref{lem:varconst} is stated in~\cite{Bellman1963} for $\mu$ continuous but holds for $\mu\in L_{\mathrm{loc}}^1(\omega,\infty)$. Indeed, the existence of a unique solution is still valid in this more general setting (see \cite[p.14]{Hale1993}) and expression~\eqref{eq:varconst} satisfies~\eqref{dde1} almost everywhere.
\end{remark}

\medskip

As announced above, we first use Lemma~\ref{lem:residue} to prove the following

\medskip

\begin{lemma} 
\label{lem:cons}
  Assume
  \begin{align} \label{hyp:omega}
    m_0<\omega<M_0.
  \end{align}
Then there exist $b>0$ and $C_0>0$ which depend only on $m_0$ and $M_0$ and such that for all $t>0$ 
\begin{align} \label{decomp:k}
  k(t)=\frac1{1+\omega}+k_1(t)
\end{align}
with
\begin{align} \label{crois:k}
  \Abs{k_1(t)} \le C_0 \exp{-bt}.
\end{align}
Moreover, $b$ can be chosen as~\eqref{max:alpha}, for any $\eta >0$.
\end{lemma}

We immediately emphasize that the point in Lemma~\ref{lem:cons} is to show that
the prefactor~$C_0$ and the exponent~$b$ appearing in~\eqref{crois:k} do
not depend on $\omega$ itself, but can be chosen locally uniformly, that
is, depend only on the bounds $m_0$ and $M_0$ of the interval
where~$\omega$ lies. Proving~\eqref{crois:k} for a fixed $\omega$ is a simple
consequence of the classical results contained
e.g. in~\cite{Bellman1963,Hale1993}.  

\medskip

\begin{remark} \label{rem:alternate}
Using numerical experiments, we will show in Section~\ref{sec:numerics} that the rate~$b$ given by~\eqref{max:alpha} for the estimation~\eqref{crois:k} is indeed sharp. It is interesting to note that our result~\ref{lem:cons} in the present section does not explicitly require such a sharpness. A simpler alternate proof (which we owe to one of the anonymous referees) shows a similar, however non sharp estimation. That proof is based on  the observation $\displaystyle k_1(t)+\int_{t-\omega}^tk_1(s)ds=0$ which shows that the function $k_1$ necessarily cancels on any interval $(m\omega,(m+1)\omega)$.
This leads to the following induction relation on the maximum value $M_m$ of $|k_1|$ on $(m\omega,(m+1)\omega)$
\begin{align} \label{eq:recurrence}
  M_{m+1}\le \max(M_{m-1},M_m)(1-\exp{-2\omega}) .
\end{align}
Indeed, denoting $t_m\in(m\omega,m\omega+\omega)$ a real such that $k_1(t_m)=0$, $k_1$ solution of~\eqref{eq:ddek} satisfies, for all $t\in(t_m,m\omega+2\omega)$,
\begin{align*}
  \Abs{k_1(t)}=\Abs{\int_{t_m}^t \exp{s-t} k_1(s-t)ds} \le \max(M_{m-1},M_m) \int_{t_m}^t \exp{s-t} ds,
\end{align*}
hence~\eqref{eq:recurrence}. Denote $\lfloor.\rfloor$ the integer part. The induction relation~\eqref{eq:recurrence} then implies
\begin{align*}
   M_m\le  \max(M_0,M_1) (1-\exp{-2\omega})^{\lfloor\frac{m}2\rfloor}
\end{align*}
and therefore, using that for all $t\in(0,2\omega)$, $  k_1(t)=\exp{-t} + \exp{\omega-t} (t-\omega) {\rm 1\mskip-4mu l}_{[\omega,2\omega]}(t)-\frac1{1+\omega},$
\begin{align*} 
  \Abs{k_1(t)} \le \tilde{C}_0(\omega) \exp{-\tilde{b}(\omega)t}
\end{align*}
where 
$ \tilde{C}_0(\omega)=2+\omega $
 and 
$  \tilde{b}(\omega)=-\frac{\log(1-\exp{-2\omega})}{2\omega} $
are respectively decreasing and increasing functions of $\omega$. This proves Lemma~\ref{lem:cons}, with the values $b=\tilde{b}(M_0)$ and $C_0=\tilde{C}_0(M_0)$.
\end{remark}
\medskip

{\em Proof of Lemma \ref{lem:cons}.}
The proof falls in three steps. We first derive an upper bound on the
real part of the nonzero roots of the function $h$ defined by
\eqref{def:h}. This upper bound actually yields the exponent $b$ in the
exponential estimates of Lemma~\ref{lem:cons} and thus of Theorem~\ref{th:tlc}. In the second step, we apply Lemma
~\ref{lem:residue}. In the third and final step, we conclude.

\refstepcounter{step}
\subparagraph*{Step \arabic{step}: Upper bound on the real part of the nonzero roots of $h$}
The roots of the  function $h$ defined by~\eqref{def:h} are $0$ and the complex numbers $\lambda=\alpha + i \beta$ (with $\alpha,\beta\in\mathbb{R}$) that satisfy
\begin{align}
  \alpha+1&= \exp{-\omega\alpha} \cos\brk{\omega\beta}, \label{eq:alpha}\\
  \beta &=  - \exp{-\omega\alpha} \sin\brk{\omega\beta}. \label{eq:beta}
\end{align} 
It is easy to check that $\alpha=0$ implies $\beta=0$ and conversely,
so that in the following, we assume $\alpha \neq 0$ and $\beta \neq 0$.
The equation \eqref{eq:beta} rewrites $-\dfrac{\sin(\omega\beta)}{\beta}=\exp{\omega\alpha}$. Since the function $x\mapsto - \dfrac{\sin(\omega x)}{x}$ is non-positive on $\Brk{-\frac\pi\omega,\frac\pi\omega}$, $\beta$ satisfies
\begin{align} \label{maj:beta}
  \Abs\beta>\frac\pi\omega.
\end{align}
Moreover, we combine \eqref{eq:alpha} and \eqref{eq:beta} and obtain
\begin{align} \label{eq:carre}
  \exp{2\omega\alpha}\brk{(\alpha+1)^2 + \beta^2} = 1. 
\end{align}
This implies that $\alpha$ is negative and therefore, using the bounds \eqref{hyp:omega} and \eqref{maj:beta} respectively on $\omega$ and $\beta$,
\begin{align*}
 \exp{2 M_0 \alpha}\brk{(\alpha+1)^2 + \frac{\pi^2}{M_0^2}} < 1.
\end{align*}
The function 
\begin{align*}
  \zeta : x \mapsto \exp{2M_0 x}\brk{(x+1)^2 + \frac{\pi^2}{M_0^2}}
\end{align*}
is continuous, satisfies $\zeta(0)>1$ and 
$\lim_{-\infty} \zeta = 0$ so that by the intermediate value theorem, there exists
\begin{align*}
b > 0\mbox{ such that } \zeta(-b)=1 \mbox{ and } \zeta(x)\ge1 \mbox{ on } \Brk{-b,0}. 
\end{align*}
The scalar $b>0$ depends only on $M_0$. Additionally, the real part $\alpha$ of the nonzero roots of $h$ satisfies
\begin{align*}
  \alpha<-b
\end{align*}
and, combining \eqref{eq:alpha} and \eqref{eq:carre},
\begin{align*}
  \alpha +1 - \exp{-\omega \alpha} \cos(\omega\sqrt{\exp{-2\omega \alpha} - (\alpha+1)^2}) = 0
\end{align*}
Therefore $b$ can be chosen as~\eqref{max:alpha}, for any $\eta >0$.

\refstepcounter{step}
\subparagraph*{Step \arabic{step}: Applying Lemma~\ref{lem:residue}}
From the previous step, we know that the only root of $h$ with real part strictly above $-b$ is $0$.  We now apply
Lemma~\ref{lem:residue} with $\alpha_m=-b$. Since the root $0$ is a simple root of $h$, the residue of 
$\dfrac{\exp{\lambda t}}{h(\lambda)}$ at $0$ is $\dfrac1{\dot h(0)} =
\dfrac1{1+\omega}$. Equation~\eqref{form:residue} therefore writes
\begin{align}
  k(t) = \frac1{1+\omega} + \frac1{2\pi i} \lim_{T\rightarrow\infty} \int_{-T}^T \frac{\exp{(-b+iu)t}}{h(-b+iu)} du. 
\end{align}
Proving that there exists $C_0>0$ which depends only on $m_0$ and $M_0$, such that for all $t>0$,
\begin{align*}
  \lim_{T\rightarrow\infty} \Abs{\int_{-T}^T \frac{\exp{(-b+iu)t}}{h(-b+iu)} du}
\le C_0 \exp{-bt}.
\end{align*}
therefore amounts to concluding the proof of
Lemma~\ref{lem:cons}. Actually, we will show that this holds up to
changing $b$ to $b -\eta$ in the right hand side, for any positive $\eta$. This will conclude
the proof.

\refstepcounter{step}
\subparagraph*{Step \arabic{step}: Exponential bound}
We first show, for all $t>0$,
\begin{align} \label{eq:it}
  \lim_{T\rightarrow\infty} \int_{-T}^T \frac{\exp{(-b+iu)t}}{h(-b+iu)} du 
= \lim_{T\rightarrow\infty} \frac1{t} \int_{-T}^T \exp{(-b+iu)t} \frac{\dot h}{h^2}(-b+iu) du.
\end{align}
By integration by parts, we have
\begin{align} \label{ipp:cplx}
  \int_{-T}^T \frac{\exp{(-b+iu)t}}{h(-b+iu)} du 
= \frac1{t} \int_{-T}^T \exp{(-b+iu)t} \frac{\dot h}{h^2}(-b+iu) du
+\frac1{it} \Brk{\frac{\exp{(-b+iu)t}}{h(-b+iu)}}_{-T}^T.
\end{align}
Introduce $T_0>0$ such that for all $\Abs T\ge T_0$, 
\begin{align*}
  \brk{1+\frac{b^2}{T^2}}^{\frac12} - \frac1{\Abs T} \brk{1+\exp{M_0 b}} \ge \frac12.
\end{align*}
Then, for all $\Abs{T}>T_0$,
\begin{align*}
\Abs{h(-b+iT)} &= \Abs{-b+iT + 1 - \exp{-\omega(-b+iT)}}\\
&\ge \sqrt{b^2+T^2} - 1 - \exp{\omega b} \ge \sqrt{b^2+T^2} - (1 + \exp{M_0 b})\ge \frac{\Abs{T}}2
\end{align*}
so that, for all $t>0$,
\begin{align} \label{maj:ippb}
  \Abs{\frac1{it} \Brk{\frac{\exp{(-b+iu)t}}{h(-b+iu)}}_{-T}^T} \le \frac{4\exp{-bt}}{\Abs{T}t}.
\end{align}
By passing to the limit $T \to \infty$ in~\eqref{ipp:cplx}, we thus obtain~\eqref{eq:it}.

Now, for all $u\in\mathbb{R}$, 
\begin{align*}
   \Abs{\frac{\dot h}{h^2}}(-b+iu) 
&= \frac{\Abs{1+\omega\exp{-\omega(-b+iu)}}}{(1-b-\exp{\omega b} \cos(\omega u))^2 + (u+\exp{\omega b} \sin(\omega u))^2}\\
&< \frac{1+M_0 \exp{M_0 b}}{(1-b)^2 
+ u^2-2\exp{\omega b} \brk{(1-b) \cos(\omega u) - u \sin(\omega u)}}.
\end{align*}
Introduce $u_0>0$ which we may take depending only on $M_0$, such that for all $\Abs{u}\ge u_0$,
\begin{align*}
  (1-b)^2 + u^2-2\exp{M_0 b} \brk{(1-b) + \Abs{u}} > 0
\end{align*}
so that
\begin{align*}
   \Abs{\frac{\dot h}{h^2}}(-b+iu) 
< \frac{1+M_0 \exp{M_0 b}}{(1-b)^2 + u^2-2\exp{M_0 b} \brk{(1-b) + \Abs{u}}}.
\end{align*}
For $T>u_0$ and $t>0$, this implies
\begin{align} 
  \frac1{t}&\Abs{ \int_{-T}^T \exp{(-b+iu)t} \frac{\dot h}{h^2}(-b+iu) du} \nonumber\\
&\le \frac{\exp{-bt}}{t} \brk{ \int_{-u_0}^{u_0} \Abs{\frac{\dot h}{h^2}}(-b+iu) du 
+ 2 \int_{u_0}^\infty \frac{\brk{1+M_0 \exp{M_0 b}} du}{(1-b)^2 + u^2-2\exp{M_0 b} \brk{(1-b) + \Abs{u}}} }. \label{maj:it}
\end{align}

The function $\omega \mapsto \int_{-u_0}^{u_0} \Abs{\frac{\dot
    h}{h^2}}(-b+iu) du$ is continuous for $\omega\in\Brk{m_0,M_0}$ and is
therefore bounded by a constant that only depends on $m_0$ and
$M_0$. From~\eqref{eq:it} and the bound~\eqref{maj:it}, we deduce that
there exists a constant~$C_0>0$ that also only depends on $m_0$ and $M_0$ such that
\begin{align*}
   \lim_{T\rightarrow\infty} \Abs{\int_{-T}^T \frac{\exp{(-b+iu)t}}{h(-b+iu)} du}
\le C_0 \exp{-(b-\eta)t},
\end{align*}
for any positive $\eta$. This concludes the proof.\hfill\endproof
\setcounter{step}{0}

\medskip

We are now in position to turn to the

\medskip

{\em Proof of Theorem~\ref{th:tlc}.} The proof proceeds in five steps. In
step \ref{s:varconst}, we apply the above lemmata to the delay differential
equation~\eqref{eq:dde}. In steps~\ref{s:terme0}
and~\ref{s:terme1}, we derive some estimates that will be useful, in the last two steps, to show convergence of $g$, and eventually $q$.

\refstepcounter{step}
\subparagraph*{Step \arabic{step}: Applying Lemmata~\ref{lem:cons} and \ref{lem:varconst}}
\label{s:varconst}
The function $g$ defined by~\eqref{def:g} satisfies~\eqref{dde1} in
the particular case~$\mu=\dot B + B$ and $\nu(t)=B(t)$. Notice that,
by the Lemma~\ref{lem:dde}, the function~$B$ defined
by~\eqref{def:B} belongs to $W^{1,1}(0,T)$ and thus, in
particular, is continuous in time. We apply Lemma~\ref{lem:varconst}
and find (by integration by parts): for all $t>\omega$,
\begin{align} 
  g(t) &= B(\omega) k(t-\omega) + \int_0^\omega B(t_1) k(t-t_1-\omega)
  dt_1 + \int_\omega^t \brk{\dot B + B}(t_1) k(t-t_1) dt_1\nonumber\\
 &= B(\omega) k(t-\omega) + \int_0^\omega B(t_1) \left(\dot{k}+k\right)(t-t_1) dt_1 + \int_\omega^t \brk{\dot B + B}(t_1) k(t-t_1) dt_1\nonumber\\
&=B(0) k(t)+\int_0^t \brk{\dot B + B}(t_1) k(t-t_1) dt_1. \label{eq:IPP}
\end{align}
 We now recall that, in this section, the value of $\omega$ is fixed
by~\eqref{def:omega} at $\omega=\frac{\sigma_c}{{\dot\gamma}_\infty}$. We can
apply Lemma \ref{lem:cons} and insert the decomposition~\eqref{decomp:k} of~$k$
into the previous equation on $g$. We obtain
\begin{align} \label{decomp:g}
  g(t)&= \frac1{1+\omega} \brk{B(0) + \int_0^t \brk{\dot B + B}(t_1)dt_1}
\nonumber\\
&\quad+B(0) k_1(t) 
+ \int_0^t \brk{\dot B + B}(t_1) k_1(t-t_1) dt_1.
\end{align}
where $k_1$ satisfies \eqref{crois:k} with $b, C_0>0$ only depending on $\omega=\frac{\sigma_c}{{\dot\gamma}_\infty}$.
Moreover, $b$ can be chosen as~\eqref{max:alpha} for any positive $\eta$ as stated in Lemma \ref{lem:cons}.

Our next two steps consist in deriving a couple of estimates (see~\eqref{maj:terme0} and~\eqref{maj:terme1} below) on the terms of~\eqref{decomp:g}.

\refstepcounter{step}
\subparagraph*{Step \arabic{step}: Longtime convergence of $\displaystyle B(0) + \int_0^t \brk{\dot B + B}(t_1)dt_1$}
\label{s:terme0}
Using~\eqref{eq:ApointplusA}, the function $B$ defined by~\eqref{def:B} satisfies, for almost all $t>0$,   
\begin{align} \label{eq:BB}
  \dot B(t) + B(t)&= {\dot\gamma}_\infty \Big(
-q_0(-\sigma_c-{\dot\gamma}_\infty t) \exp{-\frac{1}{{\dot\gamma}_\infty} \int_{-\sigma_c-{\dot\gamma}_\infty t}^{-\sigma_c} \chi(\sigma) d\sigma}
\nonumber\\
&\quad \phantom{{\dot\gamma}_\infty \Big(}
+q_0(\sigma_c-{\dot\gamma}_\infty t) \exp{-\frac{1}{{\dot\gamma}_\infty} \int_{\sigma_c-{\dot\gamma}_\infty t}^{\sigma_c} \chi(\sigma) d\sigma}\Big).
\end{align}
Computing $B(0)$ and integrating~\eqref{eq:BB} from $0$ to $t$ yield
\begin{align*}
B(0) + \int_0^t \brk{\dot B + B}
&= \int \chi q_0 
-{\dot\gamma}_\infty \int_0^t 
q_0(-\sigma_c-{\dot\gamma}_\infty t_1) \exp{-\frac1{{\dot\gamma}_\infty} \int_{-\sigma_c-{\dot\gamma}_\infty t_1}^{-\sigma_c} \chi(\sigma) d\sigma}dt_1
\\
&\quad +{\dot\gamma}_\infty \int_0^t q_0(\sigma_c-{\dot\gamma}_\infty t_1) \exp{-\frac1{{\dot\gamma}_\infty} \int_{\sigma_c-{\dot\gamma}_\infty t_1}^{\sigma_c} \chi(\sigma) d\sigma} dt_1
\end{align*}
so that, respectively with the changes of variables $v=-\sigma_c-{\dot\gamma}_\infty
t$ and  $v=\sigma_c-{\dot\gamma}_\infty t$ in the last two integrals, we obtain
\begin{align*}
B(0) + \int_0^t \brk{\dot B + B}
&= \int \chi q_0 
-\int_{-\sigma_c-{\dot\gamma}_\infty t}^{-\sigma_c} q_0(v) \exp{-\frac1{{\dot\gamma}_\infty} \int_v^{-\sigma_c} \chi(\sigma) d\sigma} dv
\\
&\quad + \int_{\sigma_c-{\dot\gamma}_\infty t}^{\sigma_c} q_0(v) \exp{-\frac1{{\dot\gamma}_\infty} \int_v^{\sigma_c} \chi(\sigma) d\sigma} dv .
\end{align*}
For $t>2\omega$, this implies that
\begin{align*}
B(0) + \int_0^t \brk{\dot B + B}(t_1)dt_1 
&= \int \chi q_0
-\int_{-\sigma_c-{\dot\gamma}_\infty t}^{-\sigma_c} q_0(v) \exp{-\frac1{{\dot\gamma}_\infty} \int_v^{-\sigma_c} \chi(\sigma) d\sigma} dv
+ \int_{-\sigma_c}^{\sigma_c} q_0
\\&\quad
+ \int_{\sigma_c-{\dot\gamma}_\infty t}^{-\sigma_c} q_0(v) \exp{-\frac1{{\dot\gamma}_\infty} \int_v^{-\sigma_c} \chi(\sigma) d\sigma} dv
\\&
=\int q_0
- \int_{-\sigma_c-{\dot\gamma}_\infty t}^{\sigma_c-{\dot\gamma}_\infty t} q_0(v)
\exp{\frac{\sigma_c+v}{{\dot\gamma}_\infty}}dv
\\&
= - {\dot\gamma}_\infty \int_{-\omega-t}^{\omega-t} q_0({\dot\gamma}_\infty v) \exp{\omega+v}dv
\end{align*}
using that  $\displaystyle \int q_0=0$. We deduce
\begin{align} \label{maj:terme0}
\Abs{B(0) + \int_0^t \brk{\dot B + B}(t_1)dt_1}
 \le {\dot\gamma}_\infty\Norm{q_0}_{L^\infty} \exp{2\omega-t}.
\end{align}

\refstepcounter{step}
\subparagraph*{Step \arabic{step}: Longtime convergence of $B(0) k_1(t) + \int_0^t \brk{\dot B + B}(t_1) k_1(t-t_1) dt_1$}
\label{s:terme1}
Using~\eqref{eq:BB} on $\dot B + B$
and the estimate \eqref{crois:k} on $k_1$,
we have, for $t>2\omega$,
\begin{align}
  &\Abs{B(0) k_1(t) + \int_0^t \brk{\dot B + B}(t_1) k_1(t-t_1) dt_1}\nonumber\\
&\quad\le C_0 \brk{\exp{-bt}\int \chi \Abs{q_0}  
+ \int_0^t {\dot\gamma}_\infty \Abs{q_0(-\sigma_c-{\dot\gamma}_\infty t_1)} \exp{-t_1-b(t-t_1)} dt_1
\right.\nonumber\\&\qquad\left.
+ \int_0^t {\dot\gamma}_\infty \Abs{q_0(\sigma_c-{\dot\gamma}_\infty t_1)} \exp{2\omega}\exp{-t_1-b(t-t_1)} dt_1}\nonumber\\
&\quad\le C_0 {\dot\gamma}_\infty \Norm{q_0}_{L^\infty} (1+\exp{2\omega})
\brk{\exp{-bt} + \exp{-bt} \int_0^t \exp{(b-1) t_1} dt_1}\nonumber\\
&\quad\le  C_0 {\dot\gamma}_\infty \Norm{q_0}_{L^\infty} (1+\exp{2\omega})
\brk{ \frac1{\Abs{b-1}}\exp{-t} + \frac{b}{\Abs{b-1}}\exp{-bt}}.
\label{maj:terme1}
\end{align}
Here, we have used the assumption~\eqref{eq:bnequn} on $b$.

\refstepcounter{step}
\subparagraph*{Step \arabic{step}: Longtime convergence of $g(t)$}
Using the decomposition~\eqref{decomp:g} 
and the estimates~\eqref{maj:terme0} and~\eqref{maj:terme1} derived in steps~\ref{s:terme0} and \ref{s:terme1}, we have, for $t>2\omega$,
\begin{align} \label{conv:g}
  \Abs{g(t) }
\le {\dot\gamma}_\infty\Norm{q_0}_{L^\infty}
\brk{ \frac{\exp{2\omega}}{1+\omega} \exp{-t}  + C_0 \brk{1+\exp{2\omega}} 
\brk{ \frac1{\Abs{b-1}}\exp{-t} + \frac{b}{\Abs{b-1}}\exp{-bt}}
}.
\end{align}
Recall that $g$ defined by~\eqref{def:g} satisfies by linearity $g=\phi-\int \chi p_\infty$ and that  $f=\phi$ a.e. (see~\eqref{eq:fphi}). We have obtained estimate \eqref{convc2:f} on $f-\int \chi p_\infty$.

\refstepcounter{step}
\subparagraph*{Step \arabic{step}: Longtime convergence of $q$}
We now turn to $q(t,\sigma)=p(t,\sigma)-p_\infty(\sigma)$. Using~\eqref{ex:q}, we have
\begin{align*} 
  \Abs{q(t,\sigma)}\le\Norm{q_0}_{L^\infty} \exp{2\omega-t}
+  \frac1{{\dot\gamma}_\infty} \Abs{g\left(t-\frac{\sigma}{{\dot\gamma}_\infty}\right)} {\rm 1\mskip-4mu l}_{(0,{\dot\gamma}_\infty t)}(\sigma).
\end{align*}
In view of the estimate~\eqref{convc2:f} on $g$, we deduce, for almost all $\sigma\in\mathbb{R}$ and $t>0$,
\begin{align*}
  \Abs{q(t,\sigma)}\le \Norm{q_0}_{L^\infty} \exp{2\omega-t}
+
 \frac{C_2}{{\dot\gamma}_\infty}
\brk{ \exp{-\brk{t-\frac\sigma{{\dot\gamma}_\infty}}} + \exp{-b\brk{t-\frac\sigma{{\dot\gamma}_\infty}}}}{\rm 1\mskip-4mu l}_{(0,{\dot\gamma}_\infty t)}(\sigma).
\end{align*}
This concludes the proof of Theorem~\ref{th:tlc}.\hfill\endproof
\setcounter{step}{0}

\section{Longtime convergence in the case ${\dot\gamma}(\epsilon t)$ }
\label{sec:long-time}

Our main result in this section is the following.

\medskip

\begin{theorem} \label{th:tle}
Consider ${\dot\gamma}$ a Lipschitz function with Lipschitz constant
$L_{\dot\gamma}$,  which satisfies, for all $t\ge0$,
\begin{align} \label{hyp:gamma}
  m_{\dot\gamma}\le{\dot\gamma}(t)\le M_{\dot\gamma},  \mbox{ for some } m_{\dot\gamma}, M_{\dot\gamma}>0 \mbox{ constant scalars.}
\end{align}
Consider an initial condition $p_0$ which satisfies~\eqref{hyp:p0m}. 
For $\theta>0, \ \epsilon >0$ such that $\frac\theta\epsilon
>2\frac{\sigma_c}{m_{\dot\gamma}}$, consider the functions $p_\epsilon(t,\sigma)$ and
$p_\infty(\theta,\sigma)$ respectively solutions to
\begin{align} \label{edp:mic}
&  \left\{
\begin{aligned}
\pd {p_\epsilon}{t}(t,\sigma) + {\dot\gamma}(\epsilon t) \pd{p_\epsilon}{\sigma}(t,\sigma) &=
-\chi(\sigma) p_\epsilon (t,\sigma) + \brk{\int \chi(\sigma) p_\epsilon (t,\sigma) \, d\sigma} \delta_0(\sigma) \\
p_\epsilon(0,\sigma)=p_0(\sigma)
\end{aligned}
\right.\\
 & {\dot\gamma}(\theta) \pd{p_\infty(\theta,\sigma)}{\sigma}= -\chi(\sigma)
 p_\infty(\theta,\sigma) + \brk{\int \chi(\sigma) p_\infty(\theta,\sigma) \, d\sigma} \delta_0(\sigma) \label{edp:mics}
\end{align}
the existence and uniqueness of which have been respectively established in Theorem~\ref{th:exu} and Lemma~\ref{lem:ss}.

Then, there exist constants $b, C_3, C_4>0$ independent from $\theta$
and $\epsilon$ (satisfying $\frac\theta\epsilon
>2\frac{\sigma_c}{m_{\dot\gamma}}$) such that,
\begin{align} 
  \Abs{\int \chi(\sigma) \brk{p_\epsilon\brk{\frac\theta\epsilon, \sigma} -
      p_\infty(\theta,\sigma)} \, d\sigma}
\le C_3 \brk{\exp{-b\frac\theta\epsilon}+ \exp{-\frac\theta\epsilon}+\epsilon}  \label{conv:geps} 
\end{align}
and, for almost all $\sigma\in \mathbb{R}$ such that $\sigma \le
\gamma\brk{\frac\theta\epsilon}$
, 
  \begin{align}
     \Abs{p_\epsilon\brk{\frac\theta\epsilon, \sigma} - p_\infty(\theta,\sigma)} 
&\le C_4 \left[ \brk{\exp{-b\frac\theta\epsilon} + \exp{-\frac\theta\epsilon} +\epsilon} 
+  \epsilon \, {\rm 1\mskip-4mu l}_{(\sigma_c,\infty)}(\sigma) \, (\sigma - \sigma_c)^2 \right].
  \label{conv:pmp}
\end{align}
\end{theorem}

In order to prove Theorem~\ref{th:tle}, we need the following technical lemma.

\medskip

\begin{lemma} \label{lem:intt}
  Consider ${\dot\gamma}$ a function of time that satisfies~\eqref{prop:gamma}. 
Denote by $\gamma(t)=\int_0^t{\dot\gamma}(s)ds$.
Then, for all $t>\frac{2\sigma_c}{m_{\dot\gamma}}$ and almost all $\sigma\in\mathbb{R}$, we have
\begin{align} \label{min:intt}
  \int_0^t \chi\brk{\sigma-\gamma(t)+\gamma(u)}du \ge t-\frac{2\sigma_c}{m_{\dot\gamma}}.
\end{align}
\end{lemma}

\medskip

\begin{proof}
Denote by $Z(t,\sigma)=\int_0^t \chi\brk{\sigma-\gamma(t)+\gamma(u)}du$.
For all $t>\frac{2\sigma_c}{m_{\dot\gamma}}$ and almost all $\sigma\in\mathbb{R}$, we have
\begin{align*}
&Z(t,\sigma)\\
&\quad=\int_0^t \brk{{\rm 1\mskip-4mu l}_{(-\infty,-\sigma_c)} + {\rm 1\mskip-4mu l}_{(\sigma_c,\infty)}}\brk{\sigma-\gamma(t)+\gamma(u)}du\\
&\quad=\int_0^t \brk{{\rm 1\mskip-4mu l}_{(-\infty,\gamma^{-1}(-\sigma_c+\gamma(t)-\sigma))} + {\rm 1\mskip-4mu l}_{(\gamma^{-1}(\sigma_c+\gamma(t)-\sigma),\infty)}}(u) du\\
&\quad=
\begin{cases}
  t & \sigma <-\sigma_c\\
\gamma^{-1}(-\sigma_c+\gamma(t)-\sigma) & -\sigma_c<\sigma<\gamma(t)-\sigma_c\\
0&\gamma(t)-\sigma_c<\sigma
\end{cases}\\
&\qquad+
\begin{cases}
 0&\sigma<\sigma_c\\
t- \gamma^{-1}(\sigma_c+\gamma(t)-\sigma)&\sigma_c<\sigma<\gamma(t)+\sigma_c\\
t&\gamma(t)+\sigma_c<\sigma
\end{cases}\\
&\quad=
\begin{cases}
  \gamma^{-1}(-\sigma_c+\gamma(t)-\sigma) & -\sigma_c<\sigma \le \sigma_c\\
t+\gamma^{-1}(-\sigma_c+\gamma(t)-\sigma)-\gamma^{-1}(\sigma_c+\gamma(t)-\sigma)&\sigma_c<\sigma
\le \gamma(t)-\sigma_c\\
t-\gamma^{-1}(\sigma_c+\gamma(t)-\sigma)&\gamma(t)-\sigma_c < \sigma
\le \gamma(t)+\sigma_c\\
t &\sigma \le - \sigma_c \mbox{ or } \sigma > \gamma(t)+\sigma_c
\end{cases}
\end{align*}
using that $\gamma(t)>2\sigma_c$.
We now estimate the above expression depending on $\sigma$.
For almost all $\sigma\in(-\sigma_c,\sigma_c)$, the function $Z$ is decreasing in $\sigma$ so that $Z(t,\sigma) \ge \gamma^{-1}(\gamma(t)-2\sigma_c)$. Moreover, because of~\eqref{prop:gamma}, the function ${\dot\gamma}$ satisfies
\begin{align*}
  \gamma(t) - \gamma\brk{t-\frac{2\sigma_c}{m_{\dot\gamma}}}
=\int_0^t {\dot\gamma} - \int_0^{t-\frac{2\sigma_c}{m_{\dot\gamma}}} {\dot\gamma}
\ge 2\sigma_c
\end{align*}
so that $\gamma^{-1}(\gamma(t)-2\sigma_c)\ge t-\dfrac{2\sigma_c}{m_{\dot\gamma}}$, hence \eqref{min:intt} for  almost all $\sigma\in(-\sigma_c,\sigma_c)$.

Additionally, because of~\eqref{prop:gamma}, the function $\gamma$ satisfies, for all $v>u\ge0$,
\begin{align*}
  {\gamma(v)-\gamma(u)} = {\int_u^v {\dot\gamma}} \ge m_{\dot\gamma} \brk{v-u}.
\end{align*}
This yields that $\gamma^{-1}$ is Lipschitz with a Lipschitz constant $\frac1{m_{\dot\gamma}}$ on $[0,\infty)$. Therefore, for almost $\sigma\in(\sigma_c,\gamma(t)-\sigma_c)$, 
\begin{align*}
t-Z(t,\sigma)=\gamma^{-1}(\sigma_c+\gamma(t)-\sigma) - \gamma^{-1}(-\sigma_c+\gamma(t)-\sigma)\le \frac{2\sigma_c}{m_{\dot\gamma}},
\end{align*}
hence \eqref{min:intt}.

For almost all $\sigma\in(\gamma(t)-\sigma_c,\gamma(t)+\sigma_c)$, the function $Z$ is increasing in $\sigma$ so that $Z(t,\sigma) \ge t-\gamma^{-1}(2\sigma_c)$.
Moreover,
\begin{align*}
  2\sigma_c = \int_0^{\frac{2\sigma_c}{m_{\dot\gamma}}} m_{\dot\gamma} \le \int_0^{\frac{2\sigma_c}{m_{\dot\gamma}}} {\dot\gamma}(u)du = \gamma\brk{\frac{2\sigma_c}{m_{\dot\gamma}}}
\end{align*}
so that $\gamma^{-1}(2\sigma_c)\le \frac{2\sigma_c}{m_{\dot\gamma}}$, hence \eqref{min:intt} for almost all $\sigma\in(\gamma(t)-\sigma_c,\gamma(t)+\sigma_c)$.

The result \eqref{min:intt} also holds in the case $\sigma\in\mathbb{R}\backslash[-\sigma_c,\gamma(t)+\sigma_c]$ where $Z(t,\sigma)=t$. This ends the proof.
\end{proof}

\medskip

Now that we have proved the technical Lemma~\ref{lem:intt}, we turn to the

\medskip

{\em Proof of Theorem~\ref{th:tle}.}
The proof is divided into six steps. The first step establishes a
delay differential equation on a function $g_\epsilon$, for which an
explicit decomposition is known thanks to the Lemma~\ref{lem:varconst}. We then rewrite $g_\epsilon$ in a different form  whose terms are estimated in Steps \ref{s:k0} and \ref{s:k1}. In the last two steps, we use these estimates to obtain \eqref{conv:geps} and then \eqref{conv:pmp}.

Before we get to the proof we introduce some notation.
The scalars $\theta>0,\ \epsilon>0$ are fixed and satisfy $\frac\theta\epsilon>2\frac{\sigma_c}{m_{\dot\gamma}}$.
 In Section~\ref{sec:tlc}, we have introduced
\begin{align*}
  \omega_\theta &= \frac{\sigma_c}{{\dot\gamma}(\theta)}
\end{align*}
which from bounds \eqref{hyp:gamma} on ${\dot\gamma}$ satisfies
\begin{align} \label{bornes:omega}
  \frac{\sigma_c}{M_{\dot\gamma}} < \omega_\theta < \frac{\sigma_c}{m_{\dot\gamma}}.
\end{align}
We can therefore apply Lemma \ref{lem:cons} to the function $k_\theta$
satisfying~\eqref{eq:ddek} with $\omega=\omega_\theta$, so that there exist $b, C_0>0$ which depend only on $\sigma_c$, $m_{\dot\gamma}$ and $M_{\dot\gamma}$ such that \eqref{decomp:k} and \eqref{crois:k} hold for all $t>0$. Notably, $b$ and $C_0$ are independent from $\theta$ (and $\epsilon$).

\refstepcounter{step}
\subparagraph*{Step \arabic{step}: Applying Lemma~\ref{lem:varconst}}
For a fixed $\theta$, denote by
\begin{align} \label{def:Atheta}
  A_\theta(t) &= \int \chi(\sigma) p_0 (\sigma - {\dot\gamma}(\theta) t) \exp{-\frac1{{\dot\gamma}(\theta)} \int_{\sigma-{\dot\gamma}(\theta) t}^\sigma \chi(v)dv} d\sigma 
\end{align}
and by $\phi_\theta$ the solution to 
\begin{align*}
  \dot \phi_\theta(t) + \phi_\theta(t) - \phi_\theta\brk{t-\omega_\theta} = \dot A_\theta(t) + A_\theta(t)
\end{align*}
with the initial condition $\phi_\theta(t)=A_\theta(t),  0<t<\omega_\theta$.
Consistently with~\eqref{def:f}, let us also introduce $f_\epsilon(t)=\int
\chi(\sigma)p_\epsilon(t,\sigma) d\sigma$ where $p_\epsilon$ satisfies~\eqref{edp:mic}.
Then 
\begin{align} \label{def:geps}
  g_\epsilon(t) =  f_\epsilon(t) - \phi_\theta(t) 
\end{align}
 belongs to $W^{1,1}(0,T)$ (because $f_\epsilon$ and $\phi_\theta$ do, see Lemma~\ref{lem:dde}) and satisfies, for almost all $t>\omega_\theta$
\begin{align*}
    \dot g_\epsilon(t) + g_\epsilon(t) - g_\epsilon\brk{t-\omega_\theta} 
= \dot f_\epsilon(t) + f_\epsilon(t) - \dot A_\theta(t) - A_\theta(t) - f_\epsilon(t-\omega_\theta).
\end{align*}
Introduce $s>2 \frac{\sigma_c}{m_{\dot\gamma}}$. We apply
Lemma~\ref{lem:varconst} and obtain (using the same computations as
in~\eqref{eq:IPP} above and the fact that $f_\epsilon(0)=\phi_\theta(0)$),
\begin{align} \label{varconst:geps}
  g_\epsilon(s) &= (f_\epsilon - \phi_\theta)(\omega_\theta) k_\theta(s-\omega_\theta) 
+ \int_0^{\omega_\theta} (f_\epsilon - \phi_\theta)(t) k_\theta(s-t-\omega_\theta)dt \nonumber\\
&\quad+ \int_{\omega_\theta}^s \brk{\dot f_\epsilon(t) + f_\epsilon(t) - \dot A_\theta(t) - A_\theta(t) - f_\epsilon(t-\omega_\theta) } k_\theta(s-t)dt \nonumber\\
&= \int_0^s \brk{\dot f_\epsilon(t) + f_\epsilon(t) - \dot A_\theta(t) - A_\theta(t)}  k_\theta(s-t)dt - \int_{\omega_\theta}^s f_\epsilon(t-\omega_\theta)  k_\theta(s-t)dt.
\end{align}

\refstepcounter{step}
\subparagraph*{Step \arabic{step}: Rewriting $g_\epsilon$}
In order to rewrite $g_\epsilon$ we show that, for almost all $t\in (0,s)$,
\begin{align} \label{eq:fpf}
   \dot f_\epsilon(t) + f_\epsilon(t)
&=  {\dot\gamma}(\epsilon t) \brk{p_\epsilon(t,\sigma_c) - p_\epsilon(t,-\sigma_c) } .
\end{align}
First, the function $p_\epsilon$ solution to~\eqref{edp:mic} with $p_0$ as initial condition satisfies, for all $\eta \in \mathcal{D}([0,s) \times \mathbb{R})$, 
\begin{align} \label{eq:fvferme}
  - \int_0^s \int_{\mathbb{R}} p_\epsilon \brk{ \pd\eta t + {\dot\gamma}(\epsilon t) \pd\eta\sigma - \chi\eta}  
=  \int_{\mathbb{R}} p_0(\sigma) \eta(0,\sigma) d\sigma
+\int_0^s f_\epsilon(t) \eta(t,0) dt.
\end{align}
Denote $\rho$ a function of $\mathcal{D}((0,s))$, $\rho^n$ a mollifier on $\mathbb{R}$ and $\chi^n=\rho^n*\chi$.
Inserting $\eta^n(t,\sigma) = \chi^n(\sigma) \rho(t)$ in~\eqref{eq:fvferme} yields, for $n$ sufficiently large such that $\chi^n(0)=0$,
\begin{align*}
 - \int_0^s \int_{\mathbb{R}} p_\epsilon \brk{ \pd{\eta^n}t + {\dot\gamma}(\epsilon t) \pd{\eta^n}\sigma - \chi\eta^n}  
= 0
\end{align*}
which rewrites
\begin{align} \label{faible:fpf}
 - \int_0^s \dot \rho(t) \int \chi^n p_\epsilon(t,\cdot) dt 
- \int \dot \chi^n(\sigma) \int_0^s \rho(t) {\dot\gamma}(\epsilon t) p_\epsilon(t,\sigma) dt d\sigma 
+ \int_0^s \rho(t) \int \chi \chi^n p_\epsilon(t,\cdot)dt
\nonumber\\
= 0.
\end{align}
The function $t \mapsto (\sigma \mapsto p_\epsilon(t,\sigma))$ belongs to ${C}([0,s],L^1)$, see Theorem~\ref{th:exu}, so that, by the dominated convergence theorem, for all $t\in[0,s]$, 
$\int \chi^n p_\epsilon(t,\cdot)$ and $\int \chi \chi^n p_\epsilon(t,\cdot)$ converge to $f_\epsilon(t)$ defined by~\eqref{def:f} as $n$ goes to infinity. Moreover $\sigma \mapsto (t \mapsto p_\epsilon(t,\sigma))$ belongs to ${C}(\mathbb{R},L^1(0,s))$ (the proof is similar to the one in Step \ref{s:espace} of Theorem~\ref{th:exu}) so that $\int_0^s \rho(t) {\dot\gamma}(\epsilon t) p_\epsilon(t,\sigma) dt$  is continuous in $\sigma$. Passing to the limit $n\rightarrow\infty$ in the above equation yields
\begin{align*}
 - \int_0^s \dot \rho(t) f_\epsilon(t) dt  
- \int_0^s  \rho(t) {\dot\gamma}(\epsilon t) \brk{p_\epsilon(t,\sigma_c) - p_\epsilon(t,-\sigma_c) } dt
+ \int_0^s \rho(t) f_\epsilon(t) dt = 0, 
\end{align*}
hence \eqref{eq:fpf} since $f_\epsilon$ and $p_\epsilon(\cdot,\pm\sigma_c)$ belong to $L^1(0,s)$.

We then denote
\begin{align} \label{def:Qeps}
  Q_\epsilon(t) = {\dot\gamma}(\epsilon t) \brk{p_\epsilon(t,\sigma_c) - p_\epsilon(t,-\sigma_c)} - \dot A_\theta(t) - A_\theta(t) - f_\epsilon\brk{t-\omega_\theta} {\rm 1\mskip-4mu l}_{(\omega_\theta,s)}(t)
\end{align}
so that the expression \eqref{varconst:geps} on $g_\epsilon$ rewrites
\begin{align} \label{decomp:geps}
  g_\epsilon(s) 
&= \int_0^s Q_\epsilon(t) k_\theta(s-t)dt\nonumber\\
&= \frac1{1+\omega_\theta} \int_0^s Q_\epsilon(t)dt + \int_0^s Q_\epsilon(t) k_{\theta,1}(s-t)dt,
\end{align}
using the decomposition $k_\theta=\frac{1}{1+\omega_\theta} + k_{1,\theta}$ (see~\eqref{decomp:k}) that was established in Lemma~\ref{lem:cons}. We now derive estimates on the two terms of the above
expression, when $s=\frac{\theta}{\epsilon}$.

\refstepcounter{step}
\subparagraph*{Step \arabic{step}: Estimate of $\int_0^{\frac\theta\epsilon} Q_\epsilon$}
\label{s:k0}
Introduce
\begin{align} \label{def:etam}
  \eta_-(t,\sigma) = {\rm 1\mskip-4mu l}_{\Brk{-\sigma_c-{\dot\gamma}(\theta)\brk{s-t},-\sigma_c}}(\sigma) \exp{\frac{\sigma + \sigma_c}{{\dot\gamma}(\theta)}}
\end{align}
which satisfies, in $\mathcal{D}'([0,s)\times \mathbb{R})$,
\begin{align*}
  \pd{\eta_-}\sigma = 
- \delta_{-\sigma_c}(\sigma)
+\delta_{-\sigma_c-{\dot\gamma}(\theta)\brk{s-t} }(\sigma) \exp{t-s}  
+ \frac1{{\dot\gamma}(\theta)} {\rm 1\mskip-4mu l}_{\Brk{-\sigma_c-{\dot\gamma}(\theta)\brk{s-t},-\sigma_c}}(\sigma) \exp{\frac{\sigma + \sigma_c}{{\dot\gamma}(\theta)}}
\end{align*}
and
\begin{align*}
- \pd{\eta_-}t - {\dot\gamma}(\epsilon t) \pd{\eta_-}\sigma + \chi\eta_-   
&={\dot\gamma}(\epsilon t) \delta_{-\sigma_c}(\sigma)
+({\dot\gamma}(\theta)-{\dot\gamma}(\epsilon t)) \delta_{-\sigma_c-{\dot\gamma}(\theta)\brk{s-t} }(\sigma) \exp{t-s} 
\\
&\quad + \frac{{\dot\gamma}(\theta)-{\dot\gamma}(\epsilon t)}{{\dot\gamma}(\theta)} {\rm 1\mskip-4mu l}_{\Brk{-\sigma_c-{\dot\gamma}(\theta)\brk{s-t},-\sigma_c}}(\sigma) \exp{\frac{\sigma + \sigma_c}{{\dot\gamma}(\theta)}}.
\end{align*}
For $n, m \in \mathbb{N}$, take as a test function 
\begin{align*}
\eta=\eta_-^{n,m}(t,\sigma) = \rho^n *  {\rm 1\mskip-4mu l}_{\Brk{-\sigma_c-{\dot\gamma}(\theta)\brk{s-t},-\sigma_c}}(\sigma) \exp{\frac{\sigma + \sigma_c}{{\dot\gamma}(\theta)}} \zeta^m_{[0,s)}(t)
\end{align*}
in \eqref{eq:fvferme} and pass to the limit in $n$ and then $m$. Here
and in the following, $\zeta^m_{[0,s)}$ denotes a
$C^\infty([0,s),{\mathbb R})$ function with compact support in
$[0,s)$, such that $\zeta^m_{[0,s)}$ converges pointwise to ${\rm
  1\mskip-4mu l}_{[0,s)}$. 

We omit
the details, the arguments being similar to those in~\eqref{faible:fpf}. 
We obtain
\begin{equation}\label{eq:mps_bis}
\begin{aligned}
  \int_0^{s}  {\dot\gamma}(\epsilon t) p_\epsilon(t,-\sigma_c)dt
+\int_0^{s} \brk{{\dot\gamma}(\theta)-{\dot\gamma}(\epsilon t)} 
p_\epsilon\brk{t,-\sigma_c-{\dot\gamma}(\theta) \brk{s-t}}\exp{t-s}dt\\
+\int_0^{s} \int_{-\sigma_c-{\dot\gamma}(\theta)\brk{s-t}}^{-\sigma_c} 
\frac{{\dot\gamma}(\theta)-{\dot\gamma}(\epsilon t)}{{\dot\gamma}(\theta)}
  p_\epsilon(t,\sigma) \exp{\frac{\sigma+\sigma_c}{{\dot\gamma}(\theta)}} d\sigma dt\\
=  \int_{-\sigma_c-{\dot\gamma}(\theta) s}^{-\sigma_c} p_0(\sigma) \exp{\frac{\sigma+\sigma_c}{{\dot\gamma}(\theta)}} d\sigma.
 \end{aligned}
\end{equation}
With changes of variable $\sigma=-\sigma_c-{\dot\gamma}(\theta)\brk{u-t}$ and $\sigma=-\sigma_c- {\dot\gamma}(\theta) t$ in the last two integrals, this rewrites
\begin{align} \label{eq:mps}
   \int_0^{s}  {\dot\gamma}(\epsilon t) p_\epsilon(t,-\sigma_c)dt
+ \int_0^{s} \brk{{\dot\gamma}(\theta)-{\dot\gamma}(\epsilon t)} 
p_\epsilon\brk{t,-\sigma_c-{\dot\gamma}(\theta) \brk{s-t}}\exp{t-s}dt
\nonumber\\
+\int_0^{s} \int_t^{s} \brk{{\dot\gamma}(\theta)-{\dot\gamma}(\epsilon t)}
  p_\epsilon(t,-\sigma_c-{\dot\gamma}(\theta) (u-t)) \exp{t-u} du dt
  \nonumber\\
={\dot\gamma}(\theta) \int_0^{s} p_0(-\sigma_c - {\dot\gamma}(\theta) t) \exp{-t} dt.
\end{align}
Let us assume that $s > 2 \omega_\theta$ and introduce
 \begin{align} \label{def:etap}
   \eta_+(t,\sigma)  
&=- {\rm 1\mskip-4mu l}_{\Brk{\sigma_c-{\dot\gamma}(\theta)\brk{s-t},-\sigma_c}}(\sigma) \exp{\frac{\sigma+\sigma_c}{{\dot\gamma}(\theta)}} {\rm 1\mskip-4mu l}_{(0,s-2\omega_\theta)}(t)
\nonumber\\&\quad
+{\rm 1\mskip-4mu l}_{\Brk{-\sigma_c,\sigma_c-{\dot\gamma}(\theta)\brk{s-t}}}(\sigma)  {\rm 1\mskip-4mu l}_{(s-2\omega_\theta,s)}(t)
-{\rm 1\mskip-4mu l}_{[-\sigma_c,\sigma_c]}(\sigma)
 \end{align}
which satisfies, in $\mathcal{D}'((0,s)\times \mathbb{R})$,
\begin{align*}
  \pd{\eta_+}\sigma &= \delta_{\sigma_c} 
- \delta_{\sigma_c-{\dot\gamma}(\theta)\brk{s-t}}(\sigma)\exp{2\frac{\sigma_c}{{\dot\gamma}(\theta)}+t-s} {\rm 1\mskip-4mu l}_{(0,s-2\omega_\theta)}(t)
\\&\quad
- \frac1{{\dot\gamma}(\theta)} \exp{\frac{\sigma+\sigma_c}{{\dot\gamma}(\theta)}} {\rm 1\mskip-4mu l}_{\Brk{\sigma_c-{\dot\gamma}(\theta)\brk{s-t},-\sigma_c}}(\sigma) {\rm 1\mskip-4mu l}_{(0,s-2\omega_\theta)}(t)
- \delta_{\sigma_c-{\dot\gamma}(\theta)\brk{s-t}}(\sigma) {\rm 1\mskip-4mu l}_{(s-2\omega_\theta,s)}(t)
\end{align*}
and 
\begin{align*}
&  -\pd{\eta_+}t - {\dot\gamma}(\epsilon t)  \pd{\eta_+}\sigma + \chi\eta_+ 
\\&\quad
= - {\dot\gamma}(\epsilon t) \delta_{\sigma_c} 
- ({\dot\gamma}(\theta)-{\dot\gamma}(\epsilon t)) \delta_{\sigma_c-{\dot\gamma}(\theta)\brk{s-t}}(\sigma)\exp{2\frac{\sigma_c}{{\dot\gamma}(\theta)}+t-s} {\rm 1\mskip-4mu l}_{(0,s-2\omega_\theta)}(t)
\\&\qquad
- \frac{{\dot\gamma}(\theta)-{\dot\gamma}(\epsilon t)}{{\dot\gamma}(\theta)} \exp{\frac{\sigma+\sigma_c}{{\dot\gamma}(\theta)}}  {\rm 1\mskip-4mu l}_{\Brk{\sigma_c-{\dot\gamma}(\theta)\brk{s-t},-\sigma_c}}(\sigma) {\rm 1\mskip-4mu l}_{(0,s-2\omega_\theta)}(t)
\\&\qquad
-({\dot\gamma}(\theta)-{\dot\gamma}(\epsilon t)) \delta_{\sigma_c-{\dot\gamma}(\theta)\brk{s-t}}(\sigma) {\rm 1\mskip-4mu l}_{(s-2\omega_\theta,s)}(t).
\end{align*}
We again use a regularization 
\begin{align*}
\eta_+^{n,m} &= - \rho^n*{\rm 1\mskip-4mu l}_{\Brk{\sigma_c-{\dot\gamma}(\theta)\brk{s-t},-\sigma_c}}(\sigma) \exp{\frac{\sigma+\sigma_c}{{\dot\gamma}(\theta)}} \ \zeta^m_{[0,s-2\omega_\theta)}(t)
\\&\quad 
+\rho^n*{\rm 1\mskip-4mu l}_{\Brk{-\sigma_c,\sigma_c-{\dot\gamma}(\theta)\brk{s-t}}}(\sigma) \ \zeta^m_{(s-2\omega_\theta,s)}(t)
\\&\quad 
-\rho^n*{\rm 1\mskip-4mu l}_{[-\sigma_c,\sigma_c]}(\sigma) \ \zeta^m_{[0,s)}(t)
\end{align*}
and pass to the limit in \eqref{eq:fvferme} 
\begin{align} \label{eq:ps}
  -\int_0^{s} {\dot\gamma}(\epsilon t) p_\epsilon(t,\sigma_c)dt
- \int_0^{s-2\omega_\theta} \brk{{\dot\gamma}(\theta)-{\dot\gamma}(\epsilon t)}
 p_\epsilon\brk{t,\sigma_c-{\dot\gamma}(\theta)\brk{s-t}} \exp{\frac{2\sigma_c}{{\dot\gamma}(\theta)} +t-s}dt
\nonumber\\
-\int_0^{s-2\omega_\theta} \int_{t+2\omega_\theta}^{s}
\brk{{\dot\gamma}(\theta)-{\dot\gamma}(\epsilon t)}
 p_\epsilon(t,\sigma_c-{\dot\gamma}(\theta) (u-t)) \exp{\frac{2\sigma_c}{{\dot\gamma}(\theta)}+t-u}du dt
 \nonumber\\
-\int_{s-2\omega_\theta}^{s} \brk{{\dot\gamma}(\theta)-{\dot\gamma}(\epsilon t)}   p_\epsilon\brk{t,\sigma_c - {\dot\gamma}(\theta)\brk{s-t}}dt
\nonumber\\
= -{\dot\gamma}(\theta) \int_0^{s} p_0(\sigma_c-{\dot\gamma}(\theta) t) \exp{-\frac1{{\dot\gamma}(\theta)} \int_{\sigma_c-{\dot\gamma}(\theta) t}^{\sigma_c} \chi}dt
- \int_0^{s - \frac{\sigma_c}{{\dot\gamma}(\theta)}} f_\epsilon .
\end{align}
In addition, from its definition~\eqref{def:Atheta}, we know that
$A_\theta$ satisfies (see~\eqref{eq:ApointplusA})
\begin{align} \label{eq:atm}
  \int_0^s \dot A_\theta + A_\theta 
  &= 
-{\dot\gamma}(\theta) \int_0^{s} p_0(-\sigma_c - {\dot\gamma}(\theta) t) \exp{-t} dt
 \nonumber\\
&\quad + {\dot\gamma}(\theta) \int_0^{s} p_0(\sigma_c-{\dot\gamma}(\theta) t) \exp{-\frac1{{\dot\gamma}(\theta)} \int_{\sigma_c-{\dot\gamma}(\theta) t}^{\sigma_c} \chi}dt.
\end{align}
Summing up expressions~\eqref{eq:mps}, \eqref{eq:ps} and \eqref{eq:atm}, we obtain
\begin{align*}
   \int_0^{s} Q_\epsilon
&=
\int_0^{s} \brk{{\dot\gamma}(\theta)-{\dot\gamma}(\epsilon t)} 
p_\epsilon\brk{t,-\sigma_c-{\dot\gamma}(\theta) \brk{s-t}}\exp{t-s}dt\\
&\quad+\int_0^{s} \int_t^{s} \brk{{\dot\gamma}(\theta)-{\dot\gamma}(\epsilon t)}
  p_\epsilon(t,-\sigma_c-{\dot\gamma}(\theta) (u-t)) \exp{t-u} du dt\\
&\quad-\int_0^{s-2\omega_\theta} \int_{t+2\omega_\theta}^{s}
\brk{{\dot\gamma}(\theta)-{\dot\gamma}(\epsilon t)}
 p_\epsilon(t,\sigma_c-{\dot\gamma}(\theta) (v-t)) \exp{2\omega_\theta+t-v}dv dt\\
&\quad- \int_0^{s-2\omega_\theta} \brk{{\dot\gamma}(\theta)-{\dot\gamma}(\epsilon t)}
 p_\epsilon\brk{t,\sigma_c-{\dot\gamma}(\theta)\brk{s-t}} \exp{2\omega_\theta +t-s}dt\\
&\quad-\int_{s-2\omega_\theta}^{s} \brk{{\dot\gamma}(\theta)-{\dot\gamma}(\epsilon t)}   p_\epsilon\brk{t,\sigma_c - {\dot\gamma}(\theta)\brk{s-t}}dt.
\end{align*}
Taking $s=\frac\theta\epsilon$ and summing up the second and the third term (with the change of variable $u=v-2\omega_\theta$ in the third term), this rewrites 
\begin{align} \label{eq:Qc}
  \int_0^{\frac\theta\epsilon}  Q_\epsilon
&=\int_0^{\frac\theta\epsilon} \brk{{\dot\gamma}(\theta)-{\dot\gamma}(\epsilon t)} 
p_\epsilon\brk{t,-\sigma_c-{\dot\gamma}(\theta) \brk{\frac\theta\epsilon-t}
}\exp{t-\frac\theta\epsilon} dt \nonumber\\
&\quad+\int_{\frac\theta\epsilon-2\omega_\theta}^{\frac\theta\epsilon}
\int_t^{\frac\theta\epsilon}  \brk{{\dot\gamma}(\theta)-{\dot\gamma}(\epsilon t)}  p_\epsilon(t,-\sigma_c-{\dot\gamma}(\theta) (u-t)) \exp{t-u} du dt \nonumber\\
&\quad + 
\int_0^{\frac\theta\epsilon-2\omega_\theta} \int_{\frac\theta\epsilon-2\omega_\theta}^{\frac\theta\epsilon}
\brk{{\dot\gamma}(\theta)-{\dot\gamma}(\epsilon t)}  p_\epsilon(t,-\sigma_c-{\dot\gamma}(\theta) (u-t)) \exp{t-u} du dt \nonumber\\
&\quad-\int_0^{\frac\theta\epsilon-2\omega_\theta} \brk{{\dot\gamma}(\theta)-{\dot\gamma}(\epsilon t)}
 p_\epsilon\brk{t,\sigma_c-{\dot\gamma}(\theta)\brk{\frac\theta\epsilon-t}} \exp{2\omega_\theta +t-\frac\theta\epsilon}dt \nonumber\\
&\quad-\int_{\frac\theta\epsilon-2\omega_\theta}^{\frac\theta\epsilon} \brk{{\dot\gamma}(\theta)-{\dot\gamma}(\epsilon t)}  p_\epsilon\brk{t,\sigma_c - {\dot\gamma}(\theta)\brk{\frac\theta\epsilon-t}}dt.
\end{align}
Using the Lipschitz property of~${\dot\gamma}$ and $L^\infty$-bound~\eqref{maj:pli} on $p_\epsilon$, this implies
\begin{align*}
  \Abs{\int_0^{\frac\theta\epsilon}  Q_\epsilon}
&\le C_\infty L_{\dot\gamma} \int_0^{\frac\theta\epsilon}
 (\theta-\epsilon t) \exp{t-\frac\theta\epsilon} dt\\
&\quad+C_\infty L_{\dot\gamma}
\int_{\frac\theta\epsilon-2\omega_\theta}^{\frac\theta\epsilon}
\int_t^{\frac\theta\epsilon}   (\theta-\epsilon t) \exp{t-u} du dt\\
&\quad + C_\infty L_{\dot\gamma}
\int_0^{\frac\theta\epsilon-2\omega_\theta} \int_{\frac\theta\epsilon-2\omega_\theta}^{\frac\theta\epsilon}
 (\theta-\epsilon t) \exp{t-u} du dt\\
&\quad+C_\infty L_{\dot\gamma} \int_0^{\frac\theta\epsilon-2\omega_\theta} 
 (\theta-\epsilon t) \exp{2\omega_\theta +t-\frac\theta\epsilon}dt\\
&\quad+C_\infty L_{\dot\gamma} \int_{\frac\theta\epsilon-2\omega_\theta}^{\frac\theta\epsilon} 
(\theta-\epsilon t)dt.
\end{align*} 
For a constant $\alpha<0$, we have
\begin{align} \label{inteps}
  \int_0^{\frac\theta\epsilon}  (\theta-\epsilon v) \exp{\alpha\brk{\frac\theta\epsilon-v}} dv
&=\frac\epsilon{\alpha^2} \int_0^{-\frac{\alpha\theta}\epsilon} u\ \exp{-u}du \nonumber\\
&<\frac\epsilon{\alpha^2} \int_0^\infty u\ \exp{-u}du.
\end{align}
Using estimate~\eqref{inteps} or variants, we find  
\begin{align*}
 \Abs{\int_0^{\frac\theta\epsilon} Q_\epsilon} 
&\le C_\infty L_{\dot\gamma} \epsilon \bigg(
\int_0^\infty u\ \exp{-u}du + 4 \omega_\theta^2 + (1+\exp{2\omega_\theta}) \int_0^\infty u\ \exp{-u}du 
\\
&\quad \phantom{C_\infty L_{\dot\gamma} \epsilon \bigg(}
+  \exp{2\omega_\theta} \int_0^\infty u\ \exp{-u}du + 2 \omega_\theta^2
\bigg) \\
&\le  C_\infty L_{\dot\gamma} \epsilon \brk{2 + 6\omega_\theta^2 + 2 \exp{2\omega_\theta}}.
\end{align*}
Since $\omega_\theta < \dfrac{\sigma_c}{m_{\dot\gamma}}$, we obtain
\begin{align} \label{conv:c0eps}
\Abs{\int_0^{\frac\theta\epsilon} Q_\epsilon} 
&\le K C_\infty L_{\dot\gamma}  \epsilon,
 \end{align}
with $K$ a constant that is independent from $\theta$ and
$\epsilon$. Throughout the rest of the proof below, we will likewise
denote by $K$ such a constant, whose precise value may change from one
occurrence to another.

\refstepcounter{step}
\subparagraph*{Step \arabic{step}: Estimate of $\int_0^{\frac\theta\epsilon} Q_\epsilon(t) k_{\theta,1}\brk{\frac\theta\epsilon-t}dt$}
\label{s:k1}
Inserting expression \eqref{def:Atheta} of $A_\theta$
(see~\eqref{eq:atm} above for a similar computation), $Q_\epsilon$ defined by \eqref{def:Qeps} satisfies
\begin{align} \label{eq:qepsk1}
&  \int_0^{\frac\theta\epsilon} Q_\epsilon(t) k_{\theta,1}\brk{\frac\theta\epsilon-t}dt
=\int_0^{\frac\theta\epsilon} {\dot\gamma}(\epsilon t)(p_\epsilon(t,\sigma_c)-p_\epsilon(t,-\sigma_c)) k_{\theta,1}\brk{\frac\theta\epsilon-t}dt\nonumber\\
&\quad+{\dot\gamma}(\theta) \int_0^{\frac\theta\epsilon} \brk{p_0(-\sigma_c - {\dot\gamma}(\theta) t) \exp{-t}
-   p_0(\sigma_c - {\dot\gamma}(\theta) t)\exp{-\frac1{{\dot\gamma}(\theta)} \int_{\sigma_c-{\dot\gamma}(\theta) t}^{\sigma_c} \chi}} k_{\theta,1}\brk{\frac\theta\epsilon-t} dt\nonumber\\
&\quad-\int_{\omega_\theta}^{\frac\theta\epsilon} f_\epsilon(t-\omega_\theta) k_{\theta,1}\brk{\frac\theta\epsilon-t}dt.
\end{align}
For further use, notice that the integral in the second line above can
be rewritten as:
\begin{align}
& \int_0^{\frac\theta\epsilon} \brk{p_0(-\sigma_c - {\dot\gamma}(\theta) t) \exp{-t}
-   p_0(\sigma_c - {\dot\gamma}(\theta) t)\exp{-\frac1{{\dot\gamma}(\theta)}
  \int_{\sigma_c-{\dot\gamma}(\theta) t}^{\sigma_c} \chi}}
k_{\theta,1}\brk{\frac\theta\epsilon-t} dt\nonumber\\
&= \int_{-\sigma_c - {\dot\gamma}(\theta) \frac\theta\epsilon}^{-\sigma_c} p_0(\sigma) \exp{\frac{\sigma+\sigma_c}{{\dot\gamma}(\theta)}}
k_{\theta,1}\brk{\frac\theta\epsilon+
  \frac{\sigma+\sigma_c}{{\dot\gamma}(\theta)}} dt\nonumber\\
&\quad - \int_{\sigma_c - {\dot\gamma}(\theta) \frac\theta\epsilon}^{\sigma_c}  p_0(\sigma)\exp{-\frac1{{\dot\gamma}(\theta)}
  \int_{\sigma}^{\sigma_c} \chi}
k_{\theta,1}\brk{\frac\theta\epsilon +
  \frac{\sigma-\sigma_c}{{\dot\gamma}(\theta)}} dt.  \label{eq:qepsk1_bis}
\end{align}

In order to rewrite the first term of the right-hand side, we use again the functions $\eta_-$ and $\eta_+$ respectively defined by
\eqref{def:etam} and \eqref{def:etap}. Using a regularization of
$\eta_-(t,\sigma)
k_{\theta,1}\brk{\frac\theta\epsilon-t+\frac{\sigma+\sigma_c}{{\dot\gamma}(\theta)}}$
as test function in \eqref{eq:fvferme} (with $s=\theta/\epsilon$) and then passing to the limit, we obtain
\begin{align*}
   \int_0^{\frac\theta\epsilon} \int_{\mathbb{R}} \brk{ -\pd{\eta_-}t - {\dot\gamma}(\epsilon t) \pd{\eta_-}\sigma + \chi\eta_-}(t,\sigma) k_{\theta,1}\brk{\frac\theta\epsilon-t+\frac{\sigma+\sigma_c}{{\dot\gamma}(\theta)}}  p_\epsilon(t,\sigma)dt d\sigma\\
+\int_0^{\frac\theta\epsilon} \int_{\mathbb{R}} \frac{{\dot\gamma}(\theta) - {\dot\gamma}(\epsilon t)}{{\dot\gamma}(\theta)}\eta_-(t,\sigma) \dot k_{\theta,1}\brk{\frac\theta\epsilon-t+\frac{\sigma+\sigma_c}{{\dot\gamma}(\theta)}} p_\epsilon(t,\sigma) dt d\sigma\\
=  \int_{\mathbb{R}} p_0(\sigma) \eta_-(0,\sigma) k_{\theta,1}\brk{\frac\theta\epsilon+\frac{\sigma+\sigma_c}{{\dot\gamma}(\theta)}} d\sigma.
\end{align*}
This rewrites (using similar computations as in~\eqref{eq:mps_bis} above)
\begin{align} \label{eq:pmk1}
    \int_0^{\frac\theta\epsilon}  {\dot\gamma}(\epsilon t) p_\epsilon(t,-\sigma_c) k_{\theta,1}\brk{\frac\theta\epsilon-t} dt\nonumber\\
+  k_{\theta,1}(0)\int_0^{\frac\theta\epsilon} \brk{{\dot\gamma}(\theta)-{\dot\gamma}(\epsilon t)} 
p_\epsilon\brk{t,-\sigma_c-{\dot\gamma}(\theta) \brk{\frac\theta\epsilon-t}}\exp{t-\frac\theta\epsilon} dt\nonumber\\
+\int_0^{\frac\theta\epsilon} \int_{-\sigma_c-{\dot\gamma}(\theta)\brk{\frac\theta\epsilon-t}}^{-\sigma_c} 
\frac{{\dot\gamma}(\theta)-{\dot\gamma}(\epsilon t)}{{\dot\gamma}(\theta)}
  p_\epsilon(t,\sigma) \exp{\frac{\sigma+\sigma_c}{{\dot\gamma}(\theta)}} (k_{\theta,1}+\dot k_{\theta,1})\brk{\frac\theta\epsilon-t+\frac{\sigma+\sigma_c}{{\dot\gamma}(\theta)}} d\sigma dt \nonumber\\
=  \int_{-\sigma_c-{\dot\gamma}(\theta) \frac\theta\epsilon}^{-\sigma_c} p_0(\sigma) \exp{\frac{\sigma+\sigma_c}{{\dot\gamma}(\theta)}}  k_{\theta,1}\brk{\frac\theta\epsilon+\frac{\sigma+\sigma_c}{{\dot\gamma}(\theta)}}d\sigma.
\end{align}
Similarly, using a regularization of $\eta_+(t,\sigma) k_{\theta,1}\brk{\frac\theta\epsilon-t+\frac{\sigma-\sigma_c}{{\dot\gamma}(\theta)}}$ as test function in \eqref{eq:fvferme}, we obtain
\begin{align*}
    \int_0^{\frac\theta\epsilon} \int_{\mathbb{R}} \brk{ -\pd{\eta_+}t - {\dot\gamma}(\epsilon t) \pd{\eta_+}\sigma + \chi\eta_+}(t,\sigma) k_{\theta,1}\brk{\frac\theta\epsilon-t+\frac{\sigma-\sigma_c}{{\dot\gamma}(\theta)}}  p_\epsilon(t,\sigma)dt d\sigma\\
+\int_0^{\frac\theta\epsilon} \int_{\mathbb{R}} \frac{{\dot\gamma}(\theta) - {\dot\gamma}(\epsilon t)}{{\dot\gamma}(\theta)}\eta_+(t,\sigma) \dot k_{\theta,1}\brk{\frac\theta\epsilon-t+\frac{\sigma-\sigma_c}{{\dot\gamma}(\theta)}} p_\epsilon(t,\sigma) dt d\sigma\\
=  \int_{\mathbb{R}} p_0(\sigma) \eta_+(0,\sigma) k_{\theta,1}\brk{\frac\theta\epsilon+\frac{\sigma-\sigma_c}{{\dot\gamma}(\theta)}} d\sigma
+ \int_0^{\frac\theta\epsilon} f_\epsilon(t) \eta_+(t,0) k_{\theta,1}\brk{\frac\theta\epsilon-t-\frac{\sigma_c}{{\dot\gamma}(\theta)}} dt,
\end{align*}
so that (using similar computations as in~\eqref{eq:ps} above)
\begin{align} \label{eq:ppk1}
    -\int_0^{\frac\theta\epsilon} {\dot\gamma}(\epsilon t) p_\epsilon(t,\sigma_c)  k_{\theta,1}\brk{\frac\theta\epsilon-t} dt\nonumber\\
- k_{\theta,1}(0) \int_0^{\frac\theta\epsilon-2\omega_\theta} \brk{{\dot\gamma}(\theta)-{\dot\gamma}(\epsilon t)}
 p_\epsilon\brk{t,\sigma_c-{\dot\gamma}(\theta)\brk{\frac\theta\epsilon-t}} \exp{2\omega_\theta +t-\frac\theta\epsilon}dt
\nonumber\\
- k_{\theta,1}(0) \int_{\frac\theta\epsilon-2\omega_\theta}^{\frac\theta\epsilon} \brk{{\dot\gamma}(\theta)-{\dot\gamma}(\epsilon t)}   p_\epsilon\brk{t,\sigma_c - {\dot\gamma}(\theta)\brk{\frac\theta\epsilon-t}}dt
\nonumber\\
-\int_0^{\frac\theta\epsilon-2\omega_\theta} \int_{\sigma_c-{\dot\gamma}(\theta)\brk{\frac\theta\epsilon-t}}^{-\sigma_c} 
\frac{{\dot\gamma}(\theta)-{\dot\gamma}(\epsilon t)}{{\dot\gamma}(\theta)}
  p_\epsilon(t,\sigma) \exp{\frac{\sigma+\sigma_c}{{\dot\gamma}(\theta)}} (k_{\theta,1}+\dot k_{\theta,1})\brk{\frac\theta\epsilon-t+\frac{\sigma-\sigma_c}{{\dot\gamma}(\theta)}} d\sigma dt\nonumber\\
-\int_{\frac\theta\epsilon-2\omega_\theta}^{\frac\theta\epsilon} \int_{\sigma_c-{\dot\gamma}(\theta)\brk{\frac\theta\epsilon-t}}^{-\sigma_c} \frac{{\dot\gamma}(\theta)-{\dot\gamma}(\epsilon t)}{{\dot\gamma}(\theta)}
  p_\epsilon(t,\sigma)  \dot k_{\theta,1}\brk{\frac\theta\epsilon-t+\frac{\sigma-\sigma_c}{{\dot\gamma}(\theta)}} d\sigma dt\nonumber\\
-\int_0^{\frac\theta\epsilon} \int_{-\sigma_c}^{\sigma_c} 
\frac{{\dot\gamma}(\theta)-{\dot\gamma}(\epsilon t)}{{\dot\gamma}(\theta)}
  p_\epsilon(t,\sigma)  \dot k_{\theta,1}\brk{\frac\theta\epsilon-t+\frac{\sigma-\sigma_c}{{\dot\gamma}(\theta)}} d\sigma dt
 \nonumber\\
=-\int_{\sigma_c-{\dot\gamma}(\theta) \frac\theta\epsilon}^{-\sigma_c} p_0(\sigma) \exp{\frac{\sigma+\sigma_c}{{\dot\gamma}(\theta)}}  k_{\theta,1}\brk{\frac\theta\epsilon+\frac{\sigma-\sigma_c}{{\dot\gamma}(\theta)}}d\sigma
-\int_{-\sigma_c}^{\sigma_c} p_0(\sigma)  k_{\theta,1}\brk{\frac\theta\epsilon+\frac{\sigma-\sigma_c}{{\dot\gamma}(\theta)}}d\sigma \nonumber\\
- \int_0^{\frac\theta\epsilon - \omega_\theta} f_\epsilon(t)  k_{\theta,1}\brk{\frac\theta\epsilon-t-\omega_\theta} dt.
\end{align}
We now perform the linear combinations: \eqref{eq:qepsk1}  - ( 
\eqref{eq:pmk1} + \eqref{eq:ppk1} ). The last two term of the
right-hand side of \eqref{eq:qepsk1} cancel out with the right-hand
sides of \eqref{eq:pmk1} and \eqref{eq:ppk1} (using in particular~\eqref{eq:qepsk1_bis}) so that 
\begin{align*}
   \int_0^{\frac\theta\epsilon} Q_\epsilon(t) k_{\theta,1}\brk{\frac\theta\epsilon-t}dt
=k_{\theta,1}(0)\int_0^{\frac\theta\epsilon} \brk{{\dot\gamma}(\theta)-{\dot\gamma}(\epsilon t)} 
p_\epsilon\brk{t,-\sigma_c-{\dot\gamma}(\theta) \brk{\frac\theta\epsilon-t}}\exp{t-\frac\theta\epsilon} dt\nonumber\\
+\int_0^{\frac\theta\epsilon} \int_{-\sigma_c-{\dot\gamma}(\theta)\brk{\frac\theta\epsilon-t}}^{-\sigma_c} 
\frac{{\dot\gamma}(\theta)-{\dot\gamma}(\epsilon t)}{{\dot\gamma}(\theta)}
  p_\epsilon(t,\sigma) \exp{\frac{\sigma+\sigma_c}{{\dot\gamma}(\theta)}} (k_{\theta,1}+\dot k_{\theta,1})\brk{\frac\theta\epsilon-t+\frac{\sigma+\sigma_c}{{\dot\gamma}(\theta)}} d\sigma dt \nonumber\\
- k_{\theta,1}(0) \int_0^{\frac\theta\epsilon-2\omega_\theta} \brk{{\dot\gamma}(\theta)-{\dot\gamma}(\epsilon t)}
 p_\epsilon\brk{t,\sigma_c-{\dot\gamma}(\theta)\brk{\frac\theta\epsilon-t}} \exp{2\omega_\theta +t-\frac\theta\epsilon}dt
 \nonumber\\
- k_{\theta,1}(0) \int_{\frac\theta\epsilon-2\omega_\theta}^{\frac\theta\epsilon} \brk{{\dot\gamma}(\theta)-{\dot\gamma}(\epsilon t)}   p_\epsilon\brk{t,\sigma_c - {\dot\gamma}(\theta)\brk{\frac\theta\epsilon-t}}dt
\nonumber\\
-\int_0^{\frac\theta\epsilon-2\omega_\theta} \int_{\sigma_c-{\dot\gamma}(\theta)\brk{\frac\theta\epsilon-t}}^{-\sigma_c} 
\frac{{\dot\gamma}(\theta)-{\dot\gamma}(\epsilon t)}{{\dot\gamma}(\theta)}
  p_\epsilon(t,\sigma) \exp{\frac{\sigma+\sigma_c}{{\dot\gamma}(\theta)}} (k_{\theta,1}+\dot k_{\theta,1})\brk{\frac\theta\epsilon-t+\frac{\sigma-\sigma_c}{{\dot\gamma}(\theta)}} d\sigma dt\nonumber\\
-\int_{\frac\theta\epsilon-2\omega_\theta}^{\frac\theta\epsilon} \int_{\sigma_c-{\dot\gamma}(\theta)\brk{\frac\theta\epsilon-t}}^{-\sigma_c} \frac{{\dot\gamma}(\theta)-{\dot\gamma}(\epsilon t)}{{\dot\gamma}(\theta)}
  p_\epsilon(t,\sigma)  \dot k_{\theta,1}\brk{\frac\theta\epsilon-t+\frac{\sigma-\sigma_c}{{\dot\gamma}(\theta)}} d\sigma dt\nonumber\\
-\int_0^{\frac\theta\epsilon} \int_{-\sigma_c}^{\sigma_c} \frac{{\dot\gamma}(\theta)-{\dot\gamma}(\epsilon t)}{{\dot\gamma}(\theta)}  p_\epsilon(t,\sigma)  \dot k_{\theta,1}\brk{\frac\theta\epsilon-t+\frac{\sigma-\sigma_c}{{\dot\gamma}(\theta)}} d\sigma dt.
\end{align*}
Using Lemma \ref{lem:cons}, $k_{\theta,1}$ satisfies
$\dot{k}_{\theta,1}(t)+k_{\theta,1}(t) - k_{\theta,1}(t-\omega_\theta) = 0$
(this is a consequence of~\eqref{eq:ddek} and~\eqref{decomp:k}) so that (using~\eqref{crois:k}):
\begin{align} \label{crois:dotk1}
  \Abs{\dot k_{\theta,1}(t)} \le C_0 \brk{1+\exp{b \frac{\sigma_c}{m_{\dot\gamma}}}} \exp{-bt}.
\end{align}
Using the Lipschitz property of ${\dot\gamma}$, the $L^\infty$-bound \eqref{maj:pli} of $p_\epsilon$ and estimates \eqref{crois:k} and \eqref{crois:dotk1} of $k_{\theta,1}$ and $\dot k_{\theta,1}$, we obtain 
\begin{align*}
&   \Abs{\int_0^{\frac\theta\epsilon} Q_\epsilon(t) k_{\theta,1}\brk{\frac\theta\epsilon-t}dt}\\
&\le C_0 C_\infty L_{\dot\gamma} \int_0^{\frac\theta\epsilon}
 (\theta-\epsilon t) \exp{t-\frac\theta\epsilon} dt\\
&\quad+C_0 \brk{2+\exp{b \frac{\sigma_c}{m_{\dot\gamma}}}} C_\infty L_{\dot\gamma} \int_0^{\frac\theta\epsilon}
 (\theta-\epsilon t) \frac{\exp{b\brk{t-\frac\theta\epsilon}}-\exp{t-\frac\theta\epsilon}}{1-b} dt\\
&\quad+C_0 C_\infty L_{\dot\gamma} \int_0^{\frac\theta\epsilon-2\omega_\theta} 
 (\theta-\epsilon t) \exp{2\omega_\theta +t-\frac\theta\epsilon}dt\\
&\quad+C_0 C_\infty L_{\dot\gamma} \int_{\frac\theta\epsilon-2\omega_\theta}^{\frac\theta\epsilon} 
(\theta-\epsilon t)    dt\\
&\quad+C_0 \brk{2+\exp{b \frac{\sigma_c}{m_{\dot\gamma}}}} C_\infty L_{\dot\gamma} \int_0^{\frac\theta\epsilon-2\omega_\theta} (\theta-\epsilon t) \frac{\exp{b\brk{t-\frac\theta\epsilon+2\omega_\theta}}-\exp{t-\frac\theta\epsilon+2\omega_\theta}}{1-b}dt\\
&\quad+C_0 \brk{1+\exp{b \frac{\sigma_c}{m_{\dot\gamma}}}} C_\infty L_{\dot\gamma} \int_{\frac\theta\epsilon-2\omega_\theta}^{\frac\theta\epsilon} (\theta-\epsilon t) 
\frac{\exp{-2\omega_\theta b} + \exp{b\brk{t-\frac\theta\epsilon}}}{b} dt\\
&\quad+C_0 \brk{1+\exp{b \frac{\sigma_c}{m_{\dot\gamma}}}} C_\infty L_{\dot\gamma} \int_0^{\frac\theta\epsilon} (\theta-\epsilon t) \exp{b\brk{t-\frac\theta\epsilon}} \frac{1+\exp{-2\omega_\theta b}}{b}dt.
\end{align*}
Using the estimate~\eqref{inteps} or variants and the bounds
\eqref{bornes:omega} on $\omega_\theta$, one concludes
\begin{align} \label{conv:c1eps}
  \Abs{  \int_0^{\frac\theta\epsilon} Q_\epsilon(t) k_{\theta,1}\brk{\frac\theta\epsilon-t}dt} 
  &\le K C_\infty C_0 L_{\dot\gamma}  \epsilon.
\end{align}

\refstepcounter{step}
\subparagraph*{Step \arabic{step}: Estimate of $g_\epsilon\brk{\dfrac\theta\epsilon}$}
Using the decomposition~\eqref{decomp:geps} and estimates \eqref{conv:c0eps} and~\eqref{conv:c1eps}, we find
\begin{align*} 
  \Abs{g_\epsilon\brk{\frac\theta\epsilon}} 
\le K C_\infty (1+C_0) L_{\dot\gamma} \epsilon.
\end{align*}  
Moreover, the estimate~\eqref{convc2:f} established in
Theorem~\ref{th:tlc} yields, using the bounds \eqref{bornes:omega} on $\omega_\theta$
\begin{align*}
 \Abs{\phi_\theta\brk{\frac\theta\epsilon} - \int \chi p_\infty(\theta,\cdot)}\le \tilde{C_3} \brk{\exp{-b\frac\theta\epsilon} + \exp{-\frac\theta\epsilon}}
\end{align*}
where $\tilde{C_3}>0$ is independent from $\epsilon$ and $\theta$.
By the definition \eqref{def:geps} of $g_\epsilon$, we have 
$f_\epsilon - \int \chi p_\infty(\theta,\cdot) = g_\epsilon + (\phi_\theta - \int \chi p_\infty(\theta,\cdot))$ so that
\begin{align*}
   \Abs{ f_\epsilon\brk{\frac\theta\epsilon} - \int \chi p_\infty(\theta,\cdot)}
&  \le \tilde{C_3} \brk{\exp{-b\frac\theta\epsilon} + \exp{-\frac\theta\epsilon}}+ K C_\infty (1+C_0) L_{\dot\gamma} \epsilon\\
&\le C_3 \brk{\exp{-b\frac\theta\epsilon} + \exp{-\frac\theta\epsilon} +\epsilon}
\end{align*}
where $C_3>0$ is independent from $\epsilon$ and $\theta$. This
concludes the proof of~\eqref{conv:geps}.

\refstepcounter{step}
\subparagraph*{Step \arabic{step}: Estimate of $p_\epsilon\brk{\dfrac\theta\epsilon,\sigma} - p_\infty(\theta,\sigma)$}
Recall that the scalars $\theta>0,\ \epsilon>0$ are fixed and satisfy $\frac\theta\epsilon>2\frac{\sigma_c}{m_{\dot\gamma}}$. For all $\sigma\in\mathbb R$, denote
\begin{align*}
  u_{\epsilon,\theta,\sigma} = 
{\gamma^{-1}\brk{\gamma(\theta)-\epsilon\sigma}}
\end{align*}
where $\gamma(t)=\int_0^t {\dot\gamma}(s) \,ds$.
The expression~\eqref{for:p} of $p$, that was established in Theorem~\ref{th:exu}, reads, for almost all $\sigma\in \mathbb{R}$ such that $\sigma \le
\gamma\brk{\frac\theta\epsilon}$,
\begin{align} \label{for:peps}
  p_\epsilon\brk{\frac\theta\epsilon, \sigma}  
&=p_0\brk{\sigma-\frac{\gamma(\theta)}\epsilon}
\exp{-\int_0^\frac\theta\epsilon \chi\brk{\sigma-\frac{\gamma(\theta)}\epsilon + \frac{\gamma(\epsilon v)}\epsilon}dv}
\nonumber\\&\quad
+ \frac{f_\epsilon\brk{\frac{u_{\epsilon,\theta,\sigma}}\epsilon}}{{\dot\gamma}(u_{\epsilon,\theta,\sigma})} \times
 \begin{cases}
     0 & \mbox{if }  \sigma<0 \\
     1& \mbox{if } 0<\sigma\le\sigma_c \\
     \exp{-\frac\theta\epsilon + \frac1\epsilon 
       \gamma^{-1}(\gamma(\theta) - \epsilon \sigma + \epsilon \sigma_c)}
     & \mbox{if }  \sigma_c<\sigma.
\end{cases}
\end{align}
Note that the condition $\sigma \le \gamma\brk{\frac\theta\epsilon}$ is not restrictive because we are interested in the limit $\epsilon \to 0$ for a fixed $\sigma$.
The rest of the proof depends on the value of~$\sigma$.

Let us start with the case $\sigma <0$. We have
\begin{align*}
  \Abs{p_\epsilon\brk{\frac\theta\epsilon, \sigma}} 
&\le \Norm{p_0}_{L^\infty}
\exp{-\int_0^\frac\theta\epsilon \chi\brk{\sigma-\frac{\gamma(\theta)}\epsilon + \frac{\gamma(\epsilon v)}\epsilon}dv}.
\end{align*}
We now apply Lemma~\ref{lem:intt} with $t=\frac\theta\epsilon>\frac{2\sigma_c}{m_{\dot\gamma}}$ and $\gamma_\epsilon(t)=\frac1\epsilon \gamma(\epsilon t)$ and obtain
\begin{align}\label{eq:bound_peps}
   \Abs{p_\epsilon\brk{\frac\theta\epsilon, \sigma}} &\le \Norm{p_0}_{L^\infty} \exp{\frac{2\sigma_c}{m_{\dot\gamma}}-\frac\theta\epsilon}\le K {\exp{-\frac\theta\epsilon}}.
\end{align}
Notice that~\eqref{eq:bound_peps} actually holds for all $\sigma \in \mathbb{R}$ (this will be used below).
From the expression~\eqref{eq:py} of $p_\infty$,
$p_\infty(\theta,\sigma)=0$ when $\sigma<0$. This gives \eqref{conv:pmp} for almost
all $\sigma<0$. 

Let us now consider the case $\sigma \in (0,\sigma_c]$. Let us
introduce the notation $f_\infty(\theta)=\int \chi(\sigma)
p_\infty(\theta,\sigma) \, d\sigma$. We have, for
all positive $\sigma$,
\begin{align}
&  \Abs{
\frac{f_\epsilon\brk{\frac{u_{\epsilon,\theta,\sigma}}\epsilon}}{{\dot\gamma}(u_{\epsilon,\theta,\sigma})} 
- \frac1{\sigma_c+{\dot\gamma}(\theta)}} \nonumber \\
&\quad \le \frac1{m_{\dot\gamma}} \Abs{f_\epsilon\brk{\frac{u_{\epsilon,\theta,\sigma}}\epsilon} - f_\infty(u_{\epsilon,\theta,\sigma})}
+ \frac1{m_{\dot\gamma}} \Abs{f_\infty(u_{\epsilon,\theta,\sigma}) - f_\infty(\theta)}\nonumber \\
&\qquad+\Abs{\frac{f_\infty(\theta)}{{\dot\gamma}(u_{\epsilon,\theta,\sigma})} - \frac1{\sigma_c+{\dot\gamma}(\theta)}}\nonumber \\
& \quad \le K (C_3+1) \brk{\exp{-b\frac\theta\epsilon} + \exp{-\frac\theta\epsilon}+\epsilon}, \label{eq:bound_feps}
\end{align}
using the estimate~\eqref{conv:geps} on $f_\epsilon\brk{\frac\cdot\epsilon}-f_\infty$, the Lipschitz property and the boundedness of~${\dot\gamma}$ and the expression~\eqref{eq:py} which gives
$f_\infty(\theta) =
\frac{{\dot\gamma}(\theta)}{\sigma_c+{\dot\gamma}(\theta)}$. Here, we also used
the fact that $u_{\epsilon,\theta,\sigma} \ge \theta - \frac{\epsilon
  \sigma}{m_{\dot\gamma}}$, which is a consequence of the Lipschitz property
of $\gamma^{-1}$: $|u_{\epsilon,\theta,\sigma} - \theta| \le \frac{\epsilon
  \sigma}{m_{\dot\gamma}}$. For almost all $\sigma \in (0,\sigma_c]$, we deduce (using again~\eqref{eq:bound_peps})
\begin{align*}
&  \Abs{p_\epsilon\brk{\frac\theta\epsilon, \sigma} - \frac1{\sigma_c+{\dot\gamma}(\theta)} }\\
&\quad\le \Abs{ p_0\brk{\sigma-\frac{\gamma(\theta)}\epsilon}
\exp{-\int_0^\frac\theta\epsilon \chi\brk{\sigma-\frac{\gamma(\theta)}\epsilon + \frac{\gamma(\epsilon v)}\epsilon}dv}}   
+\Abs{\frac{f_\epsilon\brk{\frac{u_{\epsilon,\theta,\sigma}}\epsilon}}{{\dot\gamma}(u_{\epsilon,\theta,\sigma})} - \frac1{\sigma_c+{\dot\gamma}(\theta)} }  \\
&\quad\le K {\exp{-\frac\theta\epsilon}} + K (C_3+1)\brk{\exp{-b\frac\theta\epsilon}+\exp{-\frac\theta\epsilon} +\epsilon}. 
\end{align*}
Moreover, we have, from \eqref{eq:py},
$p_\infty(\theta,\sigma)=\frac1{\sigma_c+{\dot\gamma}(\theta)}$ for almost
all  $\sigma\in (0,\sigma_c]$. This proves \eqref{conv:pmp} in this
region of $\sigma$.

We now eventually consider the case $\sigma>\sigma_c$. Applying the Taylor-Lagrange theorem on $\gamma^{-1}$ at $\gamma(\theta)$, 
we have (using the fact that ${\dot\gamma}$ is bounded from above)
\begin{align*}
 \Abs{  \frac{\gamma^{-1}(\gamma(\theta) - \epsilon (\sigma - \sigma_c))-\theta}{-\epsilon(\sigma-\sigma_c)}
 - \frac1{{\dot\gamma}(\theta)}} \le K \epsilon (\sigma -\sigma_c) .
\end{align*}
Using that the function $x\mapsto \exp{-x}$ is 1-Lipschitz on $[0,\infty)$, this implies
\begin{align*}
  \Abs{
\exp{-\frac\theta\epsilon + \frac1\epsilon \gamma^{-1}(\gamma(\theta) - \epsilon \sigma + \epsilon \sigma_c)}
 -\exp{-\frac{\sigma-\sigma_c}{{\dot\gamma}(\theta)}}}
 \le K \epsilon (\sigma -\sigma_c)^2.
\end{align*}
For almost all $\sigma>\sigma_c$, $p_\infty$ reads (see \eqref{eq:py})
\begin{align*}
 p_\infty(\theta,\sigma)=\frac1{\sigma_c+{\dot\gamma}(\theta)} \exp{\frac{\sigma_c-\sigma}{{\dot\gamma}(\theta)}}
\end{align*}
and thus, we have (using~\eqref{eq:bound_peps} and~\eqref{eq:bound_feps})
\begin{align*}
&  \Abs{p_\epsilon\brk{\frac\theta\epsilon, \sigma}  - p_\infty(\theta,\sigma) }\\
&\quad\le \Abs{ p_0\brk{\sigma-\frac{\gamma(\theta)}\epsilon}
\exp{-\int_0^\frac\theta\epsilon \chi\brk{\sigma-\frac{\gamma(\theta)}\epsilon + \frac{\gamma(\epsilon v)}\epsilon}dv}}  
\\&\qquad+\Abs{\frac{f_\epsilon\brk{\frac{u_{\epsilon,\theta,\sigma}}\epsilon}}{{\dot\gamma}(u_{\epsilon,\theta,\sigma})} - \frac1{\sigma_c+{\dot\gamma}(\theta)} } \exp{-\frac\theta\epsilon + \frac1\epsilon   \gamma^{-1}(\gamma(\theta) - \epsilon \sigma + \epsilon \sigma_c)}\\
&\qquad+  \frac1{\sigma_c+{\dot\gamma}(\theta)} \Abs{\exp{-\frac\theta\epsilon + \frac1\epsilon   \gamma^{-1}(\gamma(\theta) - \epsilon \sigma + \epsilon \sigma_c)}
 -\exp{-\frac{\sigma-\sigma_c}{{\dot\gamma}(\theta)}} }\\ 
& \quad \le K {\exp{-\frac\theta\epsilon}} + K (C_3+1) \brk{\exp{-b\frac\theta\epsilon}+\exp{-\frac\theta\epsilon} +\epsilon}+ K \epsilon (\sigma -\sigma_c)^2. 
\end{align*}
This proves~\eqref{conv:pmp} and ends the proof.
\hfill\endproof
\setcounter{step}{0}

\section{Macroscopic limit}
\label{sec:passage}
The purpose of this section is to obtain a macroscopic limit for the
equation we have been studying. 

For $\epsilon>0$ presumably small, we first introduce the notation  $\widetilde u(\theta) =
u\brk{\dfrac\theta\epsilon}$, for any real-valued function $u$
of the real variable $\theta/\epsilon$. The small parameter~$\epsilon$ encodes the
discrepancy between the typical time of variation of the macroscopic
shear rate ${\dot\gamma}(t)$ and the typical mesoscopic time~$t$ with which
the solution to~\eqref{edp:mic} varies. The parameter~$\epsilon$ will
therefore be the small parameter
on which our macroscopic limit is performed. To obtain a macroscopic
limit of our equation~\eqref{edp:mic}, we will look at a specific macroscopic time,
denoted by $\theta$, related to the mesoscopic time~$t$ by
$\theta=\epsilon\, t$. Letting $\epsilon$ vanish, we will obtain the
corresponding macroscopic behavior of our mesoscopic quantities. In
this latter process, we will of course use the results of the previous
section, since formally, given a macroscopic time~$\theta$, the
corresponding mesoscopic time
$t=\theta/\epsilon$ is a long time limit.

\medskip

We begin by  formally multiplying~\eqref{edp:mic}
respectively by $\sigma$ and $\chi(\sigma)$ and integrating in
$\sigma$. We next evaluate at the time
$\dfrac\theta\epsilon$ the two equations obtained.  This gives the following  system of equations on $\widetilde{\tau_\epsilon}$ and $\widetilde{f_\epsilon}$, the quantities associated with $p_\epsilon$ solution to~\eqref{edp:mic},
\begin{subequations} 
\label{eq:sapprox}
  \begin{empheq}[left=\empheqlbrace]{align}
    \epsilon \deriv{\widetilde{\tau_\epsilon}}\theta (\theta) &= - \int \chi(\sigma) \sigma \widetilde{p_\epsilon}(\theta,\sigma) d\sigma  +{\dot\gamma} (\theta),\label{eq:sapproxa} \\
    \epsilon \deriv{\widetilde{f_\epsilon}} \theta (\theta)&=-\widetilde{f_\epsilon}(\theta)+ {\dot\gamma} (\theta)
    (\widetilde{p_\epsilon}(\theta,\sigma_c)-\widetilde{p_\epsilon}(\theta,-\sigma_c)).\label{eq:sapproxb}
  \end{empheq}
\end{subequations}
The difficulty is that this system is not closed in the couple of
unknown functions~$(\widetilde{\tau_\epsilon},\widetilde{f_\epsilon})$
since~$\widetilde{p_\epsilon}$ still appears. 
We next intend to ``eliminate'' $\widetilde{p_\epsilon}$ from this
system, and thereby obtain a system of ordinary differential equations,
the solution of which is an approximation, for~$\epsilon$ sufficiently
small,  of $\widetilde{\tau_\epsilon}$
and $\widetilde{f_\epsilon}$. There are indeed many options to do so. We
present two sets of equations which we can derive and that, in a sense
made precise below, are equivalent to system~\eqref{eq:sapprox}. The
precise results are contained in two theorems we now successively state
and prove, namely Theorem~\ref{th:mac1} and Theorem~\ref{th:mac2}.

\medskip

\begin{theorem} 
\label{th:mac1}
For~$\epsilon >0$, consider $\tau_\epsilon^*$ the solution to the
following 
differential equation in ``macroscopic'' time~$\theta$~:
\begin{align} \label{eq:mac1}
  \epsilon\deriv{\tau_\epsilon^*} \theta &= - \tau_\epsilon^* + \frac{\sigma_c^2}{2(\sigma_c+{\dot\gamma}(\theta))}+{\dot\gamma}(\theta)
\end{align}
supplied with any initial condition~$\tau^*(0)$ independent of $\epsilon$.
   Under the assumptions of
Theorem~\ref{th:tle}, consider the function $p_\epsilon$ solution to~\eqref{edp:mic}.
Then, there exists a constant $C$, independent from $\epsilon$ and
$\theta$ provided  $\frac\theta\epsilon
>2\frac{\sigma_c}{m_{\dot\gamma}}$ (with $m_{\dot\gamma}$ the lower bound of ${\dot\gamma}$ in~\eqref{hyp:gamma}), such that 
\begin{align} \label{conv:mac1}
  \Abs{\tau_\epsilon\brk{\frac{\theta}\epsilon} -  \tau_\epsilon^*(\theta)} 
\le C\brk{\frac1\theta+1}\epsilon,
\end{align}
where (in accordance with~\eqref{def:tau}) $\tau_\epsilon(t)=\int \sigma p_\epsilon(t,\sigma) d\sigma$
denotes the stress associated to $p_\epsilon$ solution to~\eqref{edp:mic}.
\end{theorem}

\begin{remark}
  The above result holds whatever the initial condition
  $\tau^*(0)$ for the equation~\eqref{eq:mac1}. Indeed, we
  are only considering macroscopic times such that $\frac\theta\epsilon
>2\frac{\sigma_c}{m_{\dot\gamma}}$ and, in the limit $\epsilon \to 0$, the
boundary layer in time around $\theta=0$ does not affect the result.
\end{remark}

\medskip

{\em Proof}.
We first give the arguments to derive system \eqref{eq:sapprox}. 
In the proof of Theorem \ref{th:tlc}, we established equation~\eqref{eq:fpf}, which reads
\begin{align*}
   \dot f_\epsilon(t) =- f_\epsilon(t)
+  {\dot\gamma}(\epsilon t) \brk{p_\epsilon(t,\sigma_c) - p_\epsilon(t,-\sigma_c) }.
\end{align*}
Denoting $\rho$ a function of $\mathcal{D}((0,s))$, $\rho^n$ a mollifier on $\mathbb{R}$ and $\chi^n=\rho^n*\chi$, we used the test function $\eta^n(t,\sigma) = \chi^n(\sigma) \rho(t)$ in the weak form~\eqref{eq:fvferme} of equation~\eqref{edp:mic} and then passed to the limit $n\to\infty$. We now establish an equation on $\tau_\epsilon$ with the same method.

Denote $I:\sigma \mapsto \sigma,\ I^n=\rho^n* (I {\rm 1\mskip-4mu
  l}_{[-n,n]})$ and use $\eta^n=(t,\sigma) = I^n(\sigma) \rho(t)$ as a test  function in~\eqref{eq:fvferme}:
\begin{align*}
  - \int_0^s \dot \rho(t) \int I^n p_\epsilon(t,\cdot) dt 
-  \int_0^s \rho(t) {\dot\gamma}(\epsilon t) \int \dot I^n p_\epsilon(t,\cdot) dt
+ \int_0^s \rho(t) \int \chi I^n p_\epsilon(t,\cdot)dt\\
=I^n(0) \int_0^s f_\epsilon(t) \rho(t)dt.
\end{align*}
We pass to the limit $n\to\infty$, using that the function $t \mapsto
(\sigma \mapsto p_\epsilon(t,\sigma))$ belongs to ${C}([0,s],L^1)$
(see Theorem~\ref{th:exu}), $\int p_\epsilon(t,\cdot)=1$, the dominated convergence theorem for the terms in the left-hand side, and that $I^n(0)\to I(0)=0$ for the right-hand side. We obtain, for all $s>0$,
\begin{align*}
   - \int_0^s \dot \rho(t) \tau_\epsilon(t)dt
-  \int_0^s \rho(t) {\dot\gamma}(\epsilon t)dt
+ \int_0^s \rho(t) \int \chi I p_\epsilon(t,\cdot)dt = 0,
\end{align*}
so that
\begin{align} \label{eq:taueps}
  \dot \tau_\epsilon(t) = - \int \chi(\sigma) \sigma p_\epsilon(t,\sigma)d\sigma+ {\dot\gamma}(\epsilon t).
\end{align}
Changing the variable $t$ in $\frac\theta\epsilon$ in equations~\eqref{eq:fpf} and~\eqref{eq:taueps}, we obtain system~\eqref{eq:sapprox}.

Now that we have established system~\eqref{eq:sapprox}, we rewrite~\eqref{eq:sapproxa} in the form 
\begin{align*}
 \epsilon \deriv{\widetilde{\tau_\epsilon}}\theta &= - \widetilde{\tau_\epsilon} + \int_{-\sigma_c}^{\sigma_c}\sigma \widetilde{p_\epsilon}(\theta,\sigma)d\sigma +{\dot\gamma}(\theta),
\end{align*}
using the definition of~$\chi$.
Moreover, we use the expression~\eqref{eq:py} of $p_\infty$ to
rewrite~\eqref{eq:mac1} as follows
\begin{align*}
   \epsilon \deriv{ \tau_\epsilon^*}\theta &= - \tau_\epsilon^* + \int_{-\sigma_c}^{\sigma_c}\sigma  p_\infty(\theta,\sigma)d\sigma +{\dot\gamma}(\theta).
\end{align*}
Subtracting the above two equations yields
\begin{align*}
  \deriv{ \brk{\widetilde{\tau_\epsilon}-\tau_\epsilon^*}}\theta +
  \frac{\widetilde{\tau_\epsilon}-\tau_\epsilon^*}\epsilon &=
  \int_{-\sigma_c}^{\sigma_c}\sigma
  \frac{\widetilde{p_\epsilon}-p_\infty}\epsilon d\sigma.
\end{align*}
Denote $\varsigma=2\frac{\sigma_c}{m_{\dot\gamma}}$ so that $\frac\theta\epsilon>\varsigma$. We next apply the Duhamel formula and find
\begin{align} \label{eq:duhamel}
  \brk{\widetilde{\tau_\epsilon}-\tau_\epsilon^*}(\theta) = 
 \brk{\widetilde{\tau_\epsilon}-\tau_\epsilon^*}(\varsigma\epsilon)
\exp{\varsigma-\frac\theta\epsilon}
+ \int_{\varsigma\epsilon}^\theta \exp{\frac{u-\theta}\epsilon} \int_{-\sigma_c}^{\sigma_c} \sigma \ \frac{\widetilde{p_\epsilon}-p_\infty}\epsilon   d\sigma du.
\end{align}
Using the estimate~\eqref{maj:tau} on $\tau_\epsilon$, the bounds \eqref{hyp:gamma} on ${\dot\gamma}$, we find
\begin{align} \label{maj:envarsigma}
  \Abs{\widetilde{\tau_\epsilon}-\tau_\epsilon^*}(\varsigma\epsilon)
&\le
\brk{1+M_{\dot\gamma} \varsigma} C_\tau + \brk{\Abs{\tau_0}\exp{-\varsigma}+\int_0^\varsigma  \exp{u-\varsigma}
\brk{\frac{\sigma_c^2}{2(\sigma_c+m_{\dot\gamma})}+M_{\dot\gamma}}
 du}\nonumber\\
&\le K ,
\end{align}
where we recall that $K$ denotes a constant that is independent from $\theta$ and $\epsilon$ and whose precise value may change from one
occurrence to another.
Inserting the above estimate and the estimate~\eqref{conv:pmp} on $\Abs{\widetilde{p_\epsilon}-p_\infty}$ in~\eqref{eq:duhamel} yields
\begin{align}
  \Abs{\widetilde{\tau_\epsilon}-\tau_\epsilon^*}(\theta) &\le 
K\Brk{\exp{\varsigma-\frac\theta\epsilon}
+{\int_{0}^\theta \exp{\frac{u-\theta}\epsilon}
\brk{\frac{\exp{-b\frac u\epsilon}}\epsilon + \frac{\exp{-\frac u\epsilon}}\epsilon +1} du} }\label{eq:xex}\\ 
&\le K \brk{\exp{-\frac\theta\epsilon}
+ \frac1{1-b}\brk{\exp{-b\frac\theta\epsilon}- \exp{-\frac\theta\epsilon}} + \frac\theta\epsilon \exp{-\frac\theta\epsilon} +\epsilon}\nonumber\\
&\le K\brk{\frac{b}{\Abs{1-b}}\frac{\exp{-1}}{\theta} + \frac1{\Abs{1-b}}\frac{\exp{-1}}{b\theta} + \frac{4\exp{-2}}{\theta}+1}\epsilon,\nonumber
\end{align}
using that the functions $x\mapsto x \exp{-x}$ and $x\mapsto x^2 \exp{-x}$ are  respectively bounded  by ${\exp{-1}}$ and $4{\exp{-2}}$ on $\mathbb{R}_+$ in order to derive the last line.
 This concludes the proof. $\endproof$

\medskip

\begin{theorem} 
\label{th:mac2}
For $\epsilon>0$, consider $ (\tau_\epsilon^{**}, f_\epsilon^{**})$ satisfying the following system of equations in macroscopic time~$\theta$:
 \begin{subequations} \label{eq:mac2}
   \begin{empheq}[left=\empheqlbrace]{align} 
     \epsilon\deriv{\tau_\epsilon^{**}} \theta (\theta) &=  -\kappa(\theta) f_\epsilon^{**} (\theta) \tau_\epsilon^{**} (\theta) +{\dot\gamma}(\theta), \label{eq:mac2a}\\
      \epsilon\deriv{f_\epsilon^{**}} \theta  (\theta)&=-f_\epsilon^{**}  (\theta)+ \frac{{\dot\gamma}(\theta)}{\sigma_c+{\dot\gamma}(\theta)},\label{eq:mac2b}
   \end{empheq}
 \end{subequations}
where we have introduced the notation
\begin{align} \label{def:kappa}
  \kappa(\theta)=\frac2{1+\frac1{\brk{1+\frac{\sigma_c}{{\dot\gamma}(\theta)}}^2}},
\end{align}
and where the equations are
supplied with any couple of scalars (independent of $\epsilon$)
$\brk{\tau^{**}(0),f^{**}(0)}$ as initial conditions.
Consider  $\theta>0$ such that $\frac\theta\epsilon >2\frac{\sigma_c}{m_{\dot\gamma}}$.
Under the assumptions of Theorem~\ref{th:tle}, consider $p_\epsilon$ the solution of~\eqref{edp:mic}. Then, there exists a constant $C$ independent from $\epsilon$ and $\theta$ such that,
\begin{align} \label{conv:mac2}
  \Abs{\tau_\epsilon\brk{\frac{\theta}\epsilon} -  \tau_\epsilon^{**}(\theta)} 
+  \Abs{f_\epsilon\brk{\frac{\theta}\epsilon} -  f_\epsilon^{**}(\theta)} 
\le C\brk{\frac1\theta+1}\epsilon.
\end{align}
\end{theorem}

\begin{remark}
 As in Theorem~\ref{th:mac1}, the above result holds whatever the set of initial conditions $\brk{\tau^{**}(0),f^{**}(0)}$ for the system of differential equations~\eqref{eq:mac2}.
\end{remark}

\medskip

{\em Proof}.
The proof falls in three steps. We first study $f_\epsilon$, then an auxiliary function $\beta_\epsilon$ and finally $\tau_\epsilon$.
\refstepcounter{step}
\subparagraph*{Step \arabic{step}:  Approximation of $f_\epsilon^{**}$}
Applying the Duhamel formula to~\eqref{eq:mac2b} yields
\begin{align}\label{eq:duhamelf}
  f_\epsilon^{**}(\theta) &= f_0 \exp{-\frac\theta\epsilon}
+\frac1\epsilon \int_0^\theta \frac{{\dot\gamma}(v)}{\sigma_c+{\dot\gamma}(v)} \exp{\frac{v-\theta}\epsilon}dv\\
&=f_0 \exp{-\frac\theta\epsilon} 
+ \frac{{\dot\gamma}(\theta)}{\sigma_c+{\dot\gamma}(\theta)} - \frac{{\dot\gamma}(0)}{\sigma_c+{\dot\gamma}(0)} \exp{-\frac\theta\epsilon}
- \int_0^\theta  \frac{\sigma_c {\ddot\gamma}(v)}{\brk{\sigma_c+{\dot\gamma}(v)}^2}  \exp{\frac{v-\theta}\epsilon}dv.\nonumber
\end{align}
Using the Lipschitz property of ${\dot\gamma}$ and  denoting by
$f_\infty (\theta)=\int \chi(\sigma) p_\infty(\theta,\sigma) d\sigma = \frac{{\dot\gamma}(\theta)}{\sigma_c+{\dot\gamma}(\theta)}$
(following~\eqref{def:f} and~\eqref{eq:py}), we easily obtain
\begin{align} \label{conv:odef}
  \Abs{f_\epsilon^{**}- f_\infty}(\theta) \le K\brk{\exp{-\frac\theta\epsilon}+\epsilon}.
\end{align}
Collecting the above equation and the estimate~\eqref{conv:geps} established in Theorem~\ref{th:tle}, we obtain
\begin{align}
 \Abs{\widetilde{f_\epsilon} -  f_\epsilon^{**}}(\theta) 
&\le K\brk{\exp{-b\frac\theta\epsilon} +\exp{-\frac\theta\epsilon}  + \epsilon}  \label{conv:diff}\\
&\le K\brk{\frac{\exp{-1}}{b\theta} + \frac{\exp{-1}}{\theta} +1}\epsilon,  \nonumber
\end{align}
using that the function $x\mapsto x \exp{-x}$ is bounded by ${\exp{-1}}$ on $\mathbb{R}_+$.

\refstepcounter{step}
\subparagraph*{Step \arabic{step}: Introduction of the auxiliary function $\beta_\epsilon$}
We now introduce
\begin{align}
  \beta(t) = \int \chi(\sigma) \sigma p(t,\sigma) d\sigma,
\end{align}
defined for a density probability $p$ such that $\sigma p\in L^1$. 
Denote $\rho$ a function of $\mathcal{D}((0,s))$, $\rho^n$ a mollifier
on $\mathbb{R}$ and $I:\sigma\mapsto\sigma$. Using $\eta^n(t,\sigma) =
(I\chi {\rm 1\mskip-4mu
  l}_{[-n,n]})*\rho^n(\sigma) \rho(t)$  as test function in \eqref{eq:fvferme} and passing to the limit with the same arguments as in~\eqref{faible:fpf}, we obtain
\begin{align*}
  \epsilon \deriv{\widetilde{\beta_\epsilon}}\theta(\theta)  
+\widetilde{\beta_\epsilon}(\theta)
= {\dot\gamma}(\theta) \brk{\widetilde{f_\epsilon}(\theta)+ \sigma_c\ (\widetilde{p_\epsilon}(\theta,\sigma_c)+\widetilde{p_\epsilon}(\theta,-\sigma_c))}.
\end{align*}
Consider $\beta_\epsilon^*$ the solution of the ordinary differential equation
\begin{align} \label{ode:beta}
  \epsilon \deriv{\beta_\epsilon^*}\theta + \beta_\epsilon^* = {\dot\gamma}(\theta)\brk{f_\epsilon^{**} + \frac{\sigma_c}{\sigma_c+{\dot\gamma}(\theta)}}
\end{align}
supplied with a scalar $\beta_0$ as initial condition. Subtracting the
above two equations and using that (see~\eqref{eq:py}) $p_\infty(\theta,\sigma_c)=\frac{1}{\sigma_c+{\dot\gamma}(\theta)}$ and $p_\infty(\theta,-\sigma_c)=0$, we obtain
\begin{align} \label{eq:difbeta}
  \deriv{(\widetilde{\beta_\epsilon}-\beta_\epsilon^*)}\theta + \frac{\widetilde{\beta_\epsilon}-\beta_\epsilon^*}{\epsilon}
=G_\epsilon(\theta)
\end{align}
with 
\begin{align*}
G_\epsilon(\theta)={\dot\gamma}(\theta)\frac{\widetilde{f_\epsilon} - f_\epsilon^{**}}{\epsilon}(\theta) + {\dot\gamma}(\theta) \sigma_c \frac{\widetilde{p_\epsilon}-p_\infty}{\epsilon}(\theta,\sigma_c) + {\dot\gamma}(\theta) \sigma_c \frac{\widetilde{p_\epsilon}-p_\infty}{\epsilon}(\theta,-\sigma_c).
\end{align*}
Recall that $\varsigma=2\frac{\sigma_c}{m_{\dot\gamma}}$ and $\theta>\varsigma\epsilon$. Applying the Duhamel formula yields, 
\begin{align} \label{for:betaeps}
   \brk{\widetilde{\beta_\epsilon}-\beta_\epsilon^*}(\theta)
= \brk{\widetilde{\beta_\epsilon}-\beta_\epsilon^*}(\varsigma\epsilon) \exp{\varsigma-\frac\theta\epsilon} + \int_{\varsigma\epsilon}^\theta \exp{\frac{u-\theta}\epsilon} G_\epsilon(u)du.
\end{align}
Using the upper bound~\eqref{maj:tau}, $\widetilde{\beta_\epsilon}$ satisfies
  \begin{align*}
    \Abs{\widetilde{\beta_\epsilon}(\varsigma\epsilon)} \le \int \Abs{\sigma} \chi p_\epsilon(\varsigma,\cdot) \le  \int  \Abs{\sigma} p_\epsilon(\varsigma,\cdot) \le \brk{1+M_{\dot\gamma} \varsigma} C_\tau \le K.
  \end{align*}
Moreover, $\beta_\epsilon^*$ solution of \eqref{ode:beta} satisfies
  \begin{align*}
    \Abs{\beta_\epsilon^*(\varsigma\epsilon)}&\le
{\Abs{\beta_0}\exp{-\varsigma}+\int_0^\varsigma  \exp{u-\varsigma}
\brk{\brk{\Abs{f_0} + \frac{M_{\dot\gamma}}{\sigma_c+m_{\dot\gamma}}} + \frac{\sigma_c}{\sigma_c+m_{\dot\gamma}}} du}\\
&\le K.
  \end{align*}
Using~\eqref{conv:diff}, the boundedness of ${\dot\gamma}$ and the estimate
\eqref{conv:pmp} on $\widetilde{p_\epsilon}-p_\infty$ , 
the right-hand side $G_\epsilon(\theta)$ of \eqref{eq:difbeta} satisfies, 
\begin{align*}
  G_\epsilon(\theta)
\le K\brk{\frac{\exp{-b\frac\theta\epsilon}}{\epsilon}+\frac{\exp{-\frac\theta\epsilon}}{\epsilon} +1}.
\end{align*}
Inserting the three above inequalities in~\eqref{for:betaeps} implies
\begin{align} \label{conv:difbeta}
  \Abs{\widetilde{\beta_\epsilon}-\beta_\epsilon^*}(\theta)
\le K\brk{\exp{-b\frac\theta\epsilon}+\exp{-\frac\theta\epsilon} +\epsilon}.
\end{align}
Additionally, applying the Duhamel formula to~\eqref{ode:beta} and
using the explicit formula $f_\infty(\theta)=\frac{{\dot\gamma}(\theta)}{\sigma_c+{\dot\gamma}(\theta)} $ yield
\begin{align*}
  \beta_\epsilon^*(\theta) - \int_0^\theta \frac{{\dot\gamma}(u)}\epsilon\exp{\frac{u-\theta}\epsilon}du
=\beta_0 \exp{-\frac\theta\epsilon} 
+ \int_0^\theta \frac{{\dot\gamma}(u)}\epsilon \brk{f_\epsilon^{**}(u)-f_\infty} \exp{\frac{u-\theta}\epsilon}du.
\end{align*}
Using the Lipschitz property of ${\dot\gamma}$ and the estimate~\eqref{conv:odef}, we obtain
\begin{align} \label{conv:odebeta}
  \Abs{\beta_\epsilon^*-{\dot\gamma}}(\theta)
\le K\brk{\exp{-\frac\theta\epsilon} +\epsilon}.
\end{align}
Combining~\eqref{conv:difbeta} and~\eqref{conv:odebeta} leads to
\begin{align} \label{conv:beta}
  \Abs{\widetilde{\beta_\epsilon}-{\dot\gamma}}(\theta) 
\le K\brk{\exp{-b\frac\theta\epsilon}+\exp{-\frac\theta\epsilon} +\epsilon}
\end{align}
and eventually,  
\begin{align} \label{conv:tau}
  \Abs{\widetilde{\tau_\epsilon}-\tau_\infty}(\theta) 
& \le \Abs{\int \chi \sigma \brk{\widetilde{p_\epsilon} - p_\infty} }
+ \Abs{\int (1 - \chi) \sigma  \brk{\widetilde{p_\epsilon} - p_\infty} d\sigma}
\nonumber\\
&\le K\brk{\exp{-b\frac\theta\epsilon}+\exp{-\frac\theta\epsilon}+\epsilon},
\end{align}
respectively using \eqref{conv:beta} and \eqref{conv:pmp} to estimate
the two terms of the right-hand side. Here we have used the notation $\tau_\infty(\theta)=\int \sigma
p_\infty(\theta,\sigma) \, d\sigma$, and the fact that $\int
\chi(\sigma) 
\sigma p_\infty(\theta,\sigma) d\sigma={\dot\gamma}(\theta)$.

\refstepcounter{step}
\subparagraph*{Step \arabic{step}: Approximation of $\tau_\epsilon$}
We now turn to $\brk{\widetilde{\tau_\epsilon}-\tau_\epsilon^{**}}$. Combining~\eqref{eq:sapproxa} and~\eqref{eq:mac2a} yields
\begin{align}
  \epsilon \deriv{ \brk{\widetilde{\tau_\epsilon}-\tau_\epsilon^{**}}}\theta
&= + \kappa f^{**}_\epsilon \tau^{**}_\epsilon - \int \chi \sigma \tilde{p}_\epsilon\nonumber\\
&= - \kappa f_\infty \brk{\widetilde{\tau_\epsilon}-\tau_\epsilon^{**}} 
-\kappa \tau_\epsilon^{**} \brk{f_\infty-f_\epsilon^{**}} 
\nonumber\\&\quad
- \kappa f_\infty \brk{\tau_\infty-\widetilde{\tau_\epsilon}}
- \brk{-\kappa f_\infty \tau_\infty + \int \chi \sigma p_\infty}
- \int \chi \sigma (\widetilde{p_\epsilon} - p_\infty).\nonumber
\end{align}
From the formula~\eqref{eq:py} on $p_\infty$, we compute 
$\int \chi \sigma p_\infty(\theta,\cdot)={\dot\gamma}(\theta)$,
$f_\infty(\theta)=\frac{{\dot\gamma}(\theta)}{\sigma_c+{\dot\gamma}(\theta)}$, 
$\tau_\infty(\theta)=\frac12\brk{\frac{{\dot\gamma}(\theta)^2}{\sigma_c+{\dot\gamma}(\theta)}+\sigma_c+{\dot\gamma}(\theta)}$ so that
the term $\displaystyle -\kappa f_\infty \tau_\infty + \int \chi \sigma p_\infty$ cancels out because of the
definition~\eqref{def:kappa} of $\kappa$. We therefore obtain
\begin{align} \label{reec}
 \deriv{ \brk{\widetilde{\tau_\epsilon}-\tau_\epsilon^{**}}}\theta + \frac{\kappa f_\infty}\epsilon \brk{\widetilde{\tau_\epsilon}-\tau_\epsilon^{**}} 
=H_\epsilon(\theta)
\end{align}
with
\begin{align*}
H_\epsilon(\theta)
=- \kappa \tau_\epsilon^{**} \frac{f_\infty-f_\epsilon^{**}}\epsilon(\theta)
- \kappa f_\infty \frac{\tau_\infty-\widetilde{\tau_\epsilon}}\epsilon(\theta)
- \int \chi(\sigma) \sigma \frac{\widetilde{p_\epsilon} - p_\infty}\epsilon(\theta,\sigma)d\sigma.
\end{align*}
 We have $\kappa \ge 1$ (see~\eqref{def:kappa})  so that
\begin{align}
  \kappa  f_\infty \ge \frac1{1+\frac{\sigma_c}{m_{\dot\gamma}}} .
\end{align}
The Duhamel formula then implies 
\begin{align} \label{for:taueps2}
   \Abs{\widetilde{\tau_\epsilon}-\tau_\epsilon^{**}}(\theta)
\le \Abs{\widetilde{\tau_\epsilon}-\tau_\epsilon^{**}}(\varsigma\epsilon) \exp{\frac1{1+\frac{\sigma_c}{m_{\dot\gamma}}}\brk{\varsigma-\frac\theta\epsilon}} + \int_{\varsigma\epsilon}^\theta \exp{\frac1{1+\frac{\sigma_c}{m_{\dot\gamma}}}\frac{u-\theta}\epsilon} \Abs{H_\epsilon(u)}du
\end{align}
Using the upper bound~\eqref{maj:tau}, $\widetilde{\tau_\epsilon}$ satisfies
  \begin{align*}
    \Abs{\widetilde{\tau_\epsilon}(\varsigma\epsilon)} \le  \brk{1+M_{\dot\gamma} \varsigma} C_\tau\le K.
  \end{align*}
Moreover, the solution $\tau_\epsilon^{**}$ of \eqref{eq:mac2}
satisfies (using the non negativity of $\kappa$ and $f_\epsilon^{**}$):
\begin{align*}
\tau_\epsilon^{**} (\theta) &= \exp{-\frac{1}{\epsilon}\int_0^\theta \kappa(s)
f_\epsilon^{**}(s) \, ds}
\tau^{**} (0) + \frac{1}{\epsilon}
\int_0^\theta {\dot\gamma}(s) \exp{-\frac{1}{\epsilon}\int_s^\theta \kappa(r)
f_\epsilon^{**}(r) \, dr} \, ds\\
& \le \tau^{**} (0) + M_{\dot\gamma} \frac{\theta}{\epsilon}
\end{align*}
so that
  \begin{align*}
    \Abs{\tau_\epsilon^{**}(\varsigma\epsilon)}&\le K (1+\varsigma).
  \end{align*}
Collecting~\eqref{conv:geps} established in
Theorem~\ref{th:tle} , \eqref{conv:beta} and \eqref{conv:tau},
the right-hand side $H_\epsilon(u)$ of~\eqref{reec} satisfies
\begin{align*}
H_\epsilon(u)
\le K\brk{\frac{\exp{-b\frac u\epsilon}}\epsilon+\frac{\exp{-\frac u\epsilon}}{\epsilon}+1}.
\end{align*}
Inserting the three above inequalities in~\eqref{for:taueps2} implies
\begin{align*}
  \Abs{\widetilde{\tau_\epsilon}-\tau_\epsilon^{**}}(\theta)
&\le K\brk{ \exp{-\frac1{1+\frac{\sigma_c}{m_{\dot\gamma}}}\frac\theta\epsilon}
+ \int_{0}^\theta \exp{\frac1{1+\frac{\sigma_c}{m_{\dot\gamma}}}\frac{u-\theta}\epsilon} \brk{\frac{\exp{-b\frac u\epsilon}}\epsilon+\frac{\exp{-\frac{u}\epsilon}}{\epsilon}+1} du}\\
&\le K\brk{\frac1\theta+1}\epsilon. 
\end{align*}
\hfill\endproof

\setcounter{step}{0}

We end this section with a discussion on the two macroscopic
limits~\eqref{eq:mac1} and~\eqref{eq:mac2}
we have obtained. First, as mentioned above, there are many ways to
close the system~\eqref{eq:sapprox} in the limit $\epsilon \to 0$. We have 
proposed here two possible macroscopic limits, which are indeed close up to terms of
order $O(\epsilon)$ to the original problem~\eqref{edp:mic}.

Second, we would like to argue that the system~\eqref{eq:mac2}
derived in Theorem~\ref{th:mac2} is physically more relevant. 
Indeed, up to changing the coefficient $\kappa(\theta)$ defined
by~\eqref{def:kappa} by a constant, this system belongs to a class of
equations introduced in~\cite{derec-01,Picard2002} to model the
evolution of aging fluids. These equations read (see~\cite[Eq. (1)]{Picard2002})
 \begin{subequations} \label{eq:derec}
   \begin{empheq}[left=\empheqlbrace]{align} 
     \pd{\tau}{t}&=-f\tau + {\dot\gamma} ,\label{eq:derectau}\\
     \pd{f}{t} &= - U(f) + V(f,\tau,{\dot\gamma}), \label{eq:derecf}
   \end{empheq}
 \end{subequations}
where $U$ and $V$ are positive functions.  The formal similarity between~\eqref{eq:mac2} and~\eqref{eq:derec} is clear.

For this class of systems, the
fluidity $f$ appears as the inverse of the relaxation time for the
stress $\tau$ in equation~\eqref{eq:derectau}. In equation
\eqref{eq:derecf} the evolution results from the competition between
the two terms with opposite signs. 
Aging, meaning solidification of the fluid, is modeled by the
negative term. It makes the fluidity decrease so that the relaxation
phenomenon is slower with time. The opposite effect,
flow-induced rejuvenation, is modeled by the positive term, which
makes the fluidity (the inverse relaxation time) increase.

Note that the assumption $\kappa(\theta)$ constant is a reasonable
approximation when ${\dot\gamma}$ is small. In this case,
system~\eqref{eq:mac2}  is close to system~\eqref{eq:macc}, which is
indeed of the form~\eqref{eq:derec}. In section~\ref{sec:numerics}, we
present numerical results that confirm that the solutions
to~\eqref{eq:mac2} and~\eqref{eq:macc} are indeed close when ${\dot\gamma}$
is small.

\section{Numerical experiments}
\label{sec:numerics}

This section is devoted to some numerical experiments. We consider three
different situations, depending on the value of the function ${\dot\gamma}(t)$
for $t\in [0,T]$.
\begin{itemize}
\item[(i)] In our first series of tests, we consider the \emph{constant} shear rate
${\dot\gamma}(t)\equiv {\dot\gamma}_\infty$, for different values $0.1, 0.2, \cdots, 0.8$ of ${\dot\gamma}_\infty$. In that case, the final time is $T=40$
\item[(ii)] In our second series of tests, we consider ${\dot\gamma}(t)=t$,
  and the final time  $T=\frac\theta\epsilon$ with $\theta=1$ and
  $\epsilon$ varying between the values $0.005$ and $0.05$.
\item[(iii)] In our third and final series of tests, we take
  ${\dot\gamma}(t)=0.01\cdot t$ , and the same values of $T$, $\theta$, and
  $\epsilon$ as in case (ii).
\end{itemize}
In all our tests, the reference equation, namely~\eqref{edp:p} (or more precisely~\eqref{edp:mic}), is
simulated over the time interval~$[0,T]$.  Since in theory it is posed
on the whole real line, we need to truncate the domain and thus actually
solve the equation on the bounded interval
$\sigma\in [-M_\sigma,M_\sigma]$ (with periodic boundary conditions), for $M_\sigma=10$, with a constant
space step~$\Delta\sigma=\frac{M_\sigma}{2.10^5}$. The initial condition~$p_0$
is the normal centered Gaussian density, appropriately renormalized on
the interval $[-M_\sigma,M_\sigma]$. The threshold value for the
stress is $\sigma_c=2$. The time discretization is performed using a
splitting method: over the time interval $[n \Delta t, (n+1) \Delta t]$,
 \begin{subequations} 
   \begin{empheq}[left=\empheqlbrace]{align} 
       \pd{p^{n+\frac12}}{t} &= -\chi p^{n+\frac12} + \brk{\int \chi p^{n+\frac12}} \delta_0 \label{dis:source}\\
  \pd{p^{n+1}}{t} + {\dot\gamma}(\epsilon n \Delta t) \pd{p^{n+\frac12}}{\sigma} &= 0.\label{dis:adv}
  \end{empheq}
 \end{subequations}
Equation~\eqref{dis:source} is solved explicitly, pointwise
for each~$\sigma\not=0$, while the equation for $\sigma=0$ is indeed solved
using the conservation of the total mass of the density $p$.
In short, the value of the density at zero is adjusted so that the integral of $p$ is one. See~\cite{gati} for more details.
The advection equation~\eqref{dis:adv} is solved using an upwind finite
difference scheme.
All computations are performed using C++.

\medskip

\begin{figure} [ht]
  \centering 
   \includegraphics[width=.65\textwidth]{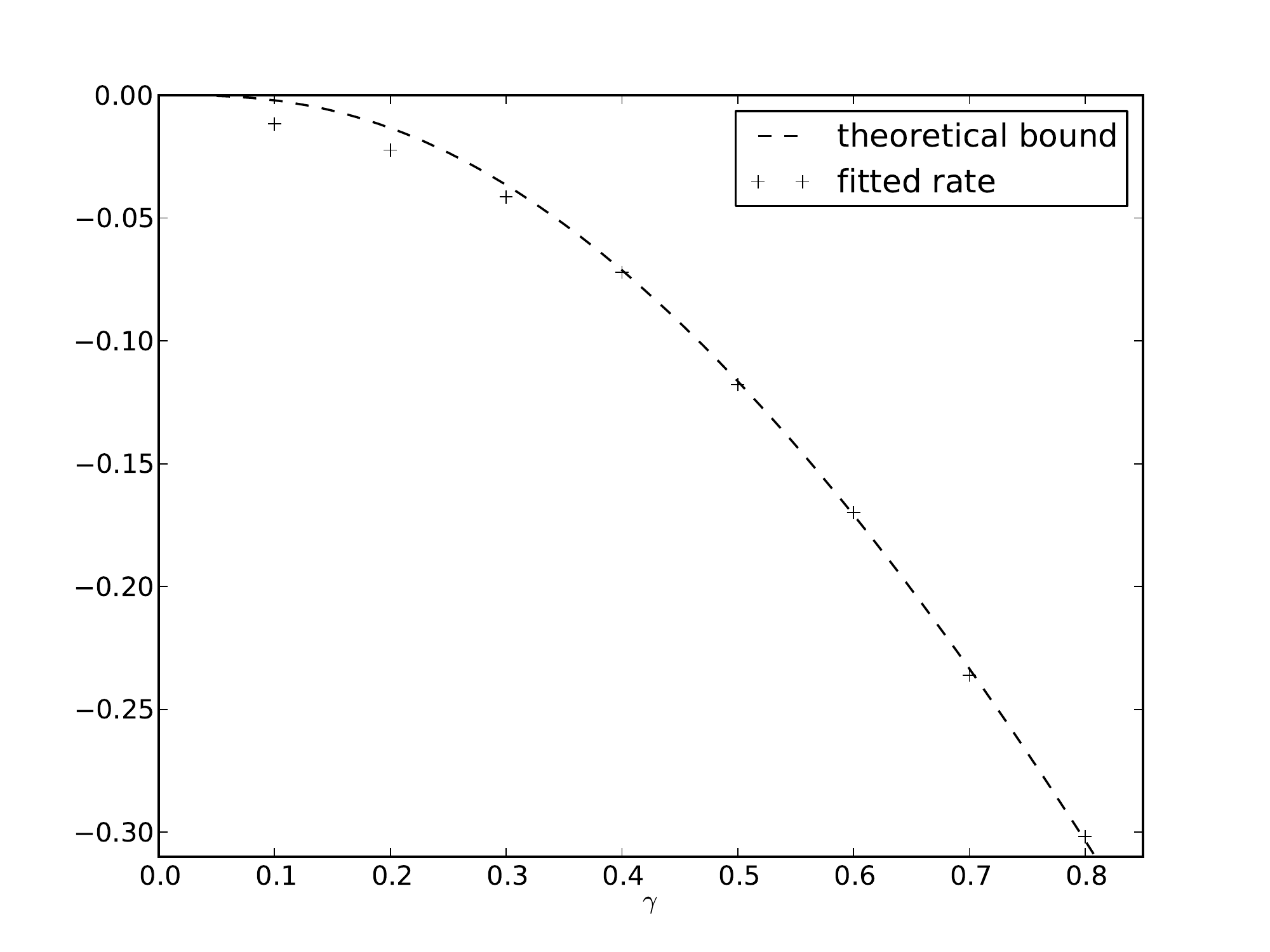}
  \caption{Exponential convergence rate function of ${\dot\gamma}(t)={\dot\gamma}_\infty$}
  \label{fig:agamma}
\end{figure}

\noindent (i) For our first series of tests, performed for the shear
rate~${\dot\gamma}(t)={\dot\gamma}_\infty$, we first check the optimality of the
long time convergence result stated in Theorem~\ref{th:tlc}. For the various
values of ${\dot\gamma}_\infty$ indicated above, we
simulate~\eqref{edp:p}. Using a least-square fit, we then estimate, in
function of ${\dot\gamma}_\infty$,  the
exponent of the exponential rate of convergence of~$\left(\int
  (p(t,\cdot)-p_\infty)^2\right)^{1/2}$  to zero as time
goes to infinity, more precisely what corresponds to the parameter~$b$
of the right-hand side of~\eqref{eq:exp-rate}. The function~$p_\infty$ is of course the 
stationary solution~\eqref{eq:py}, itself function of~${\dot\gamma}_\infty$,  explicitly determined in our theoretical
study. Note that for this practical
experiment we intentionally make a confusion between the rate of
convergence of the pointwise difference $p(t,\cdot)-p_\infty$
in~\eqref{eq:exp-rate} and its $L^2$ norm. Figure~\ref{fig:agamma} shows, as a function of ${\dot\gamma}_\infty$, a comparison between the rate of
convergence fitted on the numerical results and the theoretical value of
the inferior bound on this rate of convergence provided by our
theoretical estimate~\eqref{max:alpha}. The two sets of data agree, thereby showing the quality of our estimate~\eqref{max:alpha}.

\medskip

\begin{figure} [ht]
  \centering 
  \includegraphics[width=.65\textwidth]{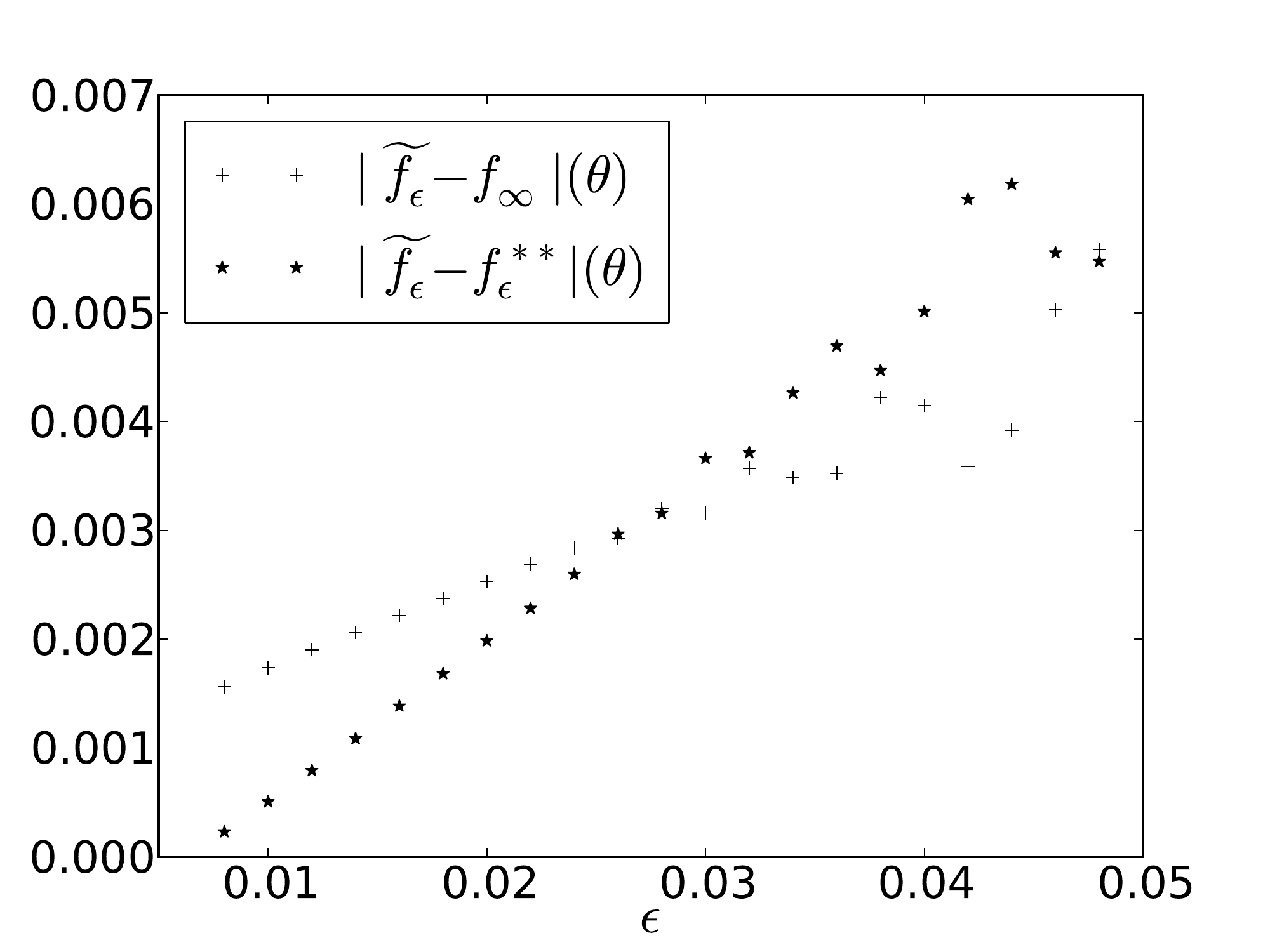}
  \caption{Convergence of $f$ }
  \label{fig:directfk_f}
\end{figure}

\begin{figure} [ht]
  \centering 
  \includegraphics[width=.65\textwidth]{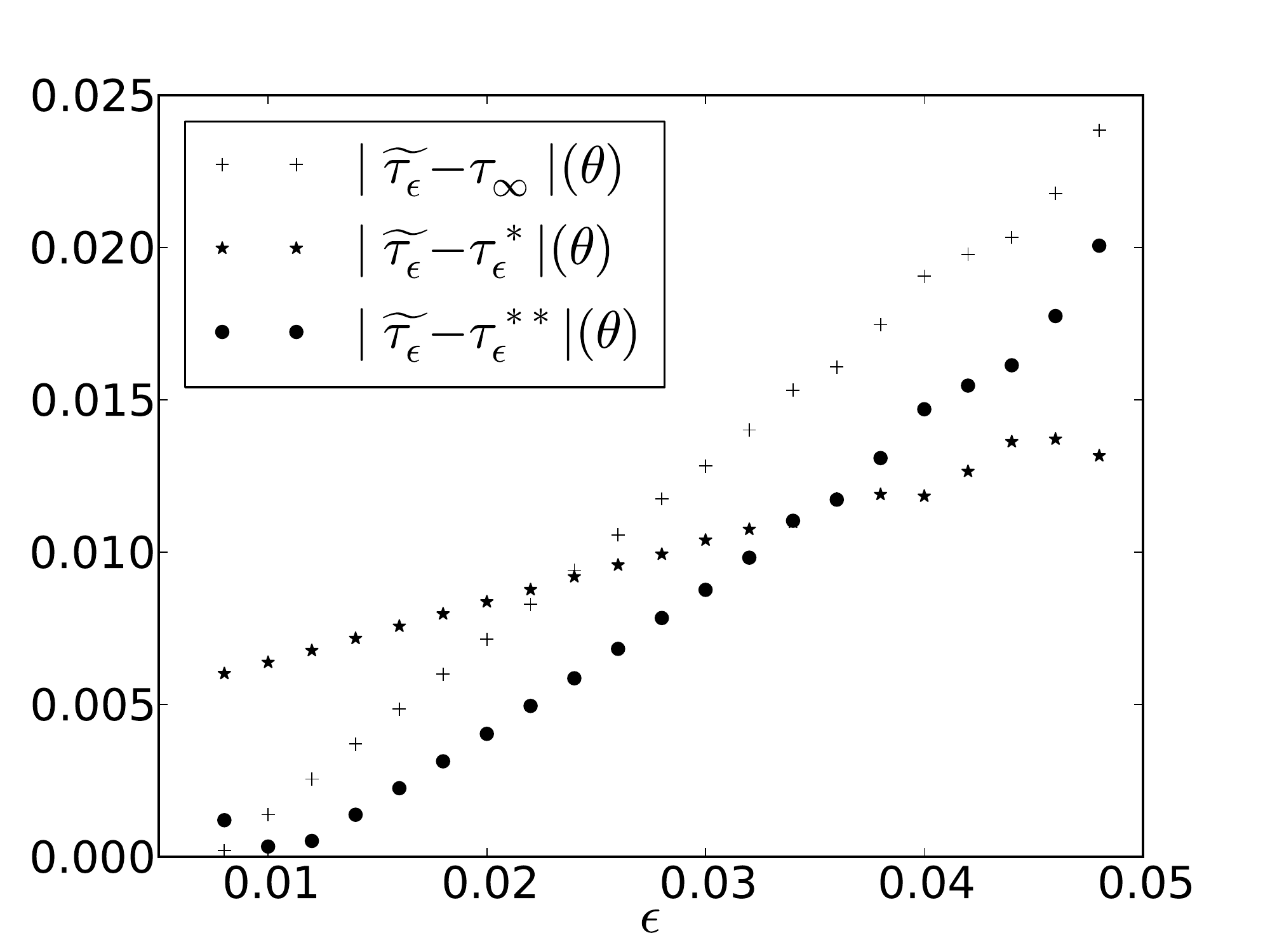}
  \caption{Convergence of $\tau$}
  \label{fig:directfk_tau}
\end{figure}

\noindent (ii) We next consider ${\dot\gamma}(t)=t$, and implement the change
of time scale by taking ${\dot\gamma}(\epsilon\,t)$ as input for
equation~\eqref{edp:p}, that is, we solve equation~\eqref{edp:mic}. We
do this for various values of the small parameter~$\epsilon$. In this
case (ii), our purpose
is twofold. First, we consider the quantities
\begin{align*}
  \Abs{\widetilde{f_\epsilon}-f_\infty}(\theta) = \Abs{\int \chi p_\epsilon\brk{\frac\theta\epsilon,\cdot} - \int \chi p_\infty(\theta,\cdot)}
\end{align*}
and
\begin{align*}
  \Abs{\widetilde{\tau_\epsilon}-\tau_\infty}(\theta) = \Abs{\int \sigma p_\epsilon\brk{\frac\theta\epsilon,\cdot} - \int \sigma p_\infty(\theta,\cdot)}
\end{align*}
where $p_\epsilon$ and $p_\infty$ are the solutions to~\eqref{edp:mic}
and \eqref{edp:mics}, respectively.
We wish to check that, as the estimate proved in
Theorem~\ref{th:tle} suggests, these two quantities behave linearly in function
of $\epsilon$ for small $\epsilon$.
The figures~\ref{fig:directfk_f} and~\ref{fig:directfk_tau} show this is
indeed the case. 

Our second purpose is to illustrate that the macroscopic equations
obtained in Theorems~\ref{th:mac1} and~\ref{th:mac2} indeed behave
similarly to the original mesoscopic equation~\eqref{edp:mic}, that is,
provide a fluidity~$f$ and a stress~$\tau$ comparable to that computed
from the solution to the latter equation. For
the different values of~$\epsilon$ mentioned above, we therefore simulate
equations~\eqref{eq:mac1} and~\eqref{eq:mac2}  on the time
interval~$\Brk{0,\theta}$, with the time step~$\Delta
t=\frac{T}{1.10^4}$. The values of
$\Abs{\widetilde{\tau_\epsilon}- \tau_\epsilon^*}(\theta)$,
$\Abs{\widetilde{f_\epsilon}- f_\epsilon^{**}}(\theta)$
et  
$\Abs{\widetilde{\tau_\epsilon}- \tau_\epsilon^{**}}(\theta)$
are displayed on Figures~\ref{fig:directfk_f} and~\ref{fig:directfk_tau}.
We observe that the convergence is linear in~$\epsilon$, as predicted by
our theoretical results Theorems~\ref{th:mac1} and~\ref{th:mac2}. The
macroscopic behavior is thus suitably reproduced, up to an error of
size~$O(\epsilon)$. 

\medskip

\begin{figure} [ht]
  \centering 
  \includegraphics[width=.65\textwidth]{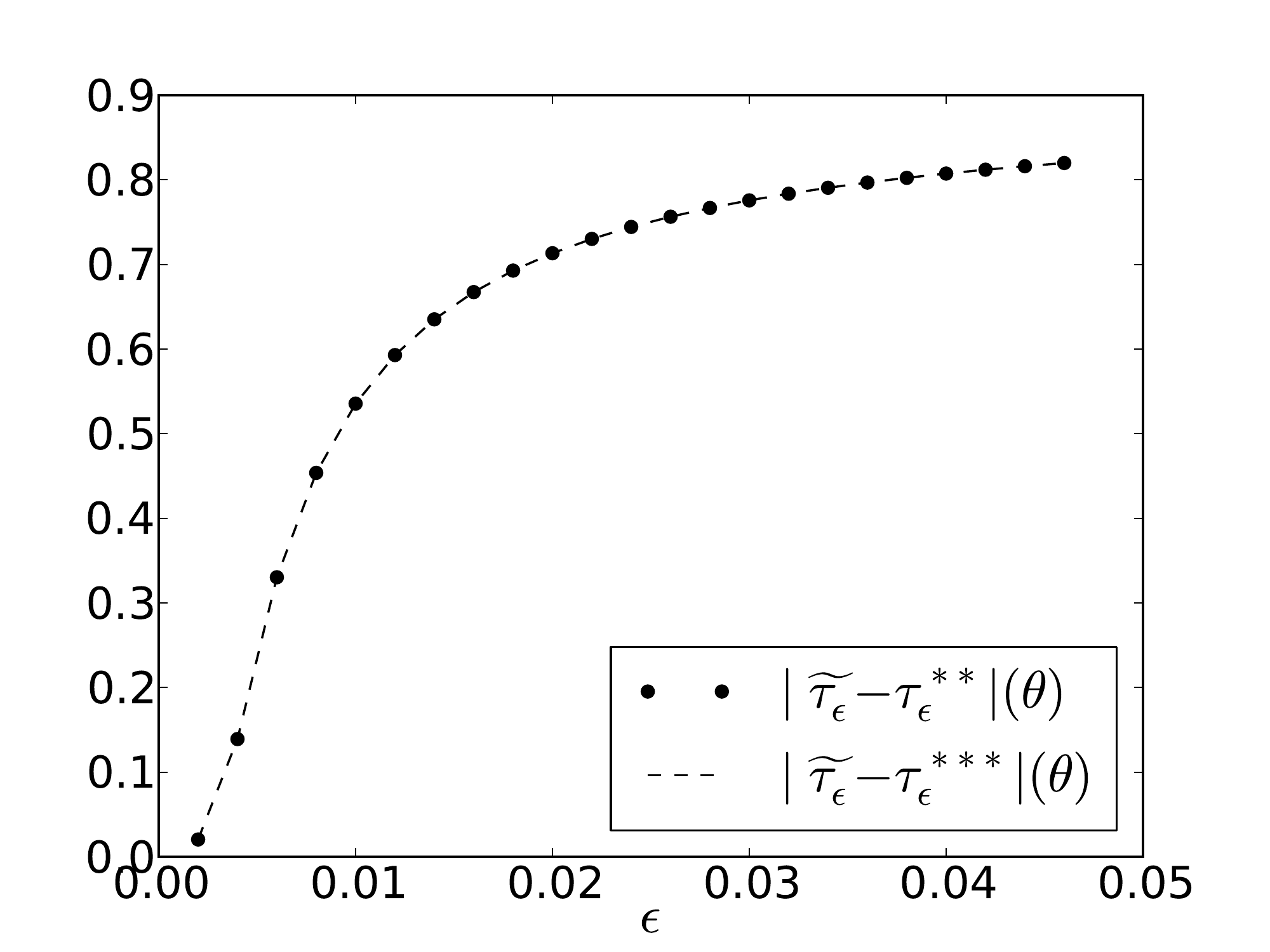}
  \caption{Convergence of $\tau$ in the case ${\dot\gamma}$ small}
  \label{fig:directfk001_tau}
\end{figure}

\noindent (iii) Our final test case addresses the case
${\dot\gamma}(t)=0.01\cdot t$. We again rescale the time and consider
${\dot\gamma}(\epsilon t)$. A similar experiment as that performed in the
previous case (ii) again shows that  equation~\eqref{eq:mac2} reproduces well
the stress tensor computed from the solution to
equation~\eqref{edp:mic}, for the different values of
$\epsilon$. Simulating~\eqref{edp:mic} 
and~\eqref{eq:mac2}, we compute  
$\Abs{\widetilde{\tau_\epsilon}-\tau_\epsilon^{**}}(\theta)$. 
The results are displayed on Figure~\ref{fig:directfk001_tau}.

But our purpose here is also to illustrate another fact. When ${\dot\gamma}$
is small, and it is indeed the case for our specific choice of ${\dot\gamma}$
in this case (iii), the value of the parameter~$\kappa(\theta)$ defined
by~\eqref{def:kappa} and appearing in the macroscopic
system~\eqref{eq:mac2} is approximately~2. System~\eqref{eq:mac2}
is thus close to the system 
 \begin{subequations} 
\label{eq:macc}
   \begin{empheq}[left=\empheqlbrace]{align} 
     \epsilon\deriv{\tau_\epsilon^{***}} \theta &=  -2 f_\epsilon^{***} \tau_\epsilon^{***} +{\dot\gamma}(\theta) \\
      \epsilon\deriv{f_\epsilon^{***}} \theta &=-f_\epsilon^{***} + \frac{{\dot\gamma}(\theta)}{\sigma_c+{\dot\gamma}(\theta)}.
   \end{empheq}
 \end{subequations}
As explained above, this system of differential equations belongs to the class of systems~\eqref{eq:derec}
explicitly suggested in~\cite{derec-01,Picard2002} as a macroscopic system of
evolution of $f$ and $\tau$ for a non Newtonian aging fluid. Our theoretical results of the previous sections can
therefore be interpreted as a \emph{derivation}, from a model at a finer
scale,  of the macroscopic
system~\eqref{eq:macc}, present in the applicative literature. On
Figure~\ref{fig:directfk001_tau}, we indeed check that the stresses solution to the
systems~\eqref{eq:mac2} and~\eqref{eq:macc} are close, up to an error of
size $O(\epsilon)$, to the stress provided by~\eqref{edp:mic},


\textbf{Acknowledgments}: We thank the two anonymous referees for their many constructive remarks, and in particular for the alternate proof outlined in Remark~\ref{rem:alternate}.


\begin{thebibliography}{99}

\bibitem{Bardet2013}
{\sc J.-B. Bardet, A. Christen, A. Guillin, F. Malrieu,
  and P.-A. Zitt}, {\em Total variation estimates for the {TCP}
  process}, Electron. J. Probab., 18 (2013), pp.~no. 10, 21.

\bibitem{Bellman1963}
{\sc R. Bellman and K.~L. Cooke}, {\em Differential-difference
  equations}, Academic Press, New York, 1963.

\bibitem{these}
{\sc D. Benoit}, {\em Various theoretical and numerical issues related to the simulation of non-Newtonian fluids}, PhD thesis, Universit\'e Paris Est, 2014, \url{http://tel.archives-ouvertes.fr/tel-00973407/en}.

\bibitem{Bocquet2009}
{\sc L.~Bocquet, A.~Colin, and A.~Ajdari}, {\em Kinetic theory of plastic flow
  in soft glassy materials}, Phys. Rev. Lett., 103 (2009), p.~036001.

\bibitem{Cances2005}
{\sc E.~Canc\`es, I.~Catto, and Y.~Gati}, {\em Mathematical analysis of a
  nonlinear parabolic equation arising in the modelling of non-Newtonian
  flows}, SIAM Journal on Mathematical Analysis, 37 (2005), pp.~60--82.

\bibitem{cances2006}
{\sc E.~Canc\`es and C.~Le~Bris}, {\em Convergence to equilibrium of a
  multiscale model for suspensions}, Discrete Contin. Dyn. Syst. Ser. B, 6
  (2006), p.~449--470.

\bibitem{derec-01}
{\sc C.~Derec, A.~Ajdari, and F.~Lequeux}, {\em Rheology and aging:
  a simple approach}, Eur. Phys. J. E, 4 (2001), pp.~355--361.

 \bibitem{LeBris}
{\sc E.~Canc\`es,  I.~Catto, Y.~Gati and C.~Le~Bris}, {\em Well-posedness of a multiscale model for concentrated suspensions}
SIAM Multiscale Modeling and Simulation, vol.~4, No.~4, pp~1041-1058,
2005.

\bibitem{gati}
{\sc Y.~Gati}, {\em  Numerical simulation of a micro-macro model of concentrated suspensions}, Int. J. Numer. Meth. Fluids, (2005), pp.~1019--1025

\bibitem{Hale1993}
{\sc J.K. Hale and S.M.V. Lunel}, {\em Introduction to Functional Differential
  Equations}, no.~vol.~99 in Applied Mathematical Sciences, Springer, 1993.

\bibitem{Hebraud1998}
{\sc P.~H\'ebraud and F.~Lequeux}, {\em Mode-coupling theory for the pasty
  rheology of soft glassy materials}, Phys. Rev. Lett., 81 (1998),
  pp.~2934--2937.

\bibitem{Picard2002}
{\sc G.~Picard, A.~Ajdari, L.~Bocquet, and F.~Lequeux}, {\em Simple model for heterogeneous
  flows of yield stress fluids}, Phys. Rev. E, 66 (2002), p.~051501.



\end{thebibliography}
\end{document}